\documentclass[a4paper,12pt]{article}
\usepackage{amsmath,amsthm,amsfonts,amssymb,bm,mathrsfs}
\usepackage[protrusion=true,expansion=true]{microtype}
\usepackage{graphicx}
\usepackage{enumerate}
\usepackage{slashed}
\usepackage{wrapfig}
\usepackage{bibentry,natbib}
\usepackage[T1]{fontenc}
\usepackage[section]{placeins}
\usepackage{footmisc}
\usepackage{sgame}
\usepackage{color}
\usepackage{multicol}
\usepackage{tikz}
\usepackage{subfloat}
\usetikzlibrary{decorations.pathreplacing}
\usepackage{subcaption}
\usepackage{fancyref}
\usepackage[colorlinks=true, linkcolor=blue, urlcolor=blue, citecolor=blue]{hyperref}
\usepackage{amsmath}
\usepackage{amsfonts}
\usepackage{amssymb}
\usepackage{lmodern}
\usepackage{enumitem}
\usepackage{setspace}
\providecommand{\U}[1]{\protect\rule{.1in}{.1in}}

\allowdisplaybreaks
\usetikzlibrary{matrix,arrows,decorations.pathmorphing}

\newtheorem{theorem}{Theorem}

\newtheorem*{axiom*}{Axiom}

\newtheorem{definition}{Definition}

\newtheorem{lemma}{Lemma}

\newtheorem{proposition}{Proposition}
\newtheorem*{proposition*}{Proposition}
\newtheorem{corollary}{Corollary}

\newtheorem{apptheorem}{Theorem}

\newcommand{\RNum}[1]{\uppercase\expandafter{\romannumeral #1\relax}}

\newcommand{\Rnum}[1]{\lowercase\expandafter{\romannumeral #1\relax}}

\usepackage{geometry}
\begin{document}

\title{\Large \textbf{Revealed Convex Preference}}
	
\author{John K.-H. Quah  
\and Rui Tang \thanks{\linespread{1} \selectfont \!We thank Peter Caradonna, Yeon-Koo Che, Rahul Deb, Federico Echenique, Qingmin Liu, Fabio Maccheroni, Luciano Pomatto,  Harry Pei, Wolfgang Pesendorfer, David Rahman, Kota Saito, Marciano Siniscalchi,  Satoru Takahashi, Norio Takeoka,  Leeat Yariv, and participants of the HEAT conference at Huazhong University of Science and Technology, 2026 PHBS Economic Theory Workshop, 2026 Hitotsubashi Summer Institute Microeconomic Theory Conference, 2026 Risk, Uncertainty and Decision Conference at Bocconi University, and seminars at Caltech, Columbia University,  Princeton University, and Hong Kong Baptist University for helpful comments. Tang acknowledges the support of the General Research Fund 16500425 by the Hong Kong Research Grants Council and the Lee Heng Fellowship. \textbf{Quah}: National University of Singapore, Department of Economics, ecsqkhj@nus.edu.sg; \textbf{Tang}:  Hong Kong University of Science and Technology, Department of Economics, ruitang@ust.hk. }}
	
\maketitle
	
\begin{abstract} \noindent  Convexity of preferences is a pervasive assumption in economic analysis, so it is important to understand how this property is revealed in data. In this paper, we develop a method to test for convexity given partial preference information in the form of weak and strict incomplete orders. We characterize the pairs that can be extended to convex preferences and provide an efficient algorithm to verify this property. Requiring the convex preference to align with an order that satisfies independence, such as the Euclidean order or stochastic dominance, does not complicate the test. We apply our results to consumer demand, choice under risk and uncertainty, and preferences over information, and establish conditions under which upper semicontinuity or continuity can also be obtained. We identify environments in which convexity imposes no additional restrictions beyond rationality, extending Afriat-type results to infinite data and co-convex budgets and establishing a convex version of exact rationalization.  \bigskip

		
	\end{abstract}

	\newpage
	\baselineskip=18pt

\section{Introduction}
	
The convexity assumption on preferences, which posits that an agent who is indifferent between two choices will prefer a mixture of the two, plays a pervasive role in economic analysis.  For example, in general equilibrium theory, this assumption on agents' preferences ensures that the demand correspondence is convex-valued, which is in turn required for equilibrium existence; in game theory, the convexity of players' preferences helps to guarantee the existence of a Nash equilibrium by guaranteeing that the best response correspondence is convex-valued. In other models, where the preference is derived from more primitive features, convexity arises as an important (and sometimes, characteristic) property.  For example, it is well-known that the reverse of a consumer's indirect preference over prices for different goods (keeping the consumer's income fixed) is necessarily a convex preference.\medskip  
	
While there is a large literature spelling out precisely when choice data are consistent with a preference, the issue of when that preference can be guaranteed to be convex has been explored a lot less. In this paper, we help to fill this lacuna by developing necessary and sufficient conditions under which choice data collected from an agent are consistent with a convex preference.\medskip         
	
We assume that the space of alternatives is a convex subset $X$ of a real vector space. We model known information about the decision maker's preference as a pair of binary relations $p=(R, P)$ on $X$, which we call an \textit{order pair}. The relation $R$ records weak preference information and $P$ records strict preference information: $x R y$ (respectively, $x P y$) means that the decision maker (DM) weakly (respectively, strictly) prefers $x$ to $y$. This framework accommodates both observed choice behavior and additional restrictions on preference. For example, in the $n$-dimensional consumption space $\mathbb{R}^n_+$, if a consumer is observed to choose a bundle $x$ from a budget set $A$ and the researcher additionally requires strict monotonicity on the consumer's preference, then the data imply $x R y$ for all $y \in A$, while strict monotonicity can be imposed by taking $P$ to be the strict Euclidean order $>_n$.\footnote{\,For any $x$ and $y$ in $\mathbb{R}^n_+$, $x>_n y$ if $x \neq y$ and for all $k \in \{1,...,n\}, x_k \ge y_k.$}  \medskip
	
A preference $\succsim$ is a complete and transitive binary relation. The preference $\succsim$ \textit{extends} an order pair $p=(R,P)$, or is a \textit{completion} of $p$, if $R \subseteq\, \succsim$ and $P \subseteq\, \succ$, where $\succ$ is the asymmetric part of $\succsim$. A preference is a \textit{convex completion} of $p$ if it is convex and  extends $p$. Our main result, Theorem \ref{thm:main}, characterizes the order pairs that admit convex completions: An order pair admits a convex completion if and only if it satisfies the \textit{separating property} (SP) on every finite menu.\footnote{\,A menu refers to a nonempty subset of alternatives in $X$.}\medskip    
	
The key idea behind SP is to use convexity to construct, for each finite menu, the largest set of alternatives that must be weakly preferred to the menu’s least preferred alternative. Suppose the DM {\em does} have a convex preference.   Then a given finite menu $A$ will contain the DM's least preferred alternative in $A$; let $x^*$ be that alternative. Since the DM's preference is convex, every alternative in the convex hull of $A$, denoted $A^0$, must be weakly preferred to $x^*$. We then use the weak component $R$ of the order pair $p$ to expand $A^0$: If $x R y$ for some $y \in A^0$, then $x$ must also be weakly preferred to $x^*$. Taking the convex hull after adding these alternatives yields a set $A^1$ which contains $A^0$. Repeating this procedure generates an increasing sequence $(A^k)_{k \ge 0}$. The set $B=\bigcup_{k=0}^{\infty} A^k$ is the maximal set whose elements must all be weakly preferred to $x^*$. Hence, we cannot have $x^*Px$ for any $x \in B$. This motivates our formulation of the separating property (SP):\, $p$ satisfies the separating property on $A$ if there exists $x^* \in A$ such that for all $x \in B$, not $x^*P x$.  We have just shown that {\em if $p$ admits a convex completion then the SP property holds on every finite menu $A$}; Theorem \ref{thm:main}  states that the converse is also true.\medskip 
	
Let $G_P=\{x\in X:xPy\mbox{ for some $y$}\}$. Notice that if $A$ contains an element that is not in $G_P$, then SP is satisfied on $A$ trivially.  Therefore, when $G_P$ is finite, checking whether $p$ admits a convex completion requires checking SP on up to $2^{|G_P|}$ menus contained in $G_P$. Theorem \ref{thm:algorithm} proposes an algorithm that simplifies this task: To verify that all finite menus contained in $G_P$ satisfy SP, it suffices to check SP along a decreasing sequence (in the sense of set inclusion) of menus in $G_P$. Consequently, verifying whether $p$ admits a convex completion requires at most $|G_P|$ SP checks, which is practicable in most empirical contexts.\medskip

In many applications, it is natural to impose additional conditions on the DM's preference besides convexity. For instance, we may require the DM's preference over consumption bundles to respect the Euclidean order; in the context of choice over risky outcomes, we may require the DM's preference over lotteries to respect first- or second-order stochastic dominance. These restrictions can be captured by appropriately chosen orders on $X$. Since these orders typically contain infinitely many comparisons, imposing them may make the relevant set of alternatives infinite and thus appear to require infinitely many SP checks. Theorem \ref{thm:porder} shows that when the imposed orders satisfy \textit{independence}, this apparent complication can be avoided. It provides a characterization for the existence of a convex completion of an order pair $p=(R,P)$ that also agrees with the imposed order. This characterization takes the form of a modified SP condition that, like the basic SP condition, is checkable in practice. \medskip

In Section \ref{subsec:non-linear-budget}, we consider a finite data set $\{(x^t, B^t)\}_{t \in T}$ in the $n$-dimensional consumption space. Each observation $(x^t, B^t)$ includes a budget set $B^t$ and a bundle $x^t$ chosen by the consumer from $B^t$. Afriat's Theorem (\cite{afriat1967construction}) and its generalizations (for example, \cite{fm09} and \cite{nishimura2017comprehensive}) provide conditions under which such a data set is {\em rationalizable}, in the sense that the consumer has a preference, strictly increasing in the Euclidean order, such that each bundle $x^t$ maximizes the preference on the budget set $B^t$. It is well known that when the sets $B^t$ are linear budget sets, any rationalizable data set is also rationalizable by a continuous and quasiconcave utility function (equivalently, by a continuous and convex preference). When budget sets are nonlinear, because of nonlinear pricing or for some other reason, there is a distinction between rationalizability and rationalizability by a convex preference, with the latter requiring strictly stronger conditions. We apply Theorem \ref{thm:porder} to formulate necessary and sufficient conditions for the rationalizability of $\{(x^t, B^t)\}_{t \in T}$ by a convex preference. Although continuity can no longer be guaranteed, we show that the rationalizing utility function can be chosen to be \textit{upper semicontinuous}. This conclusion follows from Theorem \ref{thm:convex:uppersemi}, a general mathematical result establishing conditions under which a convex preference can be replaced, while preserving the relevant comparisons, by one represented by an upper semicontinuous and quasiconcave utility function. Upper semicontinuity is empirically important because it guarantees the existence of an optimal choice from any compact budget set, including budget sets not observed in the data. \medskip

In Section \ref{subsec:risk-uncertainty}, we apply Theorem \ref{thm:porder} to test the convexity of DMs' preferences over risky or uncertain prospects. Convex preferences over risky prospects can arise when the DM entertains multiple utility functions and seeks to make cautious or conservative choices.\footnote{\,See, for instance, \cite{maccheroni2002maxmin} and \cite{cerreia2015cautious}.} Convex preferences over state-contingent acts  \citep{savage1954foundations,anscombe1963definition} are also widely used to model ambiguity aversion.\footnote{\,See \cite{gilboa1989maxmin}, \cite{klibanoff2005smooth}, \cite{maccheroni2006ambiguity}, \cite{strzalecki2011axiomatic}, and \cite{cerreia2011uncertainty}, among others.} In these settings, the DM's preference is often required to be consistent with a preorder such as first- or second-order stochastic dominance or state-wise monotonicity. A preorder can also arise naturally as the part of the preference about which the DM is confident \citep{dubra2004expected}. In all of these cases, Theorem \ref{thm:porder} provides conditions under which the relevant order pair admits a convex completion. Moreover, whenever such a completion exists, Theorem \ref{thm:convex:uppersemi} implies that there is also a convex completion represented by an upper semicontinuous and quasiconcave utility function.  When the data consist of finitely many strict comparisons and the imposed preorder admits a finite multi-utility representation, we further show that the convex completion can be represented by a \textit{continuous} and quasiconcave utility function.
\medskip

Section \ref{subsec:information} applies our results to test preferences over information structures. We consider a finite set of observed strict comparisons and characterize when they are consistent with a convex preference that respects the Blackwell order. We show that whenever such a convex rationalization exists, the data can be rationalized by a preference that admits a \textit{robust finite representation}. Under this representation, the DM evaluates each information structure according to its value in each of finitely many Bayesian decision problems and ranks it by its lowest value across these problems. Thus, although we impose only the ordinal requirement of convexity, the finiteness of the data allows us to obtain a stronger concave and continuous representation with a natural decision-theoretic interpretation. \medskip

In Afriat's Theorem, the necessary and sufficient condition for the rationalizability of a data set $\{(x^t,B^t)\}_{t\in T}$ is a no-cycling condition on revealed preference known as the {\em generalized axiom of revealed preference} (GARP). As noted earlier, when all the sets $B^t$ are linear budgets, every rationalizable data set is also rationalizable by a quasiconcave utility function. Thus, in this setting, GARP is sufficient for rationalization by a convex preference (or a {\em convex rationalization}, for short). \citet{reny15} extends Afriat's Theorem by showing that GARP remains sufficient for a convex rationalization even when the data set contains infinitely many observations, that is, when $T$ is infinite. In Section \ref{sec:garp_sp}, we use our approach to show that, whether $T$ is finite or infinite, GARP is sufficient for a convex rationalization whenever the budget sets are co-convex, meaning that their complements are convex. Such budget sets arise naturally under nonlinear pricing and have previously been studied in the revealed preference literature; see \citet{matzkin1991axioms}.  \medskip
	
Richter's Theorem (see \cite{richter1966revealed}) has a setting similar to Afriat's Theorem but uses a different notion of rationalization.  In this case, the dataset has the form $\{(A^t,B^t)\}_{t\in T}$ where $A^t$ is interpreted as the set of all the alternatives that the DM has ever chosen from $B^t$; this dataset is {\em exactly rationalizable} if there exists some preference $\succsim$ for which $A^t$ is precisely the set of $\succsim$-optimal elements in $B^t$.  Rationalization of this type is characterized by another no-cycling condition on the dataset called the {\em Congruence axiom}.  Note that Richter's Theorem does not guarantee that the rationalizing preference is convex.  We establish a variation of Richter's Theorem: We show that the congruence axiom is also sufficient for exact rationalization by a convex preference when the sets $B^t$ are co-convex and the set of choices $A^t$ is convex valued. \medskip

The remainder of the paper is organized as follows. Section \ref{sec:preliminary} introduces basic concepts and notation. Section \ref{sec:main} presents our main characterization theorem and an accompanying algorithm for empirical testing. Section \ref{sec:I-order} extends the framework by requiring convex completions to be consistent with additional orders satisfying independence. Section \ref{sec:app} applies our results to consumer choice, decisions under risk and uncertainty, and preferences over information structures, and discusses further applications. Section \ref{sec:garp_sp} identifies settings in which convexity imposes no additional testable implications beyond those imposed by basic rationality, with applications to infinite data sets and exact rationalization. The Appendix contains a general result on upper semicontinuous rationalization and proofs of the principal results in the main text. The Online Appendix contains additional proofs and supplementary results on symmetric preferences, an empirical application to choice under risk, preference identification under convexity, and a full characterization of the order pairs that admit strictly convex completions, including a generalization of the characterization in \citet{bossert2009non}.

\subsection{Related Literature}
There is a very large literature that investigates the observable properties of preference- or utility-maximization. Classical works include \cite{richter1966revealed}, \cite{afriat1967construction}, and \cite{fishburn76}. By and large, these papers focus on the rationalization of a dataset by a preference, rather than a preference with an added convexity property. There is also a closely related literature that asks when an incomplete preference admits a (not necessarily convex) completion. The early papers here include \cite{Szpilrajn1930} and \cite{suzumura1976remarks}. Subsequent research has examined conditions under which incomplete preferences can be extended to {\em continuous} preferences, as explored in \cite{levin1983}, \cite{herden1989existence}, \cite{herden2002continuous}, and \cite{nishimura2017comprehensive}, among others. Complementing this literature, we provide a general mathematical result establishing conditions under which a convex completion can be replaced, while preserving the relevant comparisons, by one represented by an \textit{upper semicontinuous} and quasiconcave utility function. This result allows us to obtain upper semicontinuous rationalizations in several of our applications. More recently, \cite{caradonna2024revealed} investigates the problem of extending an incomplete preference to a preference satisfying various invariance properties.\footnote{\,See also \cite{chambers2016revealed} for a synthetic account of the revealed preference literature.}\medskip

Among the literature on preference extension, our paper is most closely related to \cite{scapparone1999existence}, \cite{demuynck2009general}, and \cite{bossert2009non}, all of which study extensions satisfying convexity or a related strengthening of convexity.  \cite{scapparone1999existence} provides a necessary and sufficient condition under which an incomplete \textit{strict} order can be extended to a convex preference. Our main theorem generalizes this result by also considering incomplete weak orders.  This extension is important because information from choices on budget sets (as in Afriat's Theorem and related results) is typically interpreted as weak preferences; additional structure imposed on preferences may also take the form of preorders that allow for indifference.  The characterization obtained in \cite{scapparone1999existence} is a simpler version of the SP property formulated in this paper.  That paper does not provide an implementable algorithm for checking its version of the SP property.\medskip

\cite{demuynck2009general} presents a general method for extending incomplete preferences and applies it to characterize those incomplete preferences that can be extended to preferences satisfying a stronger form of the convexity property. Thus, the conditions obtained by  \cite{demuynck2009general} are sufficient but not necessary for an incomplete preference to admit a convex completion (in our sense). \cite{bossert2009non} investigate the problem of extending a weak order to a strictly convex and strictly increasing preference in the consumption space. Related to \cite{bossert2009non}, we characterize all order pairs that allow for strictly convex completions in the Online Appendix and apply our result to show that their  characterization applies in a broader setting beyond the Euclidean case.

	\section{Preliminaries}\label{sec:preliminary}
	Let $X$ be a nonempty and convex subset of a real vector space. The set $X$ is the space of alternatives. A \textit{binary relation} $R$ on $X$ is a subset of $X\times X$; we write $xRy$ for $(x,y)\in R$. The convolution of two binary relations $R$ and $\hat{R}$, denoted by $R \diamond \hat{R}$, is the binary relation defined such that $x R \diamond \hat{R} z$ if and only if $xRy\hat{R}z$ for some $y\in X$. A binary relation $R$ is \textit{reflexive} if, for all $x\in X$, $xRx$; \textit{complete} if, for all $x,y\in X$, either $xRy$ or $yRx$; \textit{transitive} if, for all $x,y,z\in X$, $xRy$ and $yRz$ imply $xRz$; and \textit{asymmetric} if, for all $x,y\in X$, $xRy$ implies that $yRx$ does not hold. Given a binary relation $R$, its symmetric part is the binary relation $R^s$ such that $xR^sy$ if and only if $xRy$ and $yRx$, and its asymmetric part is the binary relation $R^a$ such that $xR^ay$ if and only if $xRy$ and not $yRx$. For any $x\in X$, define $x^{\downarrow}_R=\{y\in X:xRy\}$ as the lower contour set of $x$ under $R$. 
	\medskip

A \textit{preorder} is a reflexive and transitive binary relation. A (strict) \textit{partial order} is an asymmetric and transitive binary relation. A \textit{preference} is a complete preorder. We usually use $\succsim$ to denote a preference and $\trianglerighteq$ to denote a preorder.  We say that a preference $\succsim$ admits a utility representation if there exists a utility function $u:X\rightarrow\mathbb{R}$ such that, for all $x,y\in X$, $x\succsim y$ if and only if $u(x)\geq u(y)$. Such a utility function $u$ is said to \textit{represent} $\succsim$. Throughout the paper, we use the notation $I_X$ to denote the minimal reflexive binary relation, i.e., $I_X=\{(x,x):x\in X\}$. A \textit{menu} is a nonempty subset of $X$. For any menu $A$ and preference $\succsim$, denote by $\max(A;\succsim)$ the set of $\succsim$-optimal alternatives in $A$, that is, $\max(A;\succsim)=\{x\in A:x\succsim y\text{~for all~}y\in A\}$. \medskip

For any $A\subseteq X$, denote by $co(A)$ the \textit{convex hull} of $A$, i.e., the smallest convex set containing $A$.\footnote{\,Equivalently, an alternative $x$ belongs to $co(A)$ if and only if there exist finitely many alternatives $\{x^k\}_{k=1}^n\subseteq A$ and numbers $\{\alpha^k\}_{k=1}^n\subseteq(0,1]$ with $\sum_{k=1}^n\alpha^k=1$ such that $x=\sum_{k=1}^n\alpha^k x^k$.} A preference $\succsim$ is said to be \textit{convex} if, for all $x,y\in X$ and $t\in[0,1]$, $x\succsim y$ implies $tx+(1-t)y\succsim y$.  There are three other equivalent characterizations of the convexity of $\succsim$: (i) for every \textit{finite menu} $A$ in $X$, if $y$ is a $\succsim$-worst alternative in $A$, then $x\succsim y$ for every $x\in co(A)$; (ii) for every $x\in X$, the upper contour set $\{y\in X:y\succsim x\}$ is convex; and (iii) for every $x\in X$, the strict upper contour set $\{y\in X:y\succ x\}$ is convex. A preference $\succsim$ is \textit{strictly convex} if, for all distinct $x,y\in X$ and $t\in(0,1)$, $x\succsim y$ implies $tx+(1-t)y\succ y$. \medskip

When the set of alternatives is a subset of $\mathbb{R}^n$, we use superscripts to distinguish different alternatives and subscripts to indicate coordinates. Thus, for an alternative $x^t \in \mathbb{R}^n$, its $k$-th coordinate is denoted by $x^t_k$. As usual, $\ge_n$ denotes the weak coordinatewise order on $\mathbb{R}^n$: For any $x,y \in \mathbb{R}^n$, $x \ge_n y$ if and only if $x_k \ge y_k$ for all $k=1,\ldots,n$. We denote by $>_n$ the asymmetric part of $\ge_n$. Finally, $\gg_n$ denotes the strong coordinatewise order: For any $x,y \in \mathbb{R}^n$, $x \gg_n y$ if and only if $x_k > y_k$ for all $k=1,\ldots,n$.


\section{Convex Completion}\label{sec:main}
In this section, we introduce our primitive, order pairs, which capture weak and strict preference information of the DM, and characterize those that can be extended to  convex preferences. We then provide an algorithm to simplify the test for convexity.

\subsection{Main Result}\label{subsec:main}
A pair of binary relations $p=(R,P)$ on $X$ is an \textit{order pair} if $R$ is reflexive. The relation $R$ captures weak preference information: $xRy$ indicates that the DM weakly prefers $x$ to $y$. The relation $P$ captures strict preference information: $xPy$ means that the DM strictly prefers $x$ to $y$.\footnote{\,A similar primitive is adopted in \cite{caradonna2024revealed}.} \medskip

To illustrate how choice data and additional assumptions on the DM's preference can be incorporated into our framework, consider the following two examples. \medskip

\noindent \textbf{Example 1.} Consider the classic revealed preference problem studied by \cite{afriat1967construction}. A data set is given by $\{(x^t, p^t)\}_{t \in T}$, where for each $t$, $x^t \in \mathbb{R}_{+}^n$ is the bundle chosen by the consumer, and $p^t \in \mathbb{R}_{++}^n$ is the price vector under which the consumer purchases $x^t$. Under the local non-satiation assumption, for each $t$, $x^t$ should be weakly preferred to any bundle $y$ that is affordable (i.e., $y \cdot p^t \le x^t \cdot p^t$) and strictly preferred to any bundle $z$ that is strictly affordable (i.e., $z \cdot p^t < x^t \cdot p^t$).\footnote{\,The local non-satiation assumption on the consumer's preference states that for any bundle $x \in \mathbb{R}^n_{+}$ and any $\epsilon > 0$, there exists another bundle $y$ with $|y - x| < \epsilon$ such that $y$ is strictly preferred to $x$.} Define
	\begin{equation*}
		\begin{split}
			& R = \{(x, y): x = x^t \text{~and~} y \cdot p^t \le x^t \cdot p^t \text{~for some~} t\}, \\
			& P = \{(x, y): x = x^t \text{~and~} y \cdot p^t < x^t \cdot p^t \text{~for some~} t\}. 
		\end{split}
	\end{equation*} The order pair $p=(R \cup I_{\mathbb{R}^n_+}, P)$ captures the information revealed by the choices. 
	\qed

	\medskip

	\noindent \textbf{Example 2.} Consider a participant in the lab who chooses option $x^t$ from the binary menu $\{x^t, y^t\}$ for each $t \in T$, and interpret each choice as revealing a strict preference for $x^t$ over $y^t$.  If, in addition, a preorder $\trianglerighteq^*$ is known to be consistent with the participant's preference \citep{nishimura2017comprehensive}, then the order pair $p$ can be constructed as 
	$$p = (\bowtie^*, \rhd^* \cup \{(x^t, y^t)\}_{t \in T}),$$
	where $\bowtie^*$ and $\rhd^*$ are the symmetric and asymmetric parts of $\trianglerighteq^*$, respectively. \qed \medskip

	Below, we define convex completions of order pairs.

	\begin{definition}
		A binary relation $\succsim$ extends an order pair $p=(R,P)$ if $R\subseteq\,\succsim$ and $P\subseteq\,\succ$, where $\succ$ is the asymmetric part of $\succsim$. The binary relation $\succsim$ is a completion of $p$ if it is a preference and extends $p$. The preference $\succsim$ is a convex (strictly convex) completion of $p$ if it is a convex (strictly convex) preference and a completion of $p$.
	\end{definition}

	If $\succsim$ is a convex completion of $p$, then  both the weak ($\succsim$) and strict ($\succ$) parts of the preference are consistent with the corresponding orders of $p$. When $p$ is constructed based on the DM's observed choices, checking whether $p$ admits a convex completion amounts to verifying the standard economic assumption that the DM optimizes a convex preference. To develop intuition about which order pairs admit convex completions,  consider the following example.
	\bigskip

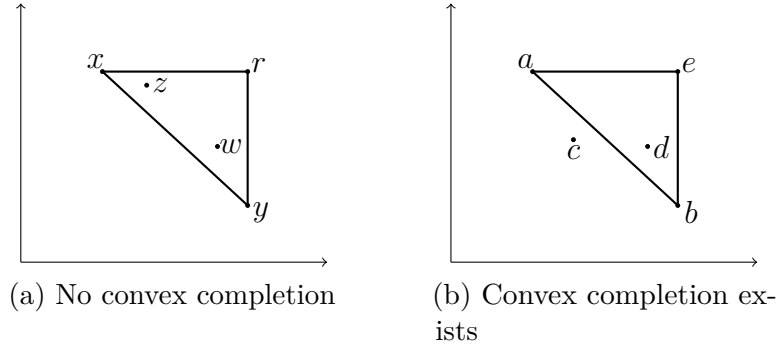
\begin{figure}[!htb]
    \centering

    \begin{subfigure}[t]{0.28\textwidth}
        \centering
        \begin{tikzpicture}[scale=9]
            \draw[->] (0,0) -- (0.45,0);
            \draw[->] (0,0) -- (0,0.38);

            \node[align=center] at (0.11,0.295) {$x$};
            \filldraw[black] (0.12,0.28) circle (0.08pt);

            \node[align=center] at (0.353,0.07333) {$y$};
            \filldraw[black] (0.33333,0.08333) circle (0.08pt);

            \node[align=center] at (0.205,0.260) {$z$};
            \filldraw[black] (0.185,0.260) circle (0.08pt);

            \node[align=center] at (0.309,0.17) {$w$};
            \filldraw[black] (0.289,0.17) circle (0.08pt);

            \draw[thick] (0.12,0.28) -- (0.33333,0.08333);
            \draw[thick] (0.12,0.28) -- (0.33333,0.28);
            \draw[thick] (0.33333,0.08333) -- (0.33333,0.28);

            \node[align=center] at (0.35,0.29) {$r$};
            \filldraw[black] (0.33333,0.28) circle (0.08pt);
        \end{tikzpicture}
        \caption{No convex completion}
    \end{subfigure}
    \hspace{0.06\textwidth}
    \begin{subfigure}[t]{0.28\textwidth}
        \centering
        \begin{tikzpicture}[scale=9]
            \draw[->] (0,0) -- (0.45,0);
            \draw[->] (0,0) -- (0,0.38);

            \node[align=center] at (0.11,0.295) {$a$};
            \filldraw[black] (0.12,0.28) circle (0.08pt);

            \node[align=center] at (0.353,0.07333) {$b$};
            \filldraw[black] (0.33333,0.08333) circle (0.08pt);

            \node[align=center] at (0.18,0.162) {$c$};
            \filldraw[black] (0.18,0.18) circle (0.08pt);

            \node[align=center] at (0.309,0.17) {$d$};
            \filldraw[black] (0.289,0.17) circle (0.08pt);

            \draw[thick] (0.12,0.28) -- (0.33333,0.08333);
            \draw[thick] (0.12,0.28) -- (0.33333,0.28);
            \draw[thick] (0.33333,0.08333) -- (0.33333,0.28);

            \node[align=center] at (0.35,0.29) {$e$};
            \filldraw[black] (0.33333,0.28) circle (0.08pt);
        \end{tikzpicture}
        \caption{Convex completion exists}
    \end{subfigure}

    \caption{Existence of convex completions}
    \label{fig_illustration}
\end{figure}

	\noindent \textbf{Example 3.} Let $X$ be the two-dimensional consumption space $\mathbb{R}^2_+$. Consider an order pair $p=(R,P)$ such that $rRx, xPz$ and $yPw$,  where the alternatives are depicted in Figure \ref{fig_illustration}(a).\footnote{\,While $R$ is reflexive (i.e., $I_X\subseteq R$), we omit the identity relation $I_X$ for notational simplicity.} 

We show that this order pair has no convex completion. Suppose, to the contrary, that a convex preference $\succsim$ extends it. Let $\underline{x}$ be the $\succsim$-worst alternative in $\{x,y,r\}$. 
By the convexity of $\succsim$, for all $\hat{x}\in co(\{x,y,r\})$, we have $\hat{x}\succsim \underline{x}$. Thus, for all $\hat{x}\in\{x,y,z,w,r\}$, we have $\hat{x}\succsim\underline{x}$. Since $\succsim$ extends $p$, $xPz$ implies $x\succ z$, and hence $\underline{x}\neq x$. Similarly, $yPw$ implies $\underline{x}\neq y$. Finally, $rRx$ and $xPz$ imply $r\succsim x$ and $x\succ z$, respectively. By transitivity, $r\succ z$, and hence $\underline{x}\neq r$. This leads to a contradiction. \medskip

	Within the same space $X$, consider another order pair $\hat{p}=(\hat{R},\hat{P})$ such that $e\hat{R}a, a\hat{P}c$ and $b\hat{P}d,$ where the alternatives are depicted in Figure \ref{fig_illustration}(b). For this pair, the previous contradiction no longer arises: If $\hat{p}$ admits a convex completion $\succsim$, then $a$ can serve as the $\succsim$-worst alternative in $\{a,b,e\}$. This is because the only alternative known to be strictly worse than $a$ is $c$, which lies outside the convex hull of $\{a,b,e\}$. \medskip

	In fact, we can explicitly construct a convex completion $\succsim$ of $\hat{p}$ as follows. Let $A=co(\{a, b, e\})$ and $B=co(\{b, e\})$. For all $x',y' \in B$ and $z' \in X\backslash B$, let $x' \sim y' \succ z'$. For all $x', y' \in A \backslash B$ and $z' \in X\backslash A$, let $x'\sim y' \succ z'$. For all $x', y' \in X\backslash A$, let $x' \sim y'$. Essentially, $\succsim$ ranks alternatives in $B$ the best, alternatives in $A \backslash B$ the second best, and alternatives in $X\backslash A$ the worst. The preference $\succsim$ is convex, because for all $x' \in X$, the upper contour set $\{y' \in X: y'\succsim x'\}$ equals either $X$, $A$, or $B$---all of which are convex. The preference $\succsim$ extends $\hat{p}$, as it induces the preference $b\sim e \succ a \sim d \succ c$, consistent with $\hat{p}$. \qed \bigskip

	The intuition from the preceding example is that if an order pair $p=(R,P)$ admits a convex completion $\succsim$, then for any finite menu $A$, every alternative in $co(A)$ is weakly preferred to the $\succsim$-worst alternative in $A$ (call it $x$). Recall that $x^{\downarrow}_P=\{y\in X:xPy\}$ denotes the lower contour set of $x$ under $P$, i.e., the set of alternatives that are strictly worse than $x$. Hence, $x^{\downarrow}_P$ must be disjoint from $co(A)$. Extending this observation, we next identify the largest set $B$ such that, for every convex completion $\succsim$ of $p$, all alternatives in $B$ are weakly preferred to the $\succsim$-worst alternative in $A$. Therefore, if $p$ admits a convex completion, then, by taking any such completion and letting $x$ be its worst alternative in $A$, we have $B\cap x^{\downarrow}_P=\emptyset$. It then follows that the existence of a convex completion requires the existence of some $x\in A$ such that $B$ is disjoint from $x^{\downarrow}_P$. \medskip

	Fix an order pair $p=(R,P)$ and a finite menu $A$. Suppose that $p$ admits a convex completion $\succsim$, and let $\underline{x}$ denote the $\succsim$-worst alternative in $A$. Define 
	$$
	\Pi^0_R(A) = co(A).
	$$
	By convexity of $\succsim$, every alternative in $\Pi^0_R(A)$ is weakly preferred to $\underline{x}$ under $\succsim$. Recursively define for each $k \in \mathbb{N}$:
	$$
	\Pi^{k+1}_R(A) = co\left( \bigcup_{x \in \Pi^{k}_R(A)} \{y: y R x\} \right).
	$$
	The set $\Pi^{k+1}_R(A)$ represents the convex hull of all alternatives that are known to be weakly preferred (under $R$) to at least one alternative in $\Pi^{k}_R(A)$. \medskip

	Observe that if $x \succsim \underline{x}$ holds for all $x \in \Pi^{k}_R(A)$, then, since $R \subseteq\, \succsim$ and $\succsim$ is convex, this property extends to all $x \in \Pi^{k+1}_R(A)$. Since $x \succsim \underline{x}$ for all $x \in \Pi^0_R(A)$, induction implies that this property holds for all $x \in \Pi^{k}_R(A)$ and every $k \in \mathbb{N}$. Define the limit set:
	\[
	\Pi_R(A) := \bigcup_{k=0}^{\infty} \Pi^k_R(A).
	\]
	This set $\Pi_R(A)$ is maximal with respect to two properties: (i) its construction relies only on the weak preference information $R$ and the convexity of $\succsim$, and (ii) every alternative $x \in \Pi_R(A)$ satisfies $x \succsim \underline{x}$. Moreover, since each $\Pi^k_R(A)$ is convex and the sequence $(\Pi^k_R(A))_{k=0}^{\infty}$ is increasing under set inclusion, $\Pi_R(A)$ is also convex. \medskip

	Since every alternative in $\Pi_R(A)$ is weakly preferred to $\underline{x}$ under $\succsim$, we have $\underline{x}^{\downarrow}_P \cap \Pi_R(A) = \emptyset$. While the convex completion $\succsim$ is unknown \textit{a priori}, the construction of $\Pi_R(A)$ depends only on $R$ and convexity, without requiring explicit knowledge of $\succsim$ itself. Similarly, although the $\succsim$-worst alternative $\underline{x}$ in $A$ depends on the specific completion $\succsim$, the finiteness of $A$ guarantees the existence of such an alternative. These observations motivate the following necessary condition for $p=(R,P)$ to admit a convex completion.

	\begin{definition}
		An order pair $p=(R,P)$ satisfies the separating property (SP) on a finite menu $A$ if there is $x \in A$ such that $x^{\downarrow}_P \cap \Pi_R(A) = \emptyset$.
	\end{definition}

	The following example demonstrates how to identify $\Pi_R(A)$ and check SP. \bigskip

\begin{figure}[!htb]
		\centering
		\begin{subfigure}[!htb]{0.45\textwidth}  
			\hspace*{\fill}  
			\begin{tikzpicture}[scale=9]
				\draw[->] (0.00, 0.00) -- (0.60, 0.00);
				\draw[->] (0.00, 0.00) -- (0.00, 0.43); 
				\node[align=center] at (0.16, 0.345) {$x$};
				\filldraw[black] (0.17,0.33) circle (0.07pt);
				\node[align=center] at (0.40, 0.12333) {$y$}; 
				\filldraw[black] (0.38333,0.13333) circle (0.07pt);
				\node[align=center] at (0.255, 0.310) {$r$};
				\filldraw[black] (0.235, 0.310) circle (0.07pt); 
				\draw[thick] (0.17, 0.33) -- (0.38333,0.13333);
				\draw[thick] (0.17, 0.33) -- (0.38333, 0.33);
				\draw[thick] (0.38333, 0.13333) -- (0.38333, 0.33);
				\node[align=center] at (0.40, 0.34) {$z$};   
				\filldraw[black] (0.38333, 0.33) circle (0.07pt);
				\node[align=center] at (0.048, 0.40) {$a$};   
				\filldraw[black] (0.07, 0.40) circle (0.07pt);
				\node[align=center] at (0.38, 0.05) {$b$}; 
				\filldraw[black] (0.39,0.07) circle (0.07pt);
				\node[align=center] at (0.57, 0.40) {$c$}; 
				\filldraw[black] (0.55,0.40) circle (0.07pt);
				\draw[thick] (0.07, 0.40) -- (0.39,0.07);
				\draw[thick] (0.07, 0.40) -- (0.55,0.40);
				\draw[thick] (0.39,0.07) -- (0.55,0.40);
				\node[align=center] at (0.50, 0.35) {$d$}; 
				\filldraw[black] (0.48,0.35) circle (0.07pt);
			\end{tikzpicture}
			\hspace*{\fill}  
		\end{subfigure} 
		\caption{Violation of SP}\label{fig_sp}
	\end{figure}
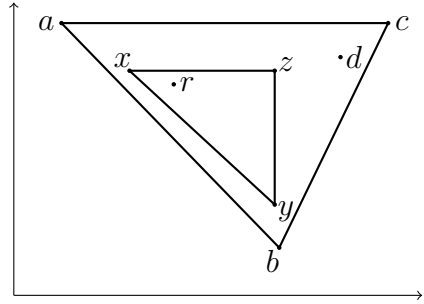

	\noindent \textbf{Example 4.} Let $X$ be the two-dimensional consumption space $\mathbb{R}^2_+$. Consider an order pair $p=(R,P)$ such that $aRr, bRr, cRr, xPd, yPd$ and $zPd$, where the alternatives are depicted in Figure \ref{fig_sp}. \medskip

	We demonstrate that $p$ violates SP on the menu $A=\{x,y,z\}$. First, $\Pi_R^0(A)=co(A)$. Since $r\in co(A)$, the set $\Pi_R^1(A)$ expands to include all alternatives that are weakly preferred to $r$ under $R$. Consequently, $\Pi_R^1(A)$ additionally contains $a$, $b$, and $c$, yielding $\Pi_R^1(A)=co(\{a,b,c\})$. As no further alternatives satisfy the inclusion criterion, $\Pi_R(A)=co(\{a,b,c\})$. However, SP is violated on $A$ because $d\in\Pi_R(A)$ and $xPd$, $yPd$, and $zPd$ all hold. \qed

	\bigskip

	To state our main characterization result, we introduce the following notation. Given an order pair $p=(R,P)$, let
	\begin{equation}\label{eq:G_p}
	G_{P}=\{x \in X: \text{~there exists~} y \in X \text{~such that~} x P y\}
	\end{equation}
	denote the set of alternatives that are better than at least one alternative according to $P$. Our main theorem states that, to determine whether $p$ admits a convex completion, it suffices to check whether $p$ satisfies SP on every finite menu \textit{contained in} $G_{P}$.

	\begin{theorem}\label{thm:main}
		For an order pair $p=(R,P)$, the following statements are equivalent. \smallskip

		\emph{(i)} $p$ admits a convex completion.

		\emph{(ii)} $p$ satisfies SP on every finite menu. 
		
		\emph{(iii)} $p$ satisfies SP on every finite  menu contained in $G_P$.  \end{theorem}

	To see why statements (ii) and (iii) are equivalent, note that for any finite menu $A$, if $A$ contains some alternative $x \notin G_P$, then $x^{\downarrow}_P=\emptyset$, which directly implies $\Pi_R(A)\cap x^{\downarrow}_P=\emptyset$. Hence, $p$ automatically satisfies SP on $A$. Notably, if $G_P=\emptyset$, meaning that the strict binary relation $P$ is trivial, then the order pair has a trivial convex completion under which all alternatives are indifferent. We defer a proof sketch of the implication from (ii) to (i) to the end of Section \ref{subsec:algorithm}, after introducing the algorithm that simplifies testing whether $p$ satisfies SP on every menu contained in a finite set $A$. We conclude this subsection with several remarks on Theorem \ref{thm:main} as follows. \medskip

	\textbf{Remark 1.} A given order pair $p=(R,P)$ admits a convex completion if and only if $q=(R\backslash P,P)$ does.\footnote{\,To ensure that $q$ is an order pair, one can alternatively define $q$ as $q=(I_X\cup(R\backslash P),P)$.} In practice, it is more convenient to check SP for $q$ than for $p$, since $q$ contains fewer weak preference relations. \medskip

	\textbf{Remark 2.} The condition stated in Theorem \ref{thm:main} \textit{cannot} be weakened to requiring that the order pair satisfy SP only on every \textit{singleton} menu: An order pair may satisfy SP on all singleton menus yet fail to admit a convex completion. For a counterexample, consider the order pair $p$ in Example 4. One can verify that for every $\bar{x}\in X$, $\bar{x}^{\downarrow}_P \cap \Pi_R(\{\bar{x}\}) =\emptyset$. Nevertheless, as demonstrated, $p$ does not admit any convex completion. \medskip

	\textbf{Remark 3.} An equivalent characterization to that in Theorem \ref{thm:main}, which avoids the existence quantifier, is that for every finite menu $A$, $A\not\subseteq\{x:xPy\text{~for some~}y\in\Pi_R(A)\}$. The condition stated in the theorem is more useful for conducting empirical tests for convexity, as will be demonstrated in Sections \ref{subsec:algorithm} and \ref{sec:I-order}. \medskip

	\textbf{Remark 4.} When the order pair is given by $p=(I_X,P)$, i.e., when the weak binary relation carries no preference information beyond reflexivity, the condition in Theorem \ref{thm:main} reduces to the requirement that for every finite menu $A$, 
	\begin{equation}\label{eq:strict_scapparone1}
		x^{\downarrow}_P \cap co(A) = \emptyset \text{~for some~} x \in A.
	\end{equation} 
	\cite{scapparone1999existence} studies such an order pair $p$ and shows that it admits a convex completion if and only if for every finite menu $B$, 
	\begin{equation}\label{eq:strict_scapparone2}
		B \not\subseteq co\left( \{x: xPy \text{~for some~} y \in B\} \right).
	\end{equation}

	We note that the two characterization conditions are equivalent in this case. First, consider a finite menu $A$ that violates condition (\ref{eq:strict_scapparone1}). For each $x\in A$, pick $d(x)\in co(A)$ such that $xPd(x)$, and define $B=\{d(x)\}_{x\in A}$. Since $A\subseteq\{x:xPy\text{~for some~}y\in B\}$ and $B\subseteq co(A)$, it follows that
	$B\subseteq co\left(\{x:xPy\text{~for some~}y\in B\}\right)$.
	Hence, $B$ violates condition (\ref{eq:strict_scapparone2}). \medskip 
	
	Conversely, consider a finite menu $B$ that violates condition (\ref{eq:strict_scapparone2}). For each $y\in B$, let $H(y)$ be a finite menu such that $y\in co(H(y))$ and, for every $x\in H(y)$, there exists $\hat{y}\in B$ with $xP\hat{y}$. Define $A=\bigcup_{y\in B}H(y)$. Then $B\subseteq co(A)$. Moreover, for every $x\in A$, there exists $\hat{y} \in B\subseteq co(A)$ such that $xP\hat{y}$. Thus, $x^{\downarrow}_P\cap co(A)\neq\emptyset$ for every $x\in A$, so $A$ violates condition (\ref{eq:strict_scapparone1}). \medskip

	\textbf{Remark 5.} \cite{suzumura1976remarks} provides a characterization condition for incomplete preferences---which contain both weak and strict components---to admit completions.\footnote{\,More concretely, \cite{suzumura1976remarks} considers a single binary relation $\trianglerighteq$ but requires the preference $\succsim$ to extend the order pair $(\trianglerighteq,\rhd)$, where $\rhd$ is the asymmetric part of $\trianglerighteq$.} Our main result, Theorem \ref{thm:main}, can be viewed as a counterpart of the extension result of \cite{suzumura1976remarks} for convex preferences. Following Theorem 3 of \cite{suzumura1976remarks}, a necessary and sufficient condition for an order pair $p=(R,P)$ to admit a completion is that $p$ is \textit{consistent}, that is, there exists \textit{no} cycle $(x^k)_{k=1}^n$ with $n\ge 2$  such that $x^{k+1}(R\cup P)x^k$ for all $k\in\{1,\dots,n-1\}$ and $x^1Px^n$. An order pair $p=(R,P)$ is also said to be \textit{consistent on a set} $A$ if it contains no cycle in $A$. \medskip

	By Theorem \ref{thm:main}, if $p$ satisfies SP on every finite menu, it must be consistent. We offer a direct proof as follows. Suppose, for contradiction, that the consistency condition fails for some sequence $(x^k)_{k=1}^n$ with $n\ge 2$. Define the set $$K=\{k\in\{2,\dots,n\}:x^kPx^{k-1}\},$$ so that for all $k\in\{2,\dots,n\}\backslash K$, we have $x^kRx^{k-1}$. Consider the menu $A=\{x^k\}_{k\in K\cup\{1\}}$. By our iterative inclusion of $R$-better alternatives, we have $\{x^k\}_{k=1}^n\subseteq\Pi_R(A)$. However, for every $k\in K\cup\{1\}$, either $x^kPx^{k-1}$ (if $k\ge 2$), or $x^kPx^n$ (if $k=1$). Hence, for every $x \in A$,   $x^{\downarrow}_P \cap \Pi_R(A) \neq \emptyset$, implying that $p$ violates SP on $A$, a contradiction. \medskip

	\textbf{Remark 6.} \cite{demuynck2009general} provides a characterization condition for an order pair $p=(R,P)$ to admit a completion that satisfies a stronger property than convexity. In particular, a preference $\succsim$ satisfies the property considered by \cite{demuynck2009general} if it is convex and additionally satisfies that for every finite menu $A=\{x^1,...,x^n\}$, alternative $y$, and $(\alpha^i)_{i=1}^n\in(0,1)^n$ with $\sum_{i=1}^n\alpha^i=1$, if $x^i\succsim y$ for all $x^i\in A$, and $x^j\succ y$ for some $x^j\in A$, then $\sum_{i=1}^n\alpha^i x^i\succ y$.\footnote{\,Note that this requirement is weaker than strict convexity, as strict convexity implies $\sum_{i=1}^n\alpha^i x^i\succ y$ without the condition that $x^j\succ y$ for some $j$.} In the Online Appendix, we demonstrate the difference between this property and convexity in terms of preference extension through an example.

	\subsection{Algorithm for Checking SP}\label{subsec:algorithm}
	In this section, we show that verifying whether $p = (R, P)$ satisfies SP on all submenus within a given finite menu $A$ reduces to checking SP on a descending sequence (in the set-inclusion sense) of menus in $A$. This reduction decreases the number of required SP checks from exponential ($2^{|A|}$) to linear (at most $|A|$) in the menu size. Consequently, when the set $G_P$ (as defined in (\ref{eq:G_p})) is finite, determining whether $p$ admits a convex completion requires checking SP at most $|G_P|$ times. Below, we present the algorithm. \bigskip

	\textbf{Algorithm I.} For any order pair $p$ and finite menu $A$, set $A^0=A$ and $k=1$.

	START. Let $A^k=\{x \in A^{k-1}: x^{\downarrow}_P \cap \Pi_R(A^{k-1}) \neq \emptyset\}$. Consider the following mutually exclusive cases: 
	
	(a) $A^k=\emptyset$: Stop and output ``SP holds''.
	
	(b) $\emptyset \neq A^k \subsetneq A^{k-1}:$ Go to START with $k=k+1$.
	
	(c) $\emptyset \neq A^k = A^{k-1}:$ Stop and output ``SP fails''.  \qed
	
	\bigskip

	In Algorithm I, observe that the procedure terminates in one of two ways: either at some iteration $k$ with $A^k = \emptyset$, or at some iteration $k$ with $A^k \neq \emptyset$ and $A^k = A^{k-1}$. Consequently, the algorithm either outputs ``SP holds'' or ``SP fails''. \medskip

	To understand the algorithm, recall that if $p$ admits a convex completion $\succsim$, then for any alternative $x$ that is $\succsim$-worst in $A^{k-1}$, it must satisfy $x^{\downarrow}_P \cap \Pi_R(A^{k-1}) = \emptyset$. The algorithm constructs $A^k$ by eliminating all such candidates for the $\succsim$-worst alternative in $A^{k-1}$. Notably, if $p$ satisfies SP on all submenus of $A$, then at each iteration $k$, the set of such candidates is not empty, and we have $A^{k} \subsetneq A^{k-1}$. Therefore, if $p$ satisfies SP on all submenus of $A$, then the algorithm outputs ``SP holds''. The following theorem establishes the converse. 
	\medskip

	\begin{theorem}\label{thm:algorithm}
		For any order pair $p=(R, P)$ and finite menu $A$, run Algorithm I for $p$ on $A$.  The following statements are equivalent. 
		\smallskip

		\emph{(i)} $p$ satisfies SP on every submenu of $A$. \smallskip
		
		\emph{(ii)} Algorithm I outputs ``SP holds''. \smallskip
		
		\emph{(iii)} There is a convex preference $\succsim$ on $X$ such that:
		\begin{equation*}
			\begin{split}
				&\forall x,y \in X, x R y \text{~implies~} x \succsim y, \\
				& \forall x \in A, y \in X, x P y \text{~implies~} x \succ y.
			\end{split}
		\end{equation*}  
	\end{theorem}

	\begin{proof}[Proof of Theorem \ref{thm:algorithm}] 
	We have demonstrated the implication from (i) to (ii). To see that (iii) implies (i), consider any submenu $B$ of $A$ and let $x$ be a $\succsim$-worst alternative in $B$. By convexity and $R\subseteq\,\succsim$, every alternative in $\Pi_R(B)$ is weakly preferred to $x$. Since $x \in A$, by (iii), $xPy$ implies $x\succ y$. Thus we have $x_P^{\downarrow}\cap\Pi_R(B)=\emptyset$, and $p$ satisfies SP on $B$. The remaining implication from (ii) to (iii) follows from an explicit construction of $\succsim$. \medskip

		By (ii), the algorithm identifies a strictly decreasing sequence $(A^k)_{k=0}^n$ of submenus of $A$ with $A^n =\emptyset$. The idea for constructing the preference $\succsim$ is to construct a sequence of nested convex sets in $X$, such that alternatives that are included in smaller sets are ranked higher. Specifically, for each $x \in X$, define 
		$$k^*(x) = 
		\begin{cases} 
			-1, & ~~\text{if } x \notin \Pi_R(A^0), \\
			\max\{k: x \in \Pi_R(A^k)\}, &~~ \text{otherwise.}
		\end{cases}$$
		Since $A^n=\emptyset$, for each $x$, we have $k^*(x) \in \{-1,...,n-1\}$. Let $\succsim$ be the preference represented by $k^*$. \medskip

		We verify that $\succsim$ satisfies the conditions in statement (iii).  Convexity holds since each upper contour set $\{y \in X : y \succsim x\}$ equals either $X$ or $\Pi_R(A^{k^*(x)})$. For $R$-consistency, if $x R y$, then $k^*(x) \geq k^*(y)$ (trivially when $k^*(y) = -1$, and otherwise because $xR y$ implies $x \in \Pi_R(A^{k^*(y)})$).  For $P$-consistency, if $x P y$ with $x \in A$, then there exists $k$ such that $x \in A^k$ and $x^{\downarrow}_P \cap \Pi_R(A^k) = \emptyset$, indicating that $y \notin \Pi_R(A^k)$ (since $y \in x^{\downarrow}_P$), and thus $k^*(x) \geq k > k^*(y)$, i.e., $x \succ y$.      
	\end{proof}

	By the proof of Theorem \ref{thm:algorithm}, when $G_P$ is finite, we can explicitly construct a quasi-concave utility function $u$, following Algorithm I, such that the preference it represents extends the order pair $p$. \medskip

With Theorem \ref{thm:algorithm}, we sketch the proof for the implication from (ii) to (i) in Theorem \ref{thm:main}. Define the collection of order pairs $\mathcal{P}$ such that $q=(R_q, P_q) \in \mathcal{P}$ if $R  \subseteq  R_q$, $P  \subseteq  P_q$, and $q$ satisfies SP on every finite menu. Define an order $\subseteq$ on $\mathcal{P}$ such that for any two order pairs $q=(R_q, P_q)$ and $\hat{q}=(R_{\hat{q}}, P_{\hat{q}})$ in $\mathcal{P}$, $q \subseteq \hat{q}$ if and only if $R_q \,\subseteq\, R_{\hat{q}}$ and $P_q \,\subseteq\, P_{\hat{q}}.$ We  apply Zorn's lemma to show that there exists a $\subseteq$-maximal element $r=(R_r, P_r)$ in $\mathcal{P}$, i.e., there is no $q\in\mathcal{P}$ with $q \neq r$ and $r \subseteq q$. We then use the maximality of $r$ and $r\in\mathcal{P}$ to show that
$R_r$ is a complete, transitive and convex binary relation whose asymmetric part is $P_r$. Thus, $R_r$ is the desired convex preference.

\section{Convex Rationalization with Independent Orders}\label{sec:I-order}
We have shown that testing whether an order pair $p=(R,P)$ admits a convex completion amounts to checking whether it satisfies SP on every finite menu in $G_P$. When $p$ is constructed based on the DM's choices observed in finitely many menus, $G_P$ is usually finite. Algorithm I then performs the test with at most $|G_P|$ rounds of SP checks. \medskip

However, in many applications, aside from convexity, we may also want the DM's preference to extend some given orders \citep{nishimura2017comprehensive}. This may result in an infinite set $G_P$. For example, consider a consumer who weakly prefers $x$ to $y$ and $z$ to $w$, where $x,y,z,w \in X=\mathbb{R}^n_+$ are consumption bundles. If we seek to test whether the consumer has a convex and \textit{strictly increasing} preference, then we should start with the order pair $$p = (\{(x,y), (z,w)\}\, \cup I_X, >_n),$$ where $>_n$ denotes the $n$-dimensional strict Euclidean order. The set $G_{>_n}$ is infinite in this example, and a direct application of Theorem \ref{thm:main} indicates that  infinitely many SP checks are needed to determine whether $p$ admits a convex completion. \medskip

In this section, we show that when the additional orders we impose on the DM's preferences satisfy  \textit{independence} (which holds for the consumer example above), we can simplify the problem of testing convexity and reduce the SP checks to finitely many rounds. We begin with the following definition.

\begin{definition}
A binary relation $\trianglerighteq$ satisfies independence if for all $x,y,z \in X$ and $t \in (0,1]$, $x \trianglerighteq y$ if and only if $tx+(1-t)z \trianglerighteq ty+(1-t)z$.  
\end{definition}

Independence is a fundamental axiom characterizing preferences over lotteries that admit expected utility representations. We refer to a binary relation (respectively, preorder, partial order) that satisfies independence as an I-binary relation (respectively, I-preorder, I-partial order). We note that the empty binary relation $\emptyset$ is both a partial order and an I-partial order.

\begin{definition}
Given two binary relations $\trianglerighteq^*$ and $\rhd$ on $X$, the pair $(\trianglerighteq^*,\rhd)$ is \textit{compatible} if for all $x,y,z\in X$, $x\trianglerighteq^* y$ and $y\rhd z$ imply $x\rhd z$.
\end{definition}

Equivalently, $(\trianglerighteq^*,\rhd)$ is compatible if
$\trianglerighteq^*\diamond\,\rhd\subseteq\rhd$. Compatibility arises as a natural property in many relevant applications:  \smallskip

\noindent (1) for any binary relation $P$, $(I_X, P)$ is compatible, where $I_X=\{(x,x): x\in X\}$;

\noindent (2) if $\trianglerighteq^*$ is a preorder, with $\bowtie^*$ and $\rhd^*$ denote its symmetric and asymmetric parts respectively, then $(\bowtie^*,\rhd^*)$ and  $(\trianglerighteq^*,\rhd^*)$ are both compatible;

\noindent (3) on the $n$-dimensional Euclidean space $\mathbb{R}^n$,  $(\ge_n, >_n)$ and $(\ge_n,\gg_n)$ are compatible.  \smallskip

To state the next theorem, for a given binary relation $R$, we  define
\begin{equation}\label{eq:Go_R}
\mathring{G}_R
=
\{x\in X:\text{ there exists }y\in X\backslash\{x\}\text{ such that }xRy\}
\end{equation}
as the set of alternatives that are $R$-better than at least one distinct alternative.

\begin{theorem}\label{thm:porder} 
Consider an order pair $p=(R \,\cup \trianglerighteq^*, P \cup \rhd)$, where $R$ is reflexive, $\rhd$ is an I-partial order, and $(\trianglerighteq^*, \rhd)$ is compatible.  Define \begin{equation}\label{eq:independence} 
 H=\mathring{G}_R \cup G_P \text{~and~} P^*=\left\{(x,y)\in H\times X:
 xPy\ \text{or}\ xP\diamond\rhd y\ \text{or}\ xR\diamond\rhd y
 \right\}. 
\end{equation} 
If $p$ violates SP on some finite menu $A$, then there exists a finite menu $B \subseteq H$ on which $q=(R \,\cup \trianglerighteq^*, P^{*})$ violates SP. The order pair $p$ admits a convex completion if and only if $q$ satisfies SP on every finite menu contained in $H$.  
\end{theorem}

We note that, in the statement of Theorem \ref{thm:porder}, the binary relation $\trianglerighteq^*$ itself need not be a preorder or satisfy independence. The only requirement on $\trianglerighteq^*$ is that the pair $(\trianglerighteq^*, \rhd)$ be compatible. \medskip

By Theorem \ref{thm:porder}, checking whether $p$ admits a convex completion reduces to verifying whether the revised order pair $q$ satisfies SP on all finite menus contained in $H$. Notably, $H$ is independent of the two orders $\trianglerighteq^*$ and $\rhd$, which makes testing for convex completions practically feasible in applications with finite data. \medskip

To illustrate how Theorem \ref{thm:porder} works, consider the special case where the order pair is given by $p=(R, \rhd)$, with $\rhd$ being an I-partial order. Write $p=(R\cup I_X,  \rhd)$. Since  $(I_X,\rhd)$ is compatible, by Theorem \ref{thm:porder}, $p$ admits a convex completion if and only if the revised order pair $q=(R, P^{*})$ satisfies SP on every finite menu contained in $\mathring{G}_R$, where   $P^{*}=\{(x,y) \in \mathring{G}_R \times X: x R\diamond \rhd y\}$.  Thus, in the consumer example introduced earlier in this section, to test whether $p=(\{(x,y), (z,w)\}\, \cup I_X, >_n)$ admits a convex completion, we need only verify that $q=(\{(x,y), (z,w)\}\, \cup I_X, P^*)$ satisfies SP on menus contained in $\{x,z\}$, where $P^*= \{(x,\hat{x}): x >_n \hat{x} \text{ or } y >_n \hat{x}\} \cup \{(z, \hat{z}): z >_n \hat{z} \text{ or } w >_n \hat{z}\}.$ Algorithm I requires at most two rounds of SP checks for this verification. \medskip

Theorem \ref{thm:porder} admits a further simplification in another special case, stated below as a corollary. The proof is in the Online Appendix.

\begin{corollary}\label{coro:porder:pure_strict}   
Let $\rhd$ be an I-partial order and $\trianglerighteq^*$ an I-preorder with $\rhd\subseteq\trianglerighteq^*$ and $(\trianglerighteq^*, \rhd)$ being compatible. The order pair $p=(\trianglerighteq^*, P \cup \rhd)$ admits a convex completion if and only if for every finite menu $A$ contained in $G_P$,  
\begin{equation}\label{eq:independence:simpler}   
\text{there is } x \in A \text{ such that } x^{\downarrow}_{P \diamond \trianglerighteq^*} \cap co(A) =\emptyset.    
\end{equation}    
\end{corollary}

Corollary \ref{coro:porder:pure_strict} is a generalization of the result of \cite{scapparone1999existence}. When $G_P$ is finite, condition (\ref{eq:independence:simpler}) can be tested on at most $|G_P|$ menus, following an iterative procedure similar to Algorithm I. To see this, note that if $x$ is the alternative in menu $A$ that satisfies condition (\ref{eq:independence:simpler}), then $x$ can serve as the one satisfying the condition in any submenu $B$ of $A$ that contains $x$. This leads to the following checking procedure. Starting with $H^0 = G_P$, we check whether $H^0$ satisfies condition (\ref{eq:independence:simpler}). If it does not, we stop; otherwise we define 
$$H^1 = \big\{ x \in H^0 : x^{\downarrow}_{P \diamond \trianglerighteq^*} \cap co(H^0) \neq \emptyset \big\}$$ 
and repeat the test on $H^1$. This yields a strictly decreasing sequence $H^0 \supseteq H^1 \supseteq \cdots \supseteq H^k$ of subsets of $G_P$, which terminates within $|G_P|$ steps.\bigskip

\noindent \textbf{More discussion on continuity.} In \cite{nishimura2017comprehensive}, they show that, under suitable regularity conditions, a finite data set that admits a rationalization consistent with a given continuous preorder also admits a continuous rationalization consistent with that preorder. Thus, in their setting, continuity can be imposed to the recovered preference without further restricting the data. An analogous conclusion does not generally hold for convex completion: The existence of a convex completion need not imply the existence of a continuous convex completion. Section \ref{subsec:non-linear-budget} demonstrates this distinction in the Euclidean setting. \medskip

Nevertheless, under suitable conditions on the imposed order, upper semicontinuity imposes no additional restrictions on the existence of a convex completion. In Appendix 1, Theorem \ref{thm:convex:uppersemi} shows that, when the imposed order satisfies appropriate topological and linearity conditions, an existing convex completion can be replaced by one represented by an upper semicontinuous and quasiconcave utility function, while preserving the imposed order and all weak and strict comparisons of finitely many designated alternatives with arbitrary alternatives in the domain. This result allows us to obtain upper semicontinuous convex completions in several of the applications studied in the next section.

\section{Applications}\label{sec:app} 
In this section, we apply our convexity tests to several specific settings: consumption bundles, lotteries, state-contingent acts, and information structures. In the first three environments, we show that under suitable finiteness conditions,  a convex completion can often be chosen to satisfy additional regularity properties, such as upper semicontinuity. For preferences over information structures, we obtain a finite robust decision-problem representation. 

\subsection{Convex Rationalization of Consumer  Choices}\label{subsec:non-linear-budget}

The literature provides conditions characterizing rational consumer behavior across a variety of settings. The classic case is studied by \cite{afriat1967construction}, where the data record consumer choices from \textit{linear} budget sets. Rationalization with general budget sets is subsequently investigated by \cite{fm09} and \cite{nishimura2017comprehensive}, among others. In this section, we allow for general budget sets and our focus is to characterize consumers' choice behavior that can be rationalized by \textit{convex} preferences.
\medskip

Throughout this section, fix the space of alternatives as $X=\mathbb{R}^n_+$ with each dimension denoting a good. An alternative $x \in X$  represents a bundle of goods, where for each $k \in \{1,...,n\}$, $x_k$ denotes the amount of good $k$.  \medskip

\begin{definition}
A  consumption data set is a non-empty collection $\mathcal{O}=\{(x^t, B^t)\}_{t \in T}$ such that for $t \in T$, $x^t \in B^t \subseteq X$.
\end{definition}

Given a consumption data set $\mathcal{O}=\{(x^t, B^t)\}_{t \in T}$, we refer to each $(x^t, B^t)$ as an observation. Observation $t$ is interpreted to mean that the consumer chooses bundle $x^t$ from budget set $B^t$.

\begin{definition}
A consumption data set $\mathcal{O}=\{(x^t, B^t)\}_{t \in T}$ can be rationalized (or is rationalizable) by a preference $\succsim$ if for every $t \in T$ and $y^t \in B^t$, $x^t \succsim y^t$.
\end{definition}

	A preference $\succsim$ rationalizes the consumption data set if, for each observation $t$, the consumer's choice $x^t$ is  $\succsim$-optimal in the budget set $B^t$. This preference is also called a rationalization of the data set. If a utility function $u$ represents the preference, the data set is also said to be rationalized by $u$.   \medskip

We are interested in the rationalizability by a \textit{convex and strictly increasing} preference. Given a consumption data set $\mathcal{O} = \{(x^t, B^t)\}_{t \in T}$, define the weak and strict revealed preference relations $R_{\mathcal{O}}$ and $P_{\mathcal{O}}$ respectively as follows:  \smallskip
	
\noindent (i) $x R_{\mathcal{O}} y$ if either $x = y$ or there exists $t \in T$ such that $x = x^t$ and $y \in B^t$, and 
	
\noindent (ii) $x P_{\mathcal{O}} y$ if there exists $t \in T$ and $z \in B^t$ such that $x = x^t$ and $z >_n y$. \smallskip
	
\noindent The weak  relation $R_{\mathcal{O}}$ is reflexive and directly revealed from the consumer's choices. The strict  relation  $P_{\mathcal{O}}$ combines the observed choices with the  Euclidean order. \medskip

In the setting of \cite{afriat1967construction}, each observation consists of a chosen bundle $x^t$ and a price vector $p^t \in \mathbb{R}^n_{++}$, and the budget set is linear and given by $B^t = \{x: x \cdot p^t \le x^t \cdot p^t\}$. It follows that $x^t R_{\mathcal{O}} y$ if $y \cdot p^t \le x^t \cdot p^t$,  and $x^t P_{\mathcal{O}} y$ if $y \cdot p^t < x^t \cdot p^t$. Our definitions of the weak and strict revealed preference relations coincide with the standard ones in the literature. We have the following theorem for the rationalization condition of $\mathcal{O}$.

\begin{theorem}\label{thm:test:consumer} 
Consider a consumption data set $\mathcal{O} = \{(x^t, B^t)\}_{t \in T}$ such that $\{x^t\}_{t \in T}$ is finite. The following statements are equivalent. \smallskip  
	
\emph{(i)} $\mathcal{O}$ can be rationalized by a strictly increasing and convex preference.  
	
\emph{(ii)} The order pair $p = (R_{\mathcal{O}}, P_{\mathcal{O}})$ satisfies SP on every finite menu $A \subseteq \{x^t\}_{t \in T}$. 
	
\emph{(iii)} $\mathcal{O}$ can be rationalized by a utility function that is strictly increasing, quasi-concave and upper semicontinuous.  
\end{theorem}

By definition, the data set $\mathcal{O}$ can be rationalized by a strictly increasing and convex preference if and only if the order pair $(R_{\mathcal{O}}, >_n)$ admits a convex completion. Since $P_{\mathcal{O}}$ coincides with the restriction of $ R_{\mathcal{O}}\, \diamond >_n$ on $\{x^t\}_{t \in T}$, and the set $\mathring{G}_{R_{\mathcal{O}}}$ coincides with $\{x^t\}_{t \in T}$, Theorem \ref{thm:porder} implies the equivalence between statements (i) and (ii) in Theorem \ref{thm:test:consumer}. This equivalence does not rely on the finiteness of $\{x^t\}_{t \in T}$, although finiteness plays an important role in establishing the equivalence between statement (iii) and the other two statements. Importantly, the upper semicontinuity conclusion does not require any topological restriction on the observed budget sets, as each $B^t$ may be arbitrary. Upper semicontinuity of the rationalizing utility function is nevertheless useful for out-of-sample predictions: It guarantees the existence of an optimal choice from any compact  set. \medskip

By Theorem \ref{thm:algorithm}, when $\{x^t\}_{t \in T}$ is a finite set, it suffices to run Algorithm I for $p$ on $\{x^t\}_{t \in T}$ to verify the rationalizability of the data set, which involves no more than $|T|$ SP checks. We illustrate this procedure in more detail through the following example. \bigskip

\begin{figure}[!htb]  
	\centering 
	\begin{tikzpicture}[scale=10.5]  
		\draw[->] (0, 0) -- (0.8, 0);  
		\draw[->] (0, 0) -- (0, 0.66);  
		
		\node[align=center] at (0.107, 0.573) {$x^1$};  
		\node[align=center] at (0.028, 0.554) {$B^1$};  
		\node[align=center] at (0.028, 0.282) {$B^2$};  
		\node[align=center] at (0.028, 0.172) {$B^3$};  
		
		\filldraw[black] (0.095, 0.55) circle (0.12pt);  
		
		\node[align=center] at (0.230, 0.286) {$x^2$};  
		\filldraw[black] (0.21, 0.2649) circle (0.12pt);  
		
		\node[align=center] at (0.716, 0.026) {$y^3$};  
		\filldraw[black] (0.7, 0) circle (0.12pt);  
		
		\node[align=center] at (0.566, 0.152) {$x^3$};  
		\filldraw[black] (0.55, 0.13) circle (0.12pt);  
		
		\draw[dashed] (0.21, 0.2649) -- (0.095, 0.55);   
		\draw[dashed] (0.21, 0.2649) -- (0.7, 0);  
		\draw[dashed] (0.21, 0.2649) -- (0.55, 0.13);   
		\draw[dashed] (0.095, 0.55) -- (0.55, 0.13);  
		
		\draw[blue, thick] (0, 0.6) -- (0.19, 0.5);  
		\draw[blue, thick] (0.19, 0) -- (0.19, 0.5);   
		\draw[red, thick, domain=0:0.555, smooth, variable=\x] 
		plot ({\x}, {0.309-\x^2});  
		\draw[yellow, thick] (0.55, 0.13) -- (0, 0.2);  
		\draw[yellow, thick] (0.55, 0.13) -- (0.7, 0);  
	\end{tikzpicture}   
	\caption{Checking SP with Consumption Data}\label{fig_budget_convex}  
\end{figure}
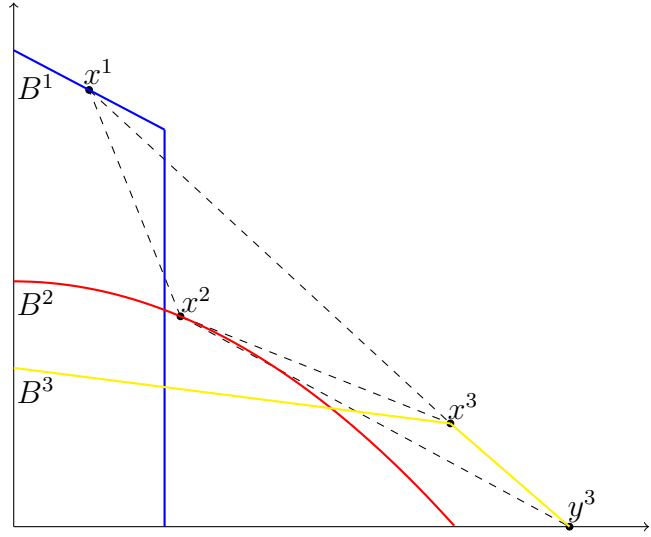

\noindent \textbf{Example 5.} Let $X = \mathbb{R}_+^2$. The consumption data set $\mathcal{O}=\{(x^t, B^t)\}_{t=1}^3$ is  depicted in Figure \ref{fig_budget_convex}: Each budget set $B^t$ contains  bundles on or below the budget line, and   $x^t$ lies on the  budget line. Applying Theorems \ref{thm:algorithm} and \ref{thm:test:consumer}, we run Algorithm I for $p = (R_{\mathcal{O}}, P_{\mathcal{O}})$ on $A=\{x^1, x^2, x^3\}$. We have $\Pi_{R_{\mathcal{O}}}(A) = co(A)$, and for each $t$, $\{y: x^t P_{\mathcal{O}} y\}$ equals the interior of $B^t$. Observe that $co(A)$  does not intersect the interiors of  $B^2$ and $B^3$, and thus  $p$ satisfies SP on $A$. By Algorithm I, it remains to check whether $p$ satisfies SP on the singleton menu $\{x^1\}$, which  holds trivially. Thus, we can find a strictly increasing, quasi-concave and upper semicontinuous utility function to rationalize this data set. \medskip

If the consumer's choice in $B^3$ is $y^3$ instead of $x^3$ (as shown in Figure \ref{fig_budget_convex}), then the modified data set $\mathcal{O}^* = \{(x^1, B^1), (x^2, B^2), (y^3, B^3)\}$ has no convex and strictly increasing rationalizations. This is because $p^* = (R_{\mathcal{O}^*}, P_{\mathcal{O}^*})$ violates SP on $\{x^2, y^3\}$: The line segment connecting $x^2$ and $y^3$ intersects the interiors of both $B^2$ and $B^3$. However, since every chosen bundle does not lie in other budget sets,  Theorem 2 of \cite{nishimura2017comprehensive} implies that $\mathcal{O}^*$ is rationalizable by a strictly increasing preference.   \qed \bigskip

\noindent \textbf{Continuous Rationalization and Connection to \cite{cherchye2014revealed}.} While upper semicontinuity of the rationalizing utility function can be obtained for free under a convex rationalization, continuity cannot be obtained automatically. We illustrate this through the following example. \bigskip

\begin{figure}[!htb]
	\centering
	\begin{tikzpicture}[scale=10.5]
		\draw[->] (0, 0) -- (0.7, 0);
		\draw[->] (0, 0) -- (0, 0.65); 
		
		\node[align=center] at (-0.018, 0.3) {$3$};
		\filldraw[black] (0,0.3) circle (0.12pt);
		
		\node[align=center] at (0.6, -0.025) {$6$};
		\node[align=center] at (-0.018, 0.6) {$6$};
		
		\node[align=center] at (0.025, 0.325) {$x^1$};
		\node[align=center] at (0.225, 0.225) {$y^1$};
		\node[align=center] at (0.625, 0.025) {$x^2$};
		\node[align=center] at (0.545, -0.03) {$y^2$};
		
		\filldraw[black] (0.6, 0) circle (0.12pt);
		\filldraw[black] (0.2, 0.2) circle (0.12pt); 
		
		\node[align=center] at (0.04, 0.515) {$B^2$};
		\node[align=center] at (0.04, 0.235) {$B^1$};
		
		\filldraw[black] (0.55, 0) circle (0.12pt);
		
		\draw[red, thick] (0, 0.3) -- (0.2,0.2);
		\draw[red, thick] (0.2,0.2) -- (0.3, 0);
		
		\draw[red, dashed] (0.2, 0.2) -- (0.6, 0);
		\draw[dashed] (0, 0.3) -- (0.55, 0);
		
		\draw[blue, thick] (0, 0.6) -- (0.6, 0);
	\end{tikzpicture} 
	
	\caption{Continuity and Convexity}\label{fig_conuter}
\end{figure}
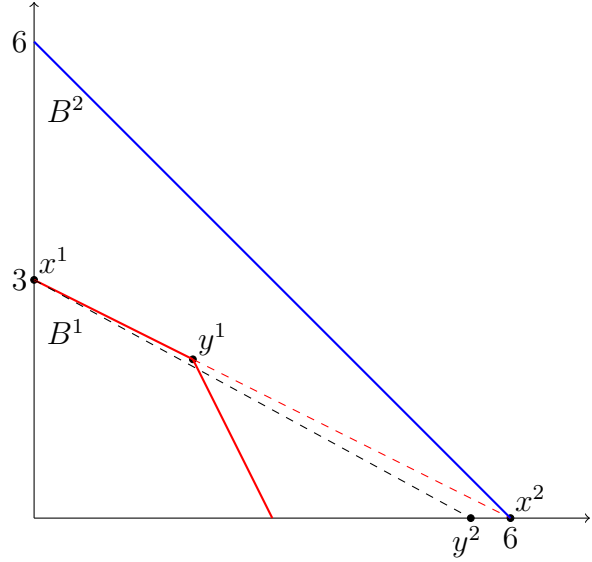

\noindent\textbf{Example 6.} Let $X=\mathbb{R}^2_+$. Consider a data set $\mathcal{O}=\{(x^t, B^t)\}_{t \in \{1,2\}}$ as depicted in Figure \ref{fig_conuter}: Each budget set contains alternatives on or below its  budget line. The chosen bundles are $x^1=(0,3)$  and $x^2=(6,0)$, respectively. The bundle $y^1$ is on the boundary of $B^1$ and lies on the line segment joining  $x^1$ and $x^2$. \medskip

	The data set has a strictly increasing, quasi-concave and upper semicontinuous utility rationalization, as it passes the SP tests in Theorem \ref{thm:test:consumer}. However, the data set cannot be rationalized by any strictly increasing, convex and continuous preference: If such a preference $\succsim$ exists, then we have $x^2 \succ x^1$. By continuity of $\succsim$, any bundle (e.g., $y^2$) sufficiently close to $x^2$ also satisfies $y^2 \succ x^1$. Thus, any convex combination of $x^1$ and $y^2$ must be weakly better than $x^1$, contradicting the fact that some of them are in the interior of  $B^1$.   \qed
	\bigskip

	\cite{cherchye2014revealed} examine the same problem studied in this section, i.e., testing whether a consumption data set $\mathcal{O}=\{(x^t, B^t)\}_{t \in T}$ is rationalizable by a strictly increasing and convex preference. They assume that $T$ is finite and each $B^t$ is closed, downward comprehensive, and has nonempty interior.\footnote{\,A budget set $B^t$ is downward comprehensive if $x \in B^t$ implies $\{y: x >_n y\} \subseteq B^t$.} Theorem 3 of \cite{cherchye2014revealed} claims that the following three statements are equivalent: (i) The data set $\mathcal{O}$ can be rationalized by a locally non-satiated and quasi-concave utility function; (ii) there exists a \textit{co-convex hull} $C^t$ for each observation $t$ such that the new consumption data set $\mathcal{O}^* = \{(x^t, C^t)\}_{t \in T}$ allows for a utility rationalization that is locally non-satiated;\footnote{\,The budget set $C^t$ is a co-convex hull for observation $t$ if it is closed and downward comprehensive, $\mathbb{R}^n_+ \backslash C^t$ is a convex set, $B^t \subseteq C^t$, and for all $x >_n x^t$, $x \notin C^t$.} (iii) the data set can be rationalized by a strictly increasing, concave and continuous utility function. \medskip

	When statement (ii) holds,  the nice properties of the co-convex hulls ensure that $\mathcal{O}^*$ has a utility rationalization with the desired properties---and so does $\mathcal{O}$. Nevertheless,  statement (i)   \textit{cannot} imply statement (ii). To see this, consider Example 6.  Assume  that $C^1$ is the co-convex hull for observation 1. Denote by $D^1$ the complement of $C^1$. The set $D^1$ contains no bundle $z$ that is strictly below the line segment $x^1x^2$, since otherwise, for some $\epsilon > 0$ that is close to $0$,    the line segment $zw^{\epsilon}$ is contained in $D^1$  (where $w^{\epsilon} =x^1+ (\epsilon, \epsilon)$) and intersects with $B^1$, which is a contradiction. Hence, all bundles strictly below the line segment $x^1x^2$ are contained in $C^1$. Since $C^1$ is closed, we have $x^2 \in C^1$. This leads to a contradiction, as $x^2$ is revealed to be strictly better than $x^1$ and should not be contained in any budget set in which $x^1$ is chosen.  \medskip


\noindent \textbf{Testing Symmetric Convex Preferences.} Our framework can also be adapted to test symmetry restrictions. These restrictions are relevant when some coordinates of an alternative are ex ante interchangeable. For example, in a contingent-consumption setting, each coordinate may represent consumption in a different state of the world; if the states are viewed symmetrically ex ante, then the preference should be invariant to relabeling the states. In a social choice application, each coordinate of an alternative may instead represent the consumption assigned to a different individual. If individuals are treated symmetrically, then relabeling them should not affect the social ranking of allocations; symmetry then captures anonymity of the planner's preference. In the Online Appendix, we show how the revealed-preference test developed here can be extended to accommodate this restriction, yielding a result analogous to Theorem \ref{thm:test:consumer}.

\subsection{Convex Completion under Risk and Uncertainty}\label{subsec:risk-uncertainty} 

Convex preferences constitute an important class for modeling DMs' behavior under risk and uncertainty. In lottery choice domains, \cite{cerreia2015cautious} propose the cautious expected utility model, where a DM evaluates each lottery using its minimal certainty equivalence across a set of expected utilities---with convexity capturing the DM's cautiousness. In choice domains with ambiguity, convexity reflects a preference for hedging \citep{gilboa1989maxmin, maccheroni2006ambiguity, strzalecki2011axiomatic}, or alternatively, aversion to ambiguity, which resolves the Ellsberg paradox \citep{ellsberg1961risk}. In this section, we investigate the testing of convex preferences in these domains.

\subsubsection{Convex Preference over Lotteries}\label{subsubsec:risk}
Let $L=[a,b]\subseteq \mathbb{R}$ be the space of monetary payoffs. Denote by $\Delta(L)$ the set of all probability distributions, or lotteries, over $L$. Equivalently, each lottery can be identified with a cumulative distribution function $F:L\to \mathbb{R}$ such that $F(a)\geq 0$, $F(b)=1$, and $F$ is nondecreasing and right-continuous. In this section, we take $\Delta(L)$ as our space of alternatives.
\medskip

A choice data set can be summarized by an order pair $(R,P)$ over $\Delta(L)$. In some experiments, a choice of $x$ over $y$ can be interpreted as indicating that $x$ is strictly preferred to $y$; in others, such behavior is interpreted only as indicating weak preference. We do not take a stand on how such choice data should be interpreted, since our framework allows the test to be conducted under either interpretation. \medskip

On top of the choice data, we consider a preorder $\trianglerighteq$ on $\Delta(L)$. This preorder represents an additional restriction imposed on the DM's preference. Hence, given the choice data $(R,P)$, the order pair for which we seek a convex completion is $(R \,\cup \bowtie,\; P \cup \rhd)$, where $\bowtie$ and $\rhd$ denote the symmetric and asymmetric parts of $\trianglerighteq$, respectively. Notably, many relevant preorders or partial orders satisfy independence. We list several examples below. \bigskip

\noindent \textbf{Example 7 (First-Order Stochastic Dominance).}  
Consider two lotteries $F$ and $G$. We say that $F$ first-order stochastically dominates $G$, denoted $F \succ^{\mathrm{FOSD}} G$, if for all $x \in L$, $F(x) \leq G(x)$, with strict inequality for some $x \in L$. The first-order stochastic dominance relation forms an I-partial order. Preferences extending this order strictly favor shifts toward higher monetary payoffs, a fundamental monotonicity requirement in domains of risk. \qed  
\bigskip

\noindent \textbf{Example 8 (Second-Order Stochastic Dominance).}  
Consider two lotteries $F$ and $G$. We say that $F$ second-order stochastically dominates $G$, denoted $F \succ^{\mathrm{SOSD}} G$, if
$$
\int_a^x F(t)\,dt \leq \int_a^x G(t)\,dt 
\quad \text{for all } x \in [a,b],
$$
with strict inequality for some $x \in [a,b]$. This relation forms an I-partial order. When two lotteries $F$ and $G$ have the same mean, $F \succ^{\mathrm{SOSD}} G$ means that $F$ is less risky than $G$ in the sense of mean-preserving spreads. Thus, consistency with $\succ^{\mathrm{SOSD}}$ captures risk aversion. \qed
\bigskip

We study a general case in which the DM's preference is required to extend a preorder admitting a \textit{multi-utility representation}, which includes the two dominance relations discussed above as special cases. To define this class of preorders, let $\mathcal{U}$ denote the collection of all continuous functions from $L$ to $[0,1]$.

\begin{definition}\label{def:multiU}
A preorder $\trianglerighteq$ on $\Delta(L)$ is said to admit a multi-utility representation if there exists a nonempty set $\mathcal{C} \subseteq \mathcal{U}$ such that, for all $F,G \in \Delta(L)$,
$$
F \trianglerighteq G 
\quad \Longleftrightarrow \quad 
E_F(u) \geq E_G(u) \text{ for every } u \in \mathcal{C}.
$$
\end{definition}

Here, $E_F(u)$ denotes the expected utility of $u$ under the distribution $F$. First-order stochastic dominance can be viewed as the asymmetric part of the preorder represented by
$
\mathcal{C}=\{u\in\mathcal{U}: u \text{ is increasing}\},
$
while second-order stochastic dominance can be viewed as the asymmetric part of the preorder represented by
$
\mathcal{C}=\{u\in\mathcal{U}: u \text{ is increasing and concave}\}.
$
Thus, both examples above are covered by the notion of multi-utility representation. Although in many applications it is natural to impose additional properties on the utilities in $\mathcal{C}$ (such as monotonicity), our results only require these functions to be continuous. \medskip 

A prominent axiomatic treatment of multi-utility representations in this setting is by \cite{dubra2004expected}. They show that a continuous I-preorder on the lottery space, endowed with the weak topology, admits a multi-utility representation by a closed and convex set of utility functions (with $\mathcal{U}$ endowed with the sup norm topology). By contrast, our framework allows for any nonempty set of utilities. \medskip

Following \cite{dubra2004expected}, a preorder that has a multi-utility representation can be interpreted as the part of the DM's preference over lotteries about which she is subjectively confident. The set $\mathcal{C}$ represents utility functions that the DM regards as plausible candidates for her own preferences. Thus, the DM behaves as if she is uncertain about which utility function to rely on, but is willing to rank $F$ above $G$ whenever $F$ yields at least as much expected utility as $G$ under every utility in $\mathcal{C}$.

\begin{theorem}\label{thm:test:uncertain}
Let $\trianglerighteq$ be an I-preorder on $\Delta(L)$ admitting a multi-utility representation $\mathcal{C}$, and let $\bowtie$ and $\rhd$ denote its symmetric and asymmetric parts, respectively. Consider an order pair $(R,P)$ on $\Delta(L)$ such that $H=\mathring{G}_R \cup G_P$ is finite. The following statements are equivalent. \smallskip

\emph{(i)} $(R\,\cup\bowtie, P\cup\rhd)$ admits a convex completion.  

\emph{(ii)} $(R\,\cup\bowtie, P^*)$ satisfies SP on every finite menu  in $H$, with $P^*$ defined by \emph{(\ref{eq:independence})}.  

\emph{(iii)} $(R\,\cup\bowtie, P\cup\rhd)$ admits a convex completion that can be represented by an upper semicontinuous and quasi-concave utility function. \smallskip

\noindent Furthermore, if $R=I_X$, $P$ is finite, and $\mathcal{C}$ is finite, then the utility function in statement \emph{(iii)} can be chosen to be continuous and quasi-concave.
\end{theorem}

By Theorem \ref{thm:test:uncertain}, the finiteness of $\mathring{G}_R \cup G_P$ ensures that upper semicontinuity is a free property of the convex preference extending the targeted order pair. This conclusion does not rely on any topological property of the set of utilities $\mathcal{C}$ representing the preorder. Moreover, when $R=I_X$ and $\mathcal{C}$ is finite, the convex preference can be chosen to further satisfy continuity. \medskip

The equivalence between statements (i) and (ii) follows directly from Theorem \ref{thm:porder}. Hence, Algorithm I provides an efficient way to test for convexity in such cases. Moreover, when $R=I_X$, the target pair has the same completions as $(\trianglerighteq, P \cup \rhd)$. Hence, by Corollary \ref{coro:porder:pure_strict}, it suffices to check a decreasing sequence of menus $(A_k)_{k=1}^m$ contained in $G_P$ and verify that, for each $k$, there exists $F \in A_k$ such that
\begin{equation}\label{eq:lottery:test}
F^{\downarrow}_{P \diamond \trianglerighteq} \cap co(A_k)=\emptyset.    
\end{equation}
We demonstrate this testing procedure through the following example.

\bigskip

\noindent \textbf{Example 9.} Let $L=[0,1]$ and $X=\Delta([0,1])$. Consider an order pair $(R,P)$ with $R=I_X$ and $P=\{(F^t,G^t)\}_{t=1}^2$. Suppose that, for all $a\in[0,1]$, $G^1(a)=G^2(a)=G(a)$ and $F^1(a)=2a-F^2(a)$, where $F^1$ and $G$ are given by:
\begin{equation*}
	\begin{split}
		& F^1(a) = 
		\begin{cases} 
			2a,~a \in [0,\frac{1}{8}], \\
			\frac{1}{4},~ a \in (\frac{1}{8}, \frac{1}{2}], \\
			2a-\frac{3}{4},~ a \in (\frac{1}{2}, \frac{7}{8}], \\
			1,~ a \in (\frac{7}{8}, 1],
		\end{cases}   
		\qquad
		G(a) = 
		\begin{cases} 
			\frac{4}{5}a,~a \in [0,\frac{1}{4}], \\
			\frac{6}{5}a-\frac{1}{10},~ a \in (\frac{1}{4}, \frac{3}{4}], \\
			\frac{4}{5}a + \frac{1}{5},~ a \in (\frac{3}{4}, 1].  
		\end{cases}
	\end{split}   
\end{equation*}
First, we claim that there exists a convex preference that extends $(R,P)$ and respects first-order stochastic dominance. Consider the menu $\{F^1,F^2\}$. At $a=\frac{1}{4}$, we have $F^1(a)>G(a)$ and $F^2(a)>G(a)$, while at $a=\frac{3}{4}$, we have $F^1(a)<G(a)$ and $F^2(a)<G(a)$. Hence, for every $t\in[0,1]$, the mixture $tF^1+(1-t)F^2$ is neither equal to $G$ nor first-order stochastically dominated by $G$. Therefore, both $F^1$ and $F^2$ satisfy condition \eqref{eq:lottery:test}, where the preorder $\trianglerighteq$ in the condition is defined as $I_X\cup \succ^{\mathrm{FOSD}}$. Thus, the targeted order pair passes the convexity test. It follows that there exists a convex preference that can be represented by an upper semicontinuous and quasi-concave utility function, extends $(R,P)$, and respects first-order stochastic dominance. \medskip

However, there does not exist a convex preference that extends $(R,P)$ and respects second-order stochastic dominance. Indeed, for the menu $\{F^1,F^2\}$, the mixture $\frac{1}{2}F^1+\frac{1}{2}F^2$ is the uniform distribution, which is second-order stochastically dominated by $G$.
\qed \bigskip

Apart from data obtained from individuals' direct comparisons between lotteries, recent studies \citep{AgranovOrtoleva2017, CDOR2019aer} suggest that random choice behavior may reflect a \textit{preference for randomization}, and such behavior can be interpreted as evidence of a locally convex preference over lotteries. \medskip

More specifically, let $A$ be a finite menu in $\Delta(L)$, and suppose that a DM's repeated choices from $A$ induce a choice distribution $\rho \in \Delta(A)$. One way to interpret $\rho$ is through its reduced lottery $G=\sum_{F\in A}\rho(F)F,$  which may be viewed as the DM's most preferred choice in $co(A)$, either in a weak or in a strict sense. While this literature provides evidence and models suggesting that stochastic choice can be rationalized by such locally convex preferences, our testing method allows one to ask whether the DM's preferences over lotteries satisfy global convexity. \medskip

In the Online Appendix, we implement our test using the recent data set of \cite{AlosFerrerBuckenmaierGaragnaniRustichini2026}. In this data set, the lotteries faced by subjects have relatively small common supports, making the data particularly suitable for our analysis. Examining individual choices, we find that convexity has no additional empirical bite when preferences are required to respect first-order stochastic dominance, whereas it does have additional bite when preferences are required to respect second-order stochastic dominance. If convex preferences are interpreted as reflecting caution and consistency with second-order stochastic dominance as reflecting risk aversion, the latter finding suggests that these two behavioral properties need not coincide empirically.

\subsubsection{Convex Preference over State-Contingent Acts}\label{subsubsec:ambiguity}  
We adopt the Anscombe--Aumann framework to study choices over ambiguity \citep{anscombe1963definition}. Let $S$ be a nonempty finite state space. An act is a function from $S$ to $\Delta(L)$. Let $\mathcal{F}$ be the set of all acts, i.e., the space of alternatives. \medskip

We assume that there exists a continuous and strictly increasing utility function $u:L\to\mathbb{R}$ such that the DM's preference between constant acts, identified with lotteries in $\Delta(L)$, depends only on their expected utilities under $u$. This assumption is standard in the ambiguity literature: Nonlinearities in preferences over acts typically arise from how state-contingent expected utilities are aggregated across states. We note that identifying the utility function $u$ is separate from the convexity problem studied here; see \cite{caradonna2024revealed} for tests and recovery from incomplete preference data.
\medskip

For any state $s\in S$ and act $f\in\mathcal{F}$, let $u(f(s))$ denote the expected utility of lottery $f(s)$. For any two acts $f,g\in\mathcal{F}$, write $f\ge_u g$ if $u(f(s))\ge u(g(s))$ for every $s\in S$, and write $f\gg_u g$ if $u(f(s))>u(g(s))$ for every $s\in S$. We impose the monotonicity condition
\begin{equation}\label{eq:ambiguity:monotone}
	\forall f,g\in\mathcal{F},\quad
	f\ge_u g \text{ implies } f\succsim g
	\quad\text{and}\quad
	f\gg_u g \text{ implies } f\succ g .
\end{equation}

Condition (\ref{eq:ambiguity:monotone}) differs from standard consumer preference monotonicity: Strict preference is guaranteed only when $f$ improves upon $g$ in every state, not merely in some state. Such a requirement is natural for many ambiguity-sensitive models. For example, under the following maxmin preference \citep{gilboa1989maxmin}, where
\[
	f\succsim g
	\quad\text{if and only if}\quad
	\min_{s\in S}u(f(s))\ge \min_{s\in S}u(g(s)),
\]
an improvement in some states need not affect the worst-case payoff, whereas an improvement in every state does. Similar to the previous section, we consider a data set represented by an order pair $(R,P)$ and impose condition \eqref{eq:ambiguity:monotone} on the convex preference to be recovered.

\begin{theorem}\label{thm:ambiguity}
Let $(R,P)$ be an order pair on $\mathcal{F}$ such that $H=\mathring{G}_R \cup G_P$ is finite. Let $u$ be a continuous and strictly increasing utility function from $L$ to $\mathbb{R}$. The following statements are equivalent.  \smallskip

\emph{(i)}  $(R\,\cup \ge_u, P\,\cup \gg_u)$ admits a convex completion.  

\emph{(ii)}  $(R\,\cup \ge_u, P^*)$ satisfies SP on every finite menu  in $H$, where 
\[P^* = \{(f,g)\in H\times \mathcal{F}:
fPg \text{ or } fP\,\diamond \gg_u g \text{ or } fR\,\diamond \gg_u g\}.
\]

\emph{(iii)}  $(R\,\cup \ge_u, P\,\cup \gg_u)$ admits a convex completion that can be represented by an upper semicontinuous and quasi-concave utility function. \smallskip

\noindent Furthermore, if $R=I_X$ and $P$ is finite, then the utility function in statement \emph{(iii)} can be chosen to be continuous and quasi-concave.
\end{theorem}

Although the strict relation $\gg_u$ is not the asymmetric part of the preorder $\ge_u$, unlike those studied in the previous sections, the desired  continuity property is preserved.

\subsection{Convex Preference over Information}\label{subsec:information}
\cite{blackwell1951comparison,blackwell1953equivalent} establish the celebrated Blackwell order, which characterizes when one information structure generates a weakly higher maximal expected payoff than another in every Bayesian decision problem. Subsequently, \cite{azrieli2008value} characterize preference orders over information structures that can be generated by a fixed Bayesian decision problem. These papers establish connections between the underlying decision problems and the resulting preferences over information. \medskip

In this section, we consider a finite set of observed comparisons between information structures and apply our method to test whether these comparisons can be rationalized by a convex preference that respects the Blackwell order. We show that our test admits a useful decision-theoretic interpretation: Whenever such a rationalization exists, the same data can be rationalized as if the DM evaluates information robustly across a finite collection of Bayesian decision problems. \medskip

Let $\Omega=\{1,\ldots,n\}$ be a finite state space and fix a prior $\mu\in\Delta(\Omega)$. An information structure can be represented by the distribution of posterior beliefs it induces. We let
\[
\mathcal{I}(\mu)
=
\left\{
\pi\in\Delta(\Delta(\Omega)):
\int_{\Delta(\Omega)}q\,\pi(dq)=\mu
\right\}
\]
be the set of Bayes-plausible distributions of posterior beliefs. The set $\mathcal{I}(\mu)$ is convex. For any $\pi,\rho\in\mathcal{I}(\mu)$, write $\pi\succsim_B\rho$ if $\pi$ Blackwell dominates $\rho$, that is, $\int_{\Delta(\Omega)}v(q)\pi(dq) \geq \int_{\Delta(\Omega)}v(q)\rho(dq)$ for every continuous convex function $v:\Delta(\Omega)\rightarrow\mathbb{R}$. For any such $v$, define
$$V_v(\pi) = \int_{\Delta(\Omega)}v(q)\pi(dq). $$
Note that $V_v$ is affine and respects the Blackwell order: $\pi\succsim_B\rho$ implies $V_v(\pi)\geq V_v(\rho)$. \medskip

We say that a preference $\succsim$ on $\mathcal{I}(\mu)$ is \textit{Blackwell monotone} if $\pi\succsim_B\rho$ implies $\pi\succsim\rho$. Consider a data set $\mathcal{D}=\{(\pi^t,\rho^t)\}_{t\in T}$, where observation $t$ means that the DM \textit{strictly} prefers $\pi^t$ to $\rho^t$. We interpret the data in a strict sense because the Blackwell order imposes only weak monotonicity; if the choice of $\pi^t$ over $\rho^t$ were interpreted as revealing only a weak preference, then the data would always admit a convex and Blackwell-monotone rationalization by taking the DM to be indifferent among all information structures. \medskip

To state our representation result, recall that a Bayesian decision problem with a finite action set consists of a finite action set $Z$ and a payoff function $w:Z\times\Omega\rightarrow\mathbb{R}$. Its value at posterior belief $q$ is $v(q) = \max_{z\in Z}
\sum_{\omega\in\Omega}q(\omega)w(z,\omega).$ Hence, $v$ is a polyhedral convex function on $\Delta(\Omega)$, i.e., $v$ can be written as the maximum of finitely many linear functions on $\Delta(\Omega)$. Conversely, every polyhedral convex function on $\Delta(\Omega)$ is the value function of some Bayesian decision problem with finite actions.

\begin{definition}\label{def:finite:robust:information}
A utility function $U:\mathcal{I}(\mu)\rightarrow\mathbb{R}$ admits a
\textit{robust finite representation} if there exist
finitely many Bayesian decision problems with finite action sets that induce the posterior value functions $\hat{\mathcal{V}}=\{v_1,\ldots,v_L\}$,  
such that for every $\pi\in\mathcal{I}(\mu)$:
\begin{equation}\label{eq:finite:robust:information}
U(\pi)
=
\min_{v \in \hat{\mathcal{V}}} V_v(\pi).
\end{equation} 
\end{definition}

We can now characterize convex rationalizability of the observed comparisons.

\begin{theorem}\label{thm:test:information}
Consider a nonempty data set
$\mathcal{D}=\{(\pi^t,\rho^t)\}_{t\in T}$ with $T$ being finite. Let
$G=\{\pi^t\}_{t\in T}$. The following statements are equivalent. \smallskip 

\emph{(i)} There exists a convex and Blackwell-monotone preference
$\succsim$ on $\mathcal{I}(\mu)$ such that, for every $t\in T$,
$\pi^t\succ\rho^t$. \smallskip

\emph{(ii)} For every nonempty $A\subseteq G$, there is $\pi\in A$ such
that, for every $\hat{\pi}\in co(A)$ and every $r\in T$ with
$\pi^r=\pi$, $\rho^r\not\succsim_B\hat{\pi}.$ 
\smallskip

\emph{(iii)} There is a utility
function $U$ that admits a robust finite representation such that for every $t \in T$, $U(\pi^t) > U(\rho^t)$.
\end{theorem}

Theorem \ref{thm:test:information} gives a decision-theoretic interpretation of convexity over information structures. The DM behaves as if she considers finitely many possible decision problems, evaluates each information structure according to its instrumental value in every problem, and uses the worst value. Note that the resulting utility function satisfies continuity and concavity, while we start with only the convexity of the preference. In this sense, the result parallels
Afriat's theorem. 
\medskip

Our result is related to \cite{shmaya2006value}, who characterizes globally defined value-of-information functions in zero-sum games. His characterization requires the value function to be concave, upper semicontinuous, and monotone with respect to the Blackwell order. In contrast, our primitive restriction is only ordinal convexity of preference, or equivalently quasi-concavity of a utility representation. The stronger concave representation obtained here is made possible by the finiteness of the data set.

\section{Rationalization and Convex Rationalization}\label{sec:garp_sp}
	
	The seminal result of \cite{afriat1967construction} establishes that for a finite consumption data set with linear budget sets, if it has a strictly increasing rationalization, then the rationalization can additionally be continuous and convex. \cite{reny15} investigates infinite data sets with linear budget sets and shows that while continuity may fail, convexity remains  an implication of strictly increasing rationalization.  \medskip

	In this section, we link our characterization of convex completions to this fundamental insight and   provide a condition under which imposing convexity on the rationalization entails no loss of generality. We then apply the theorem to  extend the results of \cite{afriat1967construction} and \cite{reny15} for consumer choice data and derive a characterization of  rationalization in the sense of \cite{richter1966revealed}.

	\subsection{Pre-Convexity} 
	In this section, we provide a condition under which convexity of the rationalization comes at no cost. We begin by introducing the following definition.  
	
	\begin{definition}
		An order pair $p=(R, P)$ is weakly pre-convex if for all $x \in X$ with $x R x'$ for some $x' \neq x$, the set $\{z: \text{not~} x R z\}$ is convex. Given a set $A$, $p$ is pre-convex on $A$ if it is weakly pre-convex, and for all $x \in A$, the set $\{z: \text{not~}  x P z\}$ is convex. The order pair $p$ is pre-convex if it is pre-convex on $X$. 
	\end{definition}

	Consider an order pair $p=(\succsim, \succ)$, where $\succsim$ is a  preference. If $\succsim$ is convex, then for all $x \in X$, both $\{z: x \not\succsim z\}= \{z: z \succ x\}$ and 
	$\{z: x \not\succ  z\}= \{z: z \succsim x\}$ are convex sets, and thus $p$ is pre-convex. Therefore, our notion of pre-convexity generalizes the standard definition of convexity to order pairs.

	\begin{theorem}\label{thm:sp=garp}
		Consider an order pair $p=(R, P)$ and a  menu $A$.  If $p$  is pre-convex on $A$ and consistent, then $p$ satisfies SP on all finite menus  in $A$. In particular, if $p$ is pre-convex, then $p$ admits a convex completion if and only if it is consistent. 
	\end{theorem}

	Recall that $p$ is consistent \citep{suzumura1976remarks} if there is no  $(x^k)_{k=1}^n$ with $n \ge 2$  such that $x^{k+1} (R \cup P) x^k$ for all $k \in \{1,\dots,n-1\}$ and $x^1 P x^n$. This condition is equivalent to the existence of a completion of $p$. Thus,  for a pre-convex order pair, it admits a completion if and only if it admits a convex completion.  \medskip

	Our pre-convexity condition is closely related in spirit to the co-convexity condition proposed by \cite{matzkin1991axioms}. Specifically, \cite{matzkin1991axioms} considers a data set $\mathcal{O}=\{(x^t,B^t)\}_{t\in T}$ where, for each $t\in T$, $B^t\subseteq \mathbb{R}_+^n$ is a budget set and $x^t\in B^t$ is the observed chosen bundle. Each $B^t$ is assumed to be \emph{co-convex}, meaning that it is closed and its complement in $\mathbb{R}_+^n$ is convex. Under co-convexity, \cite{matzkin1991axioms} shows that the existence of a preference under which each $x^t$ is the unique best bundle in $B^t$ further implies the existence of such a preference that is additionally strictly convex. In this sense, co-convexity makes strict convexity a ``free lunch.'' Note that the key feature of the co-convexity condition is that, for each $t\in T$, the set of alternatives outside the budget set $B^t$---that is, those not revealed to be strictly worse than $x^t$---forms a convex set. Our pre-convexity condition shares the same idea.

	\subsection{A Generalization of \cite{afriat1967construction} and \cite{reny15}}
	In this section, we fix $X$ to be the consumption space $\mathbb{R}_+^n$ and generalize the results of \cite{afriat1967construction} and \cite{reny15}. \medskip

	Consider a possibly infinite consumption data set $\mathcal{O}=\{(x^t, B^t)\}_{t \in T}$. We assume each budget set $B^t$ is downward comprehensive, i.e., $x \in B^t$ implies $\{y: x>_n y\} \subseteq B^t$.\footnote{\,This assumption is not necessary for our analysis, but it simplifies the notation and makes it consistent with the literature.} For each $t \in T$, define the dominated set $\hat{B}^t=\{x \in X: y >_n x \text{~for some~} y \in B^t\}$. Since  $B^t$ is downward comprehensive, we have $\hat{B}^t \subseteq B^t$. \medskip

Recall that in Section \ref{subsec:non-linear-budget}, we define the weak and strict revealed preference relations $R_{\mathcal{O}}$ and $P_{\mathcal{O}}$ such that $x R_{\mathcal{O}} y$ ($x P_{\mathcal{O}} y$) if for some $t \in T$, $x=x^t$ and $y \in B^t$ ($y \in \hat{B}^t$). We say that $\mathcal{O}$ satisfies the generalized axiom of revealed preference (GARP, \citep{afriat1967construction, varian1982nonparametric}) if the order pair $(R_{\mathcal{O}}, P_{\mathcal{O}})$ is consistent on $\{x^t\}_{t \in T}$, i.e., there exists no $x^{t_1}R_{\mathcal{O}}x^{t_2}R_{\mathcal{O}}...R_{\mathcal{O}}x^{t_n}P_{\mathcal{O}}x^{t_1}$ with $n \ge 2$.  
\medskip
	

\begin{theorem}\label{thm:generalize:AR}
Consider a data set $\mathcal{O}=\{(x^t, B^t)\}_{t \in T}$. If for each $t \in T$, the two sets $\mathbb{R}_+^n\backslash B^t$ and $\mathbb{R}_+^n\backslash \hat{B}^t$ are both  convex,  then $\mathcal{O}$ is rationalizable by a strictly increasing and convex  preference if and only if it satisfies GARP.
\end{theorem}

The primitive condition in Theorem \ref{thm:generalize:AR} implies that
\ensuremath{(R_{\mathcal{O}},P_{\mathcal{O}})} is pre-convex. By Theorem
\ref{thm:sp=garp}, it therefore satisfies SP on every finite menu in \ensuremath{\{x^t\}_{t\in T}}. The result now follows from the equivalence between convex rationalizability and the SP condition.
\medskip

Theorem \ref{thm:generalize:AR} generalizes the results in \cite{afriat1967construction} and \cite{reny15}. In their framework, each budget set $B^t$ is linear with $B^t=\{x \in \mathbb{R}^n_+: x \cdot p^t \le x^t \cdot p^t\}$ for some price vector $p^t \in \mathbb{R}^n_{++}$. It follows that $\hat{B}^t=\{x \in \mathbb{R}^n_+: x \cdot p^t < x^t \cdot p^t\}$. Therefore, the complements of $B^t$ and $\hat{B}^t$ are given respectively by  
	\begin{equation*}
		\{y: y \cdot p^t > x^t \cdot p^t\} \text{~and~}
		\{y: y \cdot p^t \ge x^t \cdot p^t\}, 
	\end{equation*}
both of which are convex. By Theorem \ref{thm:generalize:AR}, with (possibly infinitely many) linear budget sets, requiring the rationalization to be convex incurs no additional cost.  \medskip

We note that for data sets where not all budget sets  satisfy the primitive assumption in Theorem \ref{thm:generalize:AR}, GARP remains necessary but may not be sufficient for convex rationalizability of the data sets. Nevertheless, Theorem  \ref{thm:sp=garp} can still be useful to simplify the testing of convexity. We demonstrate this through an example in the Online Appendix.


	\subsection{Exact Rationalization}
	The solution concept in  \cite{afriat1967construction} requires each chosen bundle to be \textit{one of} the consumer's optima within the budget set, allowing for other optimal bundles. An alternative solution concept, investigated by \cite{richter1966revealed} and \cite{matzkin1991axioms}, among others, demands that the chosen bundles \textit{exactly coincide} with the set of optima. We adopt the second solution concept and investigate data sets that have convex rationalizations.\medskip

A \textit{general consumption data set} is a nonempty collection $\mathcal{O}=\{(A^t, B^t)\}_{t \in T}$ such that for all $t\in T$, $\emptyset \neq A^t \subseteq B^t \subseteq \mathbb{R}^n_+$, and for all distinct $t,t' \in T$, $B^t \neq B^{t'}$. Throughout this section, we also maintain the assumption that each budget set is downward comprehensive and for each $t$ and $x \in A^t$, there is no $y \in B^t$ such that $y >_n x$. To interpret, in each budget set $B^t$, bundles in $A^t$ are chosen by the consumer.  The notion of exact rationalization, defined as follows, posits that $A^t$ coincides with the set of all optimal bundles in $B^t$.

	\begin{definition}
	A general consumption data set $\mathcal{O}=\{(A^t, B^t)\}_{t \in T}$ is exactly rationalizable by a preference $\succsim$ if for all $t \in T$, $A^t=\max(B^t; \succsim)$. 
	\end{definition}
	
	We are interested in  data sets that are exactly rationalizable by strictly increasing and convex preferences. Such a data set  necessarily satisfies two conditions. First, the data set must be exactly rationalizable by a strictly increasing preference. This can be characterized by the Congruence axiom \citep{richter1966revealed}: Define $R^*_{\mathcal{O}}$ such that $x R^*_{\mathcal{O}} y$ if there is $t \in T$ with $x\in A^t$ and $y \in B^t$, and   $P^*_{\mathcal{O}}$ such that $x P^*_{\mathcal{O}} y$ if there is $t \in T$ with $x \in A^t$ and $y \in B^t\backslash A^t$. The data set $\mathcal{O}$ satisfies the Congruence axiom if for all $x^{t_1} R^*_{\mathcal{O}} x^{t_2} R^*_{\mathcal{O}} ... R^*_{\mathcal{O}} x^{t_n}$,   $x^{t_n} \in A^{t_n}$ and $x^{t_1} \in B^{t_n}$ imply $x^{t_1} \in A^{t_n}$. This axiom is satisfied if and only if the order pair $(R^*_{\mathcal{O}}, P^*_{\mathcal{O}})$ is consistent on $\cup_{t \in T} A^t$. Thus, following a similar proof to that of Theorem 2 in \cite{nishimura2017comprehensive}, we can show that the Congruence axiom is sufficient and necessary for $\mathcal{O}$ to be  exactly rationalizable by a strictly increasing preference. \medskip

	Second, the data set should be \textit{convex valued}, that is, for all $t \in T$, $co(A^t) \cap B^t =A^t$. This is implied by the convexity of the rationalization: If a convex preference $\succsim$ exactly rationalizes $\mathcal{O}$, then   bundles in $co(A^t) \cap B^t$ are weakly $\succsim$-better than some bundle in $A^t$, meaning that they are optimal in $B^t$ and should be chosen. Our next theorem states that under an analogous assumption to that in Theorem \ref{thm:generalize:AR}, the two conditions are also sufficient for the exact rationalization of $\mathcal{O}$ by a convex and strictly increasing preference.

\begin{theorem}\label{thm:generalize:exact}
Consider a general consumption data set such that for all $t \in T$,  $B^t$ is closed, and $\mathbb{R}_+^n\backslash B^t$ is a convex set. The following three statements are equivalent: \smallskip

		\emph{(i)} $\mathcal{O}$ is convex valued and satisfies the Congruence axiom.

		\emph{(ii)} $\mathcal{O}$ is convex valued and exactly rationalizable by a strictly increasing preference.
		
		\emph{(iii)} $\mathcal{O}$ is exactly rationalizable by a strictly increasing and convex preference. 
	\end{theorem}
	
When each budget set $B^t$ of $\mathcal{O}$ is linear and closed, the set $\mathbb{R}_+^n \backslash B^t$ is convex. By Theorem \ref{thm:generalize:exact}, $\mathcal{O}$ is exactly rationalizable by a strictly increasing and convex preference if and only if each $A^t$ is convex, and $(R^*_{\mathcal{O}}, P^*_{\mathcal{O}})$ is consistent. \medskip

\begin{figure}[!htb] 
	\centering 
	\begin{subfigure}{\textwidth}   
		\centering 
		\begin{tikzpicture}[scale=0.9] 
			\draw[->] (0, 0) -- (7, 0); 
			\draw[->] (0, 0) -- (0, 6.2); 
			
			\node[align=center] at (0.32, 5) {$B^2$}; 
			
			\node[align=center] at (0.32, 3.5) {$B^1$}; 
			
			\draw[blue, thick, domain=0:6] plot (\x, {6.928 - sqrt(-(\x^2) + 12*\x + 12)});

			\draw[red, thick, domain=0:3.464] plot (\x, {6 - sqrt(48.000384-(\x-6.928)^2)}); 
			
			\filldraw[black] (4, 0.295) circle (1.2pt) node[above right, inner sep=1pt] {$x^1$}; 
			\filldraw[black] (5, 0.072) circle (1.2pt) node[above right, inner sep=1pt] {$y^1$}; 
			\filldraw[black] (1, 2.414) circle (1.2pt) node[above right, inner sep=1pt] {$x^2$}; 
			\filldraw[black] (3.25, 0.137) circle (1.2pt) node[above right, inner sep=1pt] {$y^2$};  
		\end{tikzpicture}  
	\end{subfigure}  
	\caption{Exact Rationalization with Multiple Chosen Bundles}\label{fig:exact} 
\end{figure}
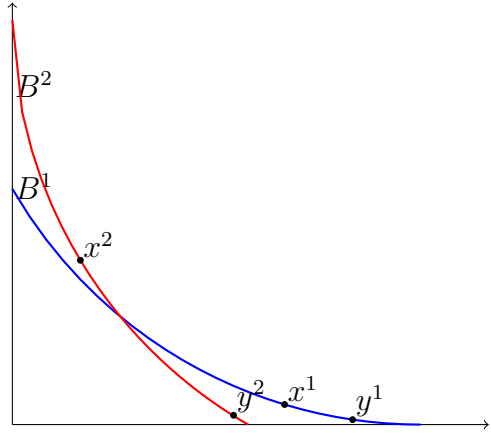

Recall that in the setting of \cite{matzkin1991axioms}, each $B^t$ is closed and has convex complement. Moreover, since each $A^t$ is a singleton contained in $B^t\backslash\hat{B}^t$, the data set is convex valued. \cite{matzkin1991axioms} shows that such a data set is exactly rationalizable by a strictly increasing, strictly convex, and continuous preference if and only if it satisfies the Congruence axiom. Theorem \ref{thm:generalize:exact} is a parallel result in which we allow an infinite number of observations (so continuity of the rationalizing preference is not guaranteed), focus on convexity rather than strict convexity, and permit the consumer to choose multiple alternatives from a given budget set. Figure \ref{fig:exact} depicts a general consumption data set with two observations where $A^1=\{x^1, y^1\}$ and $A^2=\{x^2, y^2\}$. It can be easily verified that it satisfies the conditions in Theorem \ref{thm:generalize:exact} and thus is exactly rationalizable in our sense.

\section*{Appendix 1: Convexity and Upper Semicontinuity}

In this section, we present a general result showing that upper semicontinuity comes for free from the existence of a convex preference
when the preference is required to (1) extend an order pair $(\trianglerighteq_K,\rhd_\gamma)$ satisfying suitable continuity and linearity properties, and (2) rationalize the choices of finitely many alternatives, in either the weak or strict sense, against possibly arbitrary alternatives in their respective choice sets.

\begin{apptheorem}\label{thm:convex:uppersemi} Let $X$ be a nonempty and convex subset of a locally convex Hausdorff topological
vector space $Y$, and let $K$ be a closed convex cone in $Y$. Suppose that there exists a compact and convex set $B^0\subseteq Y$
such that $X\subseteq B^0+K.$ Let $\gamma:K\rightarrow\mathbb{R}_+$ be continuous, concave, and
positively homogeneous. Define two binary relations on $X$ such that $x\trianglerighteq_K y$ if $x-y\in K$ and $x\rhd_\gamma y$ if $x \trianglerighteq_K y$ and $\gamma(x-y)>0.$  \smallskip

For any convex completion  $\succsim_0$ of $(\trianglerighteq_K, \rhd_\gamma)$ and any finite set $D \subseteq X$, there is an upper semicontinuous and quasi-concave
utility function $u:X\rightarrow\mathbb{R}$ such that \smallskip 

\emph{(i)} for all $x,y \in X$,  $x\trianglerighteq_K y$ implies $u(x)\ge u(y)$ and $x \rhd_\gamma y$ implies $u(x)>u(y)$,  and  \smallskip 

\emph{(ii)} for all $x\in D$ and $y\in X$, $x\succsim_0 y$ implies 
$u(x)\geq u(y)$, and $x\succ_0 y$ implies $u(x)>u(y).$ 
\end{apptheorem}

\begin{proof}[Proof of Theorem \ref{thm:convex:uppersemi}]
Concavity and positive homogeneity of $\gamma$ imply that $\gamma$ is superadditive, i.e.,   for all  $x,y\in K,$ we have $\gamma(x+y)\geq\gamma(x)+\gamma(y).$  
Divide $D$ into indifference blocks $D^1,\ldots,D^m$ under
$\succsim_0$, ordered such that every alternative in
$D^k$ is strictly preferred to every alternative in $D^l$ when $k>l$. Here, we assume $D\neq\emptyset$ and thus $m>0$. The case where $D=\emptyset$ will be discussed at the end of the proof. Let $M=\{1,...,m\}$, $M^-=\{0,...,m-1\}$, $C^0=X$ and $C^{m+1}=\emptyset$.  For every $k\in M$, define
\[
A^k=\bigcup_{l=k}^mD^l, ~B^k=co(A^k), \text{~and~} C^k=(B^k+K)\cap X.
\] 
The sets $(C^k)_{k=0}^{m+1}$ are decreasing with respect to set inclusion, convex and closed with respect to the relative topology on $X$: $C^0$ and $C^{m+1}$ are obviously closed, and for each $k \in M$, the closedness of $C^k$ is by the compactness of $B^k$ and closedness of $K$. \medskip

Let $\succ_0$ denote the asymmetric part of $\succsim_0$. For any  $k \in M$ and $x \in D^k$,  by convexity of $\succsim_0$ and $\trianglerighteq_K \subseteq \succsim_0$, we have
\begin{equation}\label{eq:C:weak}
C^k\subseteq\{y:y\succsim_0x\},~ C^{k+1}\subseteq\{y:y\succ_0x\} \text{~and~} D^k\subseteq C^k\backslash C^{k+1}.
\end{equation} 
For any $k \in M^-$ and $a\geq0$, let $H^k(a)$ be the set of
$x\in X$ for which there exist $\alpha\in[0,1], y\in B^{k+1}, z\in B^k$ (note that $B^0$ is given as a primitive) and
$d\in K$  such that
\begin{equation}\label{eq:H:representation}
x=\alpha y+(1-\alpha)z+d \text{~and~} \gamma(d)\geq(1-\alpha)a.
\end{equation} 
For $k=m$, define
\begin{equation}\label{eq:H:m}
H^m(a)
=
\left\{
x\in X:
 x=z+d
\text{ for some }z\in B^m,\ d\in K,
\text{~and~} \gamma(d)\geq a
\right\}.
\end{equation} By the construction, for all $k \in M^- \cup \{m\}$, $H^k(0)=C^k$ and $H^k(b)\subseteq H^k(a)$ whenever $b>a$. Here, $H^0(0)=X$ follows from the primitive condition $X\subseteq B^0+K$. Moreover, for all $k \in M^- \cup \{m\}$, we have $C^{k+1}\subseteq H^k(a)$ for every $a \ge 0$. \medskip  

Each $H^k(a)$ is convex. To see this, take any $k\in M^-$. Consider
$x_i\in H^k(a)$ with representations
$(\alpha_i,y_i,z_i,d_i)$ as in (\ref{eq:H:representation}), $i=1,2$, and let $t\in[0,1]$. Let $\alpha=t\alpha_1+(1-t)\alpha_2,$ and $d=td_1+(1-t)d_2.$ 
Consider the nontrivial case where $\alpha\in(0,1)$. Let
$$y=
\frac{t\alpha_1y_1+(1-t)\alpha_2y_2}{\alpha}
\text{~and~} z=
\frac{t(1-\alpha_1)z_1+(1-t)(1-\alpha_2)z_2}{1-\alpha}.$$ We have $y \in B^{k+1}$, $z \in B^{k}$ and $tx_1+(1-t)x_2=\alpha y+(1-\alpha)z+d.$
Since $\gamma(d_1) \ge (1-\alpha_1)a$ and $\gamma(d_2) \ge (1-\alpha_2)a$, the concavity of $\gamma$ implies $\gamma(d) \ge (1-\alpha)a$. Thus, $tx_1+(1-t)x_2\in H^k(a)$. 
The argument for $H^m(a)$ is immediate.\medskip

Each $H^k(a)$ is closed. Consider first $k \in M^-$. Let
$(x_i)_{i \in I}$ in $H^k(a)$ be a net converging to $x$. Choose a representation
$(\alpha_i,y_i,z_i,d_i)$ for each $x_i$ as in (\ref{eq:H:representation}). By compactness of
$[0,1]\times B^{k+1}\times B^k$, there is a subnet with $(\alpha_{i_j})_{j \in J} \rightarrow\alpha$, $(y_{i_j})_{j \in J} \rightarrow y\in B^{k+1}$ and $(z_{i_j})_{j \in J} \rightarrow z\in B^{k}$. 
It follows that $(d_{i_j})_{j \in J}\rightarrow d=x-\alpha y-(1-\alpha)z.$ Since $K$ is closed, we have $d\in K$, and continuity of $\gamma$ implies
$\gamma(d)\geq(1-\alpha)a$. Hence $x\in H^k(a)$. The argument for
$H^m(a)$ is the same. \medskip 

Next, we show that for every $k \in \{0\} \cup M$, 
\begin{equation}\label{eq:H:intersection}
\bigcap_{a\geq0}H^k(a)=C^{k+1}, 
\end{equation} where $C^{m+1}=\emptyset$. For $k<m$, by the construction of $H^k(a)$, we have $C^{k+1} \subseteq H^k(a)$ for each $a \ge 0$. Thus, $C^{k+1} \subseteq \bigcap_{a\geq0}H^k(a)$.  Conversely, suppose that $x\in H^k(a)$ for every
$a\geq0$. Take a sequence $(a_j)_{j=1}^{+\infty} \rightarrow +\infty$ and write for each $a_j$, $x=\alpha_jy_j+(1-\alpha_j)z_j+d_j$, which is a representation of $x$ in $H^k(a_j)$ according to (\ref{eq:H:representation}). Since $d_j=x-\alpha_jy_j-(1-\alpha_j)z_j$, the set $\{d_j\}_{j=1}^{+\infty}$ is contained in a compact set. Since $K$ is closed,$\gamma$ is
continuous on $K$, and each $d_j$ is in $K$, $\{\gamma(d_j)\}_{j=1}^{+\infty}$ is bounded. Since for each $j$, $\gamma(d_j)\geq(1-\alpha_j)a_j$, it follows that $(\alpha_j)_{j=1}^{+\infty}$ converges to $1$. Consider a convergent subnet $(y_{j_t})$ of $(y_j)_{j=1}^{+\infty}$ and $(z_{j_t})$ of $(z_j)_{j=1}^{+\infty}$ that converge to $y$ and $z$ respectively. It follows that $(d_{j_t})$ converges to $x-y$. Since $K$ is closed, $x-y\in K$, and thus
\[
x\in(B^{k+1}+K)\cap X=C^{k+1}.
\]
For $k=m$, if $x\in H^m(a)$, then $x=z+d$ for some $z\in B^m$ and
$d\in K$ with $\gamma(d)\geq a$. The possible values of $d=x-z$
form a compact set, on which $\gamma$ is bounded. Hence no fixed $x$
can belong to $H^m(a)$ for every $a\geq0$. This proves
(\ref{eq:H:intersection}). \medskip

The last property we show for $H^k(a)$ is that for each $k \in \{0\}\cup M$, $x \in H^k(a)$ and $\hat{d} \in K$, if $x+\hat{d} \in X$, then $x+\hat{d}\in H^k(a+\gamma(\hat{d})).$ 
For $k<m$, let $x$ be represented by $(\alpha, y,z, d)$ as in (\ref{eq:H:representation}). By the superadditivity of $\gamma$, we have $\gamma(d+\hat d) \geq\gamma(d)+\gamma(\hat d) \geq(1-\alpha)a+\gamma(\hat d) \geq(1-\alpha)(a+\gamma(\hat d)).$ The case $k=m$ follows directly from the definition of $H^m(a)$.\medskip

We proceed to define the utility function. For every $k \in \{0\}\cup M$ and $x\in C^k\backslash C^{k+1}$, define
\[
\lambda_k(x)
=
\max\{a\geq0:x\in H^k(a)\}.
\]
This maximum is well defined. To see this, note that the set $\{a \ge 0: x \in H^k(a)\}$  is nonempty because
$x\in H^k(0)$, and it is bounded due to (\ref{eq:H:intersection}). It is
also closed: If $(a_j)_{j=1}^{+\infty} \rightarrow a$ and $x\in H^k(a_j)$ for
every $j$, we can apply the same compactness argument used to establish closedness
of $H^k(a)$ to show that 
$x\in H^k(a)$. \medskip

Since the sets $(C^k)_{k=0}^{m+1}$ are nested with $C^0=X$ and $C^{m+1}=\emptyset$, every $x\in X$ belongs to a unique
layer $C^k\backslash C^{k+1}$ for some $k \in \{0\} \cup M$. Let $k(x)$ denote this index and define
\begin{equation}\label{eq:usc:utility}
u(x) = k(x)+
\frac{\lambda_{k(x)}(x)}{1+\lambda_{k(x)}(x)}.
\end{equation}
For every $k\in \{0\} \cup M$ and $r\in[0,1)$, we have $\{y\in X:u(y)\geq k+r\}
=
H^k\left(\frac{r}{1-r}\right)$, which is a convex and closed set. This shows that every upper level set of $u$ is closed and
convex. Hence $u$ is upper semicontinuous and quasi-concave. \medskip

The last step is to verify that $u$ satisfies statements (i) and (ii). Suppose that
$x\trianglerighteq_Ky$ and let $d=x-y\in K$. Since every $C^k$ is
upward closed under $K$, we have $k(x)\geq k(y)$. If $k(x)>k(y)$,
then $u(x)>u(y)$. If $k(x)=k(y)=k$, then we have $\lambda_k(x) \geq \lambda_k(y)+\gamma(d)
\geq \lambda_k(y).$ 
Hence $u(x)\geq u(y)$, with strict inequality whenever $\gamma(d)>0$. Thus, statement (i) holds. \medskip 

To verify statement (ii), fix any $k \in M$ and $x\in D^k$. Since $x \in C^{k}\backslash C^{k+1}$, we have $k=k(x)$. We show that for every $a>0$, 
$$H^k(a)\subseteq\{y:y\succ_0x\}.$$ For $k<m$, take $y\in H^k(a)$ and write $y=\alpha y'+(1-\alpha)z+d$ as in (\ref{eq:H:representation}).
Since $y' \succ_0 x$, $z \succsim_0 x$ and $d \in K$ with $\gamma(d) \ge (1-\alpha)a$, we have either $1-\alpha=0$, in which case $y = y'+d \trianglerighteq_K y'$ implies $y \succ_0 x$, or $1-\alpha > 0$, in which case $y \rhd_{\gamma} \alpha y' + (1-\alpha)z \succsim_0 x$. In both cases, $y  \succ_0 x$. The case $k=m$ follows in the same way. Thus, we have $\lambda_k(x)=0$ and $u(x)=k$. If $x\succsim_0 y$ and $u(y)>k$, then either $y\in C^{k+1}$ or
$y\in H^k(a)$ for some $a>0$. In either case, we have a contradiction. Hence $u(x)=k\geq u(y)$. If $x\succ_0y$ and $y\in C^k$, then condition
 (\ref{eq:C:weak}) implies
$y\succsim_0x$, a contradiction. Thus, $y \notin C^k$, implying $k(y)<k$. Thus, $u(x)>u(y)$. \medskip

In the case $D=\emptyset$, note that we just need to define $H^0(a)$ according to condition (\ref{eq:H:m}). The utility function $u(x)= \frac{\lambda_0(x)}{1+\lambda_0(x)}$ can be shown to satisfy all the requirements. 
\end{proof} 

\section*{Appendix 2: Omitted Proofs} 
\begin{proof}[Proof of Theorem \ref{thm:main}]   	
We have shown the equivalence between statements (ii) and (iii), and that (i) implies (ii). It remains to show that statement (ii) implies (i). Consider an order pair $p = (R, P)$ such that $p$ satisfies SP on every finite menu. We say that an order pair $q = (R_q, P_q)$ extends $p$, denoted by $p \subseteq q$, if $R \subseteq R_q$ and $P \subseteq P_q$. Let $\mathcal{P}$ be the set of all order pairs that extend $p$ and satisfy SP on every finite menu. Endowed with the extension order $\subseteq$, $(\mathcal{P}, \subseteq)$ forms a partially ordered set. A subset $\mathcal{Q} \subseteq \mathcal{P}$ is called a chain of $(\mathcal{P}, \subseteq)$ if for all $q, \hat{q} \in \mathcal{Q}$, either $q \subseteq \hat{q}$ or $\hat{q} \subseteq q$.

\begin{lemma}\label{lm:finite1} 
Consider an order pair $q=(\hat{R}, \hat{P})$ and a finite menu $A$. For any nonempty $B\subseteq A$, if for all $x \in B$, $x^{\downarrow}_{\hat{P}} \cap \Pi_{\hat{R}}(A) \neq \emptyset$, then there exists a finite sequence of finite menus $(A^k)_{k=0}^n$ such that \emph{(i)} $A^0=A$, \emph{(ii)} for all $k \in \{1,...,n\}$ and $x \in A^k$, there exists $y \in co(\cup_{l=0}^{k-1} A^l)$ with $x \hat{R} y$, and \emph{(iii)} for all $x \in B$, $x^{\downarrow}_{\hat{P}} \cap co(\cup_{k=0}^{n} A^k) \neq \emptyset$.   
\end{lemma}

\begin{proof} 
It suffices first to consider the case $B=\{x\}$. Since $x^{\downarrow}_{\hat{P}} \cap \Pi_{\hat{R}}(A) \neq \emptyset$, there exists $y^* \in \Pi_{\hat{R}}(A)$ such that $x \hat{P} y^*$. Therefore, it suffices to show that for all $y \in \Pi_{\hat{R}}(A)$, there exists a finite sequence of finite menus $(A^k)_{k=0}^n$ such that $A^0 = A$, $y \in A^n$, and for all $k \in \{1, \ldots, n\}$ and $z \in A^k$, there exists $w \in co\left(\bigcup_{l=0}^{k-1} A^l\right)$ with $z \hat{R} w$. We call such a sequence a \textit{finite path} from $A$ to $y$. If $y \in \Pi_{\hat{R}}^0(A)=co(A)$, then the sequence $(A^k)_{k=0}^1$ with $A^0=A$ and $A^1=\{y\}$ is a finite path from $A$ to $y$. Inductively, suppose that for all $y \in \Pi_{\hat{R}}^k(A)$, a finite path from $A$ to $y$ exists. We will show that it is also true for each $y \in \Pi_{\hat{R}}^{k+1}(A)$. Since $y \in \Pi_{\hat{R}}^{k+1}(A)$, we can find $(\alpha^i)_{i=1}^m \in (0,1]^m$ with $\sum_{i=1}^m \alpha^i = 1$, $\{z^1, \ldots, z^m\} \subseteq X$  and $\{y^1, \ldots, y^m\} \subseteq \Pi_{\hat{R}}^{k}(A)$ such that   $y = \sum_{i=1}^m \alpha^i z^i,$  and for all $i \in \{1, \ldots, m\}$, $z^i \hat{R} y^i$. For each $y^i$, by our induction hypothesis, let $(A^{i,j})_{j=0}^{l_i}$ be the finite path from $A$ to $y^i$. Pick an integer $L$ such that for all $i \in \{1, \ldots, m\}$, $L \ge l_i$. For each $i \in \{1, \ldots, m\}$ and $j > l_i$, let $A^{i,j} = A^{i,l_i}$. By our construction, the sequence $(A^{i,j})_{j=0}^{L}$ remains to be a finite path from $A$ to $y^i$. It  follows that $(A^k)_{k=0}^{L+2}$ is a finite path from $A$ to $y$, where $A^{L+2} = \{y\}$, $A^{L+1} = \{z^1, \ldots, z^{m}\}$, and for all $k \le L$, $A^k = \bigcup_{i=1}^m A^{i,k}$.  \medskip

For general finite $B$, choose $y_x\in x^{\downarrow}_{\hat{P}}\cap\Pi_{\hat{R}}(A)$ for each $x\in B$. For each $y_x$, construct a finite path from $A$ to $y_x$. By extending shorter paths if necessary and taking the union of the menus at each step, we obtain a finite sequence satisfying conditions (i)--(iii). This proves the lemma.
\end{proof}

	\begin{lemma}\label{lm:chain1}
		Every nonempty chain $\mathcal{Q}$ of $(\mathcal{P}, \subseteq)$ has a least upper bound. 
	\end{lemma}
	\begin{proof}
		Consider a chain $\{(R_q, P_q)\}_{q \in \mathcal{Q}}$. Define $o=(R_o, P_o)$ such that $R_o=\cup_{q \in \mathcal{Q}} R_q$ and $P_o=\cup_{q \in \mathcal{Q}} P_q$. It suffices to show that $o=(R_o, P_o)$ satisfies SP on every finite menu. Suppose to the contrary that there is some finite menu $A$ such that $o$ violates SP on $A$. It follows that for all $x \in A$, we have $x^{\downarrow}_{P_o} \cap \Pi_{R_o}(A) \neq \emptyset$. By Lemma \ref{lm:finite1}, we can find a finite sequence of finite menus $(A^k)_{k=0}^n$ such that   (i) $A^0=A$,  (ii) for all $k \in \{1,...,n\}$ and $x \in A^k$, there exists $y \in co(\cup_{l=0}^{k-1} A^l)$ with $x R_o y$, and (iii) for all $x \in A$, there exists $y \in co(\cup_{k=0}^{n} A^k)$ such that $x P_o y$. Without loss of generality, assume $n \ge 1$. Condition (ii) is equivalent to that for all $k \in \{1,...,n\}$ and $x \in A^k$, there exists $y \in co(\cup_{l=0}^{k-1} A^l)$ and $q(x) \in \mathcal{Q}$ with $x R_{q(x)} y$. Condition (iii) is equivalent to that for all $x \in A$, there exists $y \in co(\cup_{k=0}^{n} A^k)$ and $r(x) \in \mathcal{Q}$ such that $x P_{r(x)} y$. Since $\mathcal{Q}$ is a chain, the finite set $\{r(x)\}_{x \in A} 
		\cup \{q(x)\}_{x \in \cup_{k=1}^n A^k}$ contains a $\subseteq$-maximal order pair $q^*$. Therefore, for all $k \in \{1,...,n\}$ and $x \in A^k$, there exists $y \in co(\cup_{l=0}^{k-1} A^l)$ with $x R_{q^*} y$, and for all $x \in A$, there exists $y \in co(\cup_{k=0}^{n} A^k)$ with $x P_{q^*} y$. We conclude that $co(\cup_{k=0}^{n} A^k)  \subseteq \Pi_{R_{q^*}}(A)$, and for all $x \in A$, there exists $y \in \Pi_{R_{q^*}}(A)$ such that $x P_{q^*} y$. This contradicts the fact that $q^*$ satisfies SP on every finite menu.  \end{proof}

	We say that a preference $\succsim$ extends an order pair $(R, P)$ \textit{on a menu} $A$ if for all $x,y$, $x R y$ implies $x \succsim y$, and for all $x,y$ with $x \in A$, $x P y$ implies $x \succ y$. \smallskip

	\begin{lemma}\label{lm:complete1}
		Let $q=(R_q, P_q)$ be an element of $\mathcal{P}$. If for some $x, y \in X$, neither $x R_q y$ nor $y R_q x$, then either $r=(R_r, P_r)$ or $w=(R_w, P_w)$ is an element in $\mathcal{P}$, where   $R_r= R_q \cup\, \{(x,y)\}$, $R_w= R_q \cup\, \{(y,x)\}$ and $P_r=P_w=P_q$.
	\end{lemma}

	\begin{proof}
		Suppose to the contrary that $r$ violates SP on a finite menu $A$, and $w$ violates SP on a finite menu $B$. Let $C = A \cup B \cup \{x, y\}$. Since $q$ satisfies SP on every finite menu, and in particular on every menu contained in $C$, by Theorem \ref{thm:algorithm}, there exists a convex preference $\succsim$ that extends $q$ on $C$. It follows that $\succsim$ extends either $r$ or $w$ on $C$. By Theorem \ref{thm:algorithm}, either $r$ or $w$ satisfies SP on every menu contained in $C$, which leads to a contradiction. Therefore, either $r$ or $w$ satisfies SP on every finite menu, that is, either $r \in \mathcal{P}$ or $w \in \mathcal{P}$. 
	\end{proof}

	The next lemma can be proved similarly to Lemma \ref{lm:complete1}, and its proof is omitted.

	\begin{lemma}\label{lm:complete2}
		Let $q=(R_q, P_q)$ be an element of $\mathcal{P}$. If for some $x, y \in X$, $x R_q y$,  not $y R_q x$ and not $x P_q y$, then either $(R_q  \cup\, \{(y,x)\}, P_q)$ or $(R_q, P_q \cup\, \{(x,y)\})$ is an element of $\mathcal{P}$.
	\end{lemma}

	\begin{lemma}\label{lm:transitive}
		If $q=(R_q, P_q)$ is an element of $\mathcal{P}$, then $r=(R_r, P_r)$ is also an element of $\mathcal{P}$, where $P_r = P_q$, and $R_r$ is the transitive closure of $R_q$.  
	\end{lemma}
	
	\begin{proof}
	For every finite menu $A$, since $R_r$ is the transitive  closure of $R_q$, we have  $\Pi_{R_q}(A)=\Pi_{R_r}(A)$. Thus, $r$ satisfies SP on every finite menu.  
	\end{proof}

	\begin{lemma}\label{lm:asymmetry}
		Let $q=(R_q, P_q)$ be an element of $\mathcal{P}$. If $y P_q x$, then not $x R_q y$.
	\end{lemma}
	
	\begin{proof}
		Suppose to the contrary that $y P_q x$ and $x R_q y$. It follows that $x \in \Pi_{R_q}(\{y\})$. Thus, $q$ violates SP on $\{y\}$, which is a contradiction. 
	\end{proof}

Since every nonempty chain in $\mathcal{P}$ has a least upper bound, by Zorn's lemma, there exists a $\subseteq$-maximal element $p^* = (R^*, P^*)$ in $\mathcal{P}$ that extends $p = (R, P)$. By Lemmas \ref{lm:complete1} and \ref{lm:transitive}, along with the $\subseteq$-maximality of $p^*$, $R^*$ must be complete and transitive.  By Lemma \ref{lm:complete2} and the $\subseteq$-maximality of $p^*$, if not $x R^* y$, then $y P^* x$. By Lemma \ref{lm:asymmetry}, if $y P^* x$, then not $x R^* y$. Therefore, $P^*$ coincides with the asymmetric part of $R^*$. This means that $R^*$ is a completion of $p$. To show that $R^*$ is convex, note that for all $x \in X$, its upper contour set $\{y : y R^* x\}$ is a subset of $\Pi_{R^*}(\{x\})$. Moreover, since $\Pi_{R^*}(\{x\}) \cap \{y : x P^* y\} = \emptyset$ by SP, we have $\Pi_{R^*}(\{x\}) \subseteq \{y : y R^* x\}$. Hence, $\{y : y R^* x\} = \Pi_{R^*}(\{x\})$ is convex. Therefore, $R^*$ is a convex preference.\end{proof}

\begin{proof}[Proof of Theorem \ref{thm:porder}]
First, assume that $p$ admits a convex completion $\succsim$. It follows that
$\succsim$ must also extend $q$. Thus, $q$ admits a convex completion and satisfies SP on every finite menu, including those contained in $H$. \medskip

It remains to prove that if $p$ violates SP on some finite menu $A$, then $q$ violates SP on some finite menu $B\subseteq H$. We begin with two lemmas. The first follows directly from standard arguments using independence and transitivity, and we therefore omit its proof.

\begin{lemma}\label{lm:linear:basic}
Let $\rhd$ be an I-partial order. For all
$\{x^k,y^k\}_{k=1}^n\subseteq X$ and
$\{\lambda^k\}_{k=1}^n\subseteq(0,1]$ with
$\sum_{k=1}^n\lambda^k=1$, if for all $k\in\{1,\ldots,n\}$, either
$x^k\rhd y^k$ or $x^k=y^k$, with the former holding for at least one $k$,
then $\sum_{k=1}^n\lambda^kx^k
\rhd \sum_{k=1}^n\lambda^ky^k.$
\end{lemma}

\begin{lemma}\label{lm:linear:reduction}
Let $\rhd$ be an I-partial order. Consider a finite menu $A$ and
$x\in A$ such that, for some $y\in co(A)$, $x\rhd y$. Let
$B=A\backslash\{x\}$. For all $z\in X$ and $w\in co(A)$,

\emph{(i)} if $z\rhd w$, then there is $\hat{w}\in co(B)$ such that
$z\rhd\hat{w}$;

\emph{(ii)} if $zPw$, then there is $\hat{w}\in co(B)$ such that
$zP\hat{w}$ or $zP\diamond\rhd\hat{w}$;

\emph{(iii)} if $zP\diamond\rhd w$, then there is
$\hat{w}\in co(B)$ such that $zP\diamond\rhd\hat{w}$;

\emph{(iv)} if $zR\diamond\rhd w$, then there is
$\hat{w}\in co(B)$ such that $zR\diamond\rhd\hat{w}$;

\emph{(v)} if $z\hat{P}w$, then there is $\hat{w}\in co(B)$ such that $z\hat{P}\hat{w},$ where $\hat{P}=P\cup(P\diamond\rhd)\cup(R\diamond\rhd)$. 
\end{lemma}

\begin{proof}[Proof of Lemma \ref{lm:linear:reduction}] Since $x\rhd y\in co(A)$ and $\rhd$ is asymmetric, we have $B \neq \emptyset$ and can find
$\lambda\in[0,1)$ and $\hat{y}\in co(B)$ such that $
y=\lambda x+(1-\lambda)\hat{y}.$
By the independence of $\rhd$, we have $x\rhd\hat{y}$. Next, consider any
$z\in X$ and $w\in co(A)$. Pick
$\bar{w}\in co(B)$ and $t\in[0,1]$ such that $
w=tx+(1-t)\bar{w}.$
If $t=0$, then all the statements hold. If
$t>0$, then pick $\hat{w} \in co(B)$ such that $
\hat{w}=t\hat{y}+(1-t)\bar{w}.$ By the independence of $\rhd$, we have $w\rhd\hat{w}$. Statements (i)--(iv) then follow immediately, using the transitivity of $\rhd$ where needed, and thus statement (v) holds.
\end{proof}

We construct a finite menu $B\subseteq H$ on which $q$
violates SP. The trivial case is when $A\subseteq H$, and we can take $B=A$. Consider the non-trivial case where $A\not\subseteq H$. Define the binary
partition $\{A^H,A^N\}$ of $A$ by $
A^H=A\cap H \text{~and~} 
A^N=A\backslash H\neq\emptyset.$  Since $p$ violates SP on $A$, by Lemma \ref{lm:finite1}, we can construct a
finite collection of finite menus $(D^l)_{l=0}^L$ (with $D:=\cup_{l=0}^LD^l$) such that: \smallskip

(i) $D^0=A$, and for all $l\in\{0,\ldots,L-1\}$ and
$x\in D^{l+1}$, we have $$x\notin
co\left(\bigcup_{j=0}^lD^j\right) \text{~and~}   x\,(R\,\cup\trianglerighteq^*)\,y ~\text{for some~} y\in co\left(\bigcup_{j=0}^lD^j\right), \text{~and}$$ 

(ii) for every $x\in A$, there exists $y\in co(D)$ such that $x\,(P\cup\rhd)\,y.$ 
\medskip

\noindent Note that we can remove each  $x \in D^{l+1} \cap co\left(\bigcup_{j=0}^lD^j\right)$ from $D^{l+1}$ until we have $D^{l+1} \cap co\left(\bigcup_{j=0}^lD^j\right) = \emptyset$, thereby ensuring that (i) holds. For every $x\in A^N$, since $x\notin G_P$, condition (ii) implies that
there exists $f(x)\in co(D)$ such that $
x\rhd f(x).$  For every $x\in A^H$, fix $g(x)\in co(D)$ such that $xPg(x)$ or  $x\rhd g(x).$  In either case, we have $xP^*g(x)$. \medskip

We now construct a partition $D^l=E^l\cup F^l\cup N^l$ for every $l\geq1$, where the three sets are mutually disjoint and satisfy:
\smallskip

 (a) for every $x\in E^l$, we have $x\in H$ and    $xP^*g(x)$ for some $g(x)\in co(D)$; \smallskip

 (b) for every $x\in F^l$, 
\[
\text{there exists~}y\in
co\left(
A^H\cup
\left(\bigcup_{j=1}^{l-1}(E^j\cup F^j)\right)\right) \text{~such that~} x\,(R\,\cup\trianglerighteq^*)\,y;
\]

(c) for every $x\in N^l$, we have $x\notin H$ and $x\rhd f(x)$ for some
$f(x)\in co(D)$. \medskip

To see why such a partition exists, fix $l\geq1$ and $x\in D^l$. Assume by the induction hypothesis that such a partition exists for $j \le l-1$. Define
$$M^{l-1}
=
A^H\cup
\left(\bigcup_{j=1}^{l-1}(E^j\cup F^j)\right)
\text{~and~}
N^{l-1}
=
A^N\cup
\left(\bigcup_{j=1}^{l-1}N^j\right).$$
If $x (R\,\cup\trianglerighteq^*)y$ for some $y \in co(M^{l-1})$, put $x$ in $F^l$. Otherwise, by condition (i), there exist 
$\lambda\in(0,1]$, $z\in co(N^{l-1})$ and $w\in co(M^{l-1})$ such that $x (R\,\cup\trianglerighteq^*) y$ with $y=\lambda z+(1-\lambda)w.$  Write $z=\sum_{v\in N^{l-1}}\alpha^vv$. Note that for every
$v\in N^{l-1}$, we have a selection $f(v)\in co(D)$ such
that $v\rhd f(v)$: For $v\in A^N$, this follows from the construction
above; for $v\in N^j$ with $j<l$, this follows from the induction
hypothesis. Let $z'=\sum_{v\in N^{l-1}}\alpha^vf(v).$ By Lemma \ref{lm:linear:basic} and the independence of $\rhd$, we have $y \rhd y'=\lambda z'+(1-\lambda)w \in co(D)$.  \medskip 

If $xRy$, then $x\neq y$ by condition (i), and hence
$x\in\mathring{G}_R\subseteq H$. Moreover, $xR y\rhd y'$ implies $xP^*y'$. We put
$x\in E^l$ and set $g(x)=y'$.  It remains to consider the case where $x\trianglerighteq^*y.$ 
Since $y\rhd y'$ and $(\trianglerighteq^*,\rhd)$ is compatible, we have $x\rhd y'.$
If $x\in H$, the reflexivity of $R$ implies $xR x\rhd y'$ and hence $xP^*y'$. We then put $x\in E^l$ and set $g(x)=y'$. If
$x\notin H$, put $x\in N^l$ and set $f(x)=y'$. This completes the
construction of the partition. \medskip

To proceed, define $$B=A^H\cup
\left(\bigcup_{l=1}^LE^l\right), M
=
A^H\cup
\left(\bigcup_{l=1}^L(E^l\cup F^l)\right) \text{~and~} N
=
A^N\cup
\left(\bigcup_{l=1}^LN^l\right).$$ 
Note that $B\subseteq H$ and 
$D=M\cup N.$ Moreover, for every $x\in B$, there exists $g(x)\in co(D)$ such that
$xP^*g(x)$, and for every $x\in N$, there exists
$f(x)\in co(D)$ such that $x\rhd f(x)$. We first show that
\begin{equation}\label{eq:Mclosure}
M\subseteq\Pi_{R\cup\trianglerighteq^*}(B).
\end{equation}
To show (\ref{eq:Mclosure}), it suffices to consider any $x\in F^l$ and show that $x\in\Pi_{R\cup\trianglerighteq^*}(B)$. By condition (b), there is
\[
y\in
co\left(
A^H\cup
\bigcup_{j=1}^{l-1}(E^j\cup F^j)
\right)
\]
such that $x\,(R\cup\trianglerighteq^*)\,y.$ A simple induction on $l$, using the convexity and
$(R\cup\trianglerighteq^*)$-upward closure of
$\Pi_{R\cup\trianglerighteq^*}(B)$, establishes
(\ref{eq:Mclosure}). 
\medskip

We next show that $B\neq\emptyset$. Suppose, to the contrary, that
$B=\emptyset$. Then $A^H=\emptyset$ and $E^l=\emptyset$ for every
$l\geq1$. By condition (b), when
$A^H=\emptyset$, we have $F^1=\emptyset$. A simple induction using
condition (b) implies that $F^l=\emptyset$ for every $l\geq1$. Thus, $M=\emptyset$ and $D=N.$ 
Label $N=\{x^1,\ldots,x^n\}.$
For every $x^i\in N$, there is $f(x^i)\in co(N)$ such that
$x^i\rhd f(x^i)$. Applying Lemma \ref{lm:linear:reduction}(i)
successively to $x^1,\ldots,x^{n-1}$, we obtain $x^n\rhd x^n,$ contradicting the asymmetry of $\rhd$. Hence, $B\neq\emptyset$. \medskip

Finally, we show that $B$ is the desired finite menu on which $q$ violates SP. Note that by our construction, $B \subseteq H$. By condition (\ref{eq:Mclosure}), if $N=\emptyset$, then $D=M$, and we are done. 
Otherwise, label $
N=\{x^1,\ldots,x^n\}.$ Recall that for every $x\in N$, there is $f(x) \in co(M\cup N)$ such that $x \rhd f(x)$, while for every $z\in B$, there is a
$g(z) \in co(M\cup N)$ such that $z P^* g(z)$. Applying Lemma
\ref{lm:linear:reduction} to $x^1$, we have that for every $x^i \in N\backslash\{x^1\}$, there is $y^i \in co(M\cup (N\backslash\{x^1\}))$ such that $x^i \rhd y^i$, and for every $z \in B$, there is $w \in co(M\cup (N\backslash\{x^1\}))$ such that $z P^* w$. Repeating this argument for $x^2,\ldots,x^n$, we conclude that, for every
$x\in B$, there exists $y\in co(M)$ such that $xP^*y.$
By (\ref{eq:Mclosure}), we have $co(M)
\subseteq
\Pi_{R\cup\trianglerighteq^*}(B)$, and thus for every $x\in B$, $x_{P^*}^{\downarrow}
\cap
\Pi_{R\cup\trianglerighteq^*}(B)
\neq\emptyset$. That is, $q$ violates SP on the finite menu $B$ with $B \subseteq H$. \end{proof}

\begin{proof}[Proof of Theorem  \ref{thm:test:consumer}]
The equivalence between (i) and (ii) is implied by Theorem \ref{thm:porder}. Clearly, (iii) implies (i). To see that (i) implies (iii), apply Theorem \ref{thm:convex:uppersemi} by taking $\succsim_0$ as a preference that satisfies (i), $Y=\mathbb{R}^n$, $X=K=\mathbb{R}^n_+$ and $D=\{x^t\}_{t \in T}$, $B^0=\{0\}$, and defining $\gamma: \mathbb{R}^n_+ \rightarrow \mathbb{R}_+$ such that $\gamma(d)=\sum_{k=1}^n d_k$.  
\end{proof}

\begin{proof}[Proof of Theorem \ref{thm:test:uncertain}] 
For the equivalence among (i), (ii) and (iii), it suffices to show that (i) implies (iii). Let $\trianglerighteq$ be represented by $\mathcal{C}$. Note that each $u \in \mathcal{C}$ has sup norm $\|u\|_{\infty} \le 1$. Since $L$ is a compact metric space, the space of all continuous functions from $L$ to $[0,1]$ is separable under the sup norm metric, and so is $\mathcal{C}$. Choose a countable dense subset
$\{u_j\}_{j\geq1}$ of $\mathcal C$ and define $v=\sum_{j=1}^{\infty}2^{-j}u_j$. It is easy to show that $v$ is also a continuous function from $L$ to $[0,1]$. It follows that $F \trianglerighteq G$ implies $E_{F}(v) \ge E_{G}(v)$, and $F \rhd G$ implies $E_{F}(v) > E_{G}(v)$. The latter implication is by the fact that if there is $u \in \mathcal{C}$ such that  $E_{F}(u) > E_{G}(u)$, then the denseness of $\{u_j\}_{j\geq1}$ under the sup norm metric implies that the same holds for some $u_j$ and thus for $v$. Now, apply Theorem \ref{thm:convex:uppersemi} as follows. Take any preference $\succsim_0$ that satisfies (i). Let $Y$ be the space of all finite signed Borel measures on $L$, equipped with the weak$^*$ topology. 
Let $X=\Delta(L)$, $K \subseteq Y$ such that $\mu \in K$ if and only if $\int_L u d\mu \ge 0$ for every $u \in \mathcal{C}$, $B^0=X$, and $D=H$. Note that $K$ is a closed convex cone, and $B^0=X$ is weak$^*$ compact. Define $\gamma: K \rightarrow \mathbb{R}_+$ such that for each $\mu \in K$, $\gamma(\mu)=\int_L v d\mu$. This proves the implication from (i) to (iii).  \medskip

When $R=I_X$, and $P$ and $\mathcal{C}$ are finite, let $P=\{(F^t, G^t)\}_{t \in T}$ and $n=|\mathcal{C}|$. We can transfer each $F^t$ and $G^t$ to an $n$-dimensional vector in $\mathbb{R}_{++}^n$ by defining $x^t = (E_{F^t}(u))_{u \in \mathcal{C}} + a$ and  $y^t = (E_{G^t}(u))_{u \in \mathcal{C}} + a$ for each $t \in T$, where $a \in \mathbb{R}_{++}^n$ is a constant vector to ensure that for all $F \in \Delta(L)$, $(E_{F}(u))_{u \in \mathcal{C}} + a \in \mathbb{R}^n_{++}$. By Corollary \ref{coro:porder:pure_strict}, for all nonempty $A \subseteq \{x^t\}_{t \in T}$, there is $x \in A$ such that for all $r \in T$ with $x^r = x$,  $\{y\in \mathbb{R}^n_{++}: y^r \ge_n y\} \cap co(A) = \emptyset$. By Corollary 2 of \cite{trip}, there is a continuous, strictly increasing, and quasi-concave utility function $u$ on $\mathbb{R}_{++}^n$ such that for all $t \in T$, $u(x^t) > u(y^t)$. It follows that the function $U:\Delta(L) \rightarrow \mathbb{R}$ defined such that $U(F)=u((E_{F}(u))_{u \in \mathcal{C}} + a)$ is the desired one. 
\end{proof}

\begin{proof}[Proof of Theorem \ref{thm:ambiguity}]
The equivalence between (i) and (ii) is implied by Theorem \ref{thm:porder}. Clearly, (iii) implies (i). To see that (i) implies (iii), note that we can transfer the space $\mathcal{F}$ to a subset of $\mathbb{R}^{n}$ by $u$, where $n=|S|$, such that each $f$ is transferred to $x_f=(u(f(s)))_{s \in S}$. It follows that $\ge_u$ can be transferred to $\ge_n$, and $\gg_u$ can be transferred to $\gg_n$. Apply Theorem \ref{thm:convex:uppersemi} by taking $X=\{x_f\}_{f \in \mathcal{F}}$, $\succsim_0$ as the induced preference over $X$ by a preference that satisfies (i), $Y=\mathbb{R}^n$, $K=\mathbb{R}_+^n$, $D=\{x_f\}_{f \in H}$, and $B^0=\{\underline{x}\}$ with $\underline{x}_k=\min_{a \in L} u(a)$ for each $k=1,...,n$. Let $\gamma: K \rightarrow \mathbb{R}_+$ be such that $\gamma(d)=\min_{k \in \{1,...,n\}} d_k$. \medskip

When $R=I_X$ and $P$ is finite, observe that by Corollary \ref{coro:porder:pure_strict}, (i) implies that for each nonempty $A \subseteq G_P$, there exists $x \in A$ such that $x^{\downarrow}_{P \diamond \ge_u} \cap co(A) = \emptyset$. Note that $\gg_u$ plays no role here, and this condition reduces to the same one in Corollary 2 of \cite{trip} when we transfer each act $f$ to $x_f$. Thus, following the same argument as that for Theorem \ref{thm:test:uncertain}, we can show that the utility function in (iii) can be strengthened to be continuous. 
\end{proof}

\begin{proof}[Proof of Theorem \ref{thm:test:information}]
Note that (i) is equivalent to the statement that $(\succsim_B, P \cup \rhd)$ admits a convex completion, where $P=\{(\pi^t, \rho^t)\}_{t \in T}$ and $\rhd=\emptyset$.  The equivalence between (i) and (ii) is by Corollary \ref{coro:porder:pure_strict}. Since (iii) trivially implies (i), we prove that (i) implies (iii).  \medskip

Let $\mathcal{V}$ be the set of all polyhedral convex functions from $\Delta(\Omega)$ to $\mathbb{R}$. We first show that for any
nonempty finite $A\subseteq\mathcal I(\mu)$ and
$\eta$ with $\eta_{\succsim_B}^{\downarrow} \cap co(A)=\emptyset$, there exists $v \in \mathcal{V}$ such that $V_v(\eta) < \min_{\hat{\eta} \in A}V_v(\hat{\eta}).$ Note that it suffices to show that a continuous and convex $v$ exists, and then we can pick $\hat{v} \in \mathcal{V}$ close to $v$ to maintain the strict inequality. Suppose to the contrary that such a $v$ does not exist. Then for every continuous convex $v$, we have $\min_{\hat{\eta} \in co(A)}V_v(\hat{\eta}) \leq V_v(\eta).$ For each such $v$, let $C_v=\{\hat{\eta}\in co(A):V_v(\hat{\eta})\leq V_v(\eta)\}.$ We prove that $\{C_v\}_v$ has the finite intersection property: Otherwise,  $\cap_{j=1}^m C_{v_j}=\emptyset$ for some $v_1,\ldots,v_m$, and we have 
$$ \min_{\hat{\eta} \in co(A)}
\max_{j=1,\ldots,m}
\left\{
V_{v_j}(\hat{\eta})-V_{v_j}(\eta)
\right\}
>0 \Rightarrow \min_{\hat{\eta}\in co(A)}
\sum_{j=1}^m\gamma_j
\left(
V_{v_j}(\hat{\eta})-V_{v_j}(\eta)
\right)
>0,$$ where $\gamma\in\Delta(\{1,\ldots,m\})$, and the implication is by the minimax theorem. The function $v=\sum_j\gamma_jv_j$ is continuous and convex, which contradicts the preceding inequality. Hence $\{C_v\}_v$ has the finite
intersection property. Since $co(A)$ is compact, there exists
$\eta^*\in co(A)$ such that $V_v(\eta^*)\leq V_v(\eta)$ for every continuous convex $v$, that is $\eta\succsim_B\eta^*$, a contradiction. 
\medskip

By (i), we can find a convex and Blackwell-monotone preference $\succsim$ over $\mathcal{I}(\mu)$ such that for each $t \in T$, $\pi^t \succ \rho^t$. Let $A=\{\pi^t\}_{t \in T} \cup \{\rho^t\}_{t \in T}$. Partition $A$ into indifference blocks $\{A^k\}_{k=1}^m$ ($m \ge 2$) under $\succsim$ such that for all $k<l$, alternatives in $A^k$ are strictly better than those in $A^l$ under $\succsim$. Hence, for every $\eta \in A^l$ with $l \ge 2$, we can find $v \in \mathcal{V}$ such that $V_v(\eta) < \min_{k \in \{1,...,l-1\}} \left( \min_{\hat{\eta} \in A^k} V_v(\hat{\eta}) \right)$. We construct the desired representation by induction. For notational convenience, we label alternatives in each $A^k$ by $\eta^{k,i}$, with $i\in \{1,...,|A^k|\}$. When $m=2$, for each $\eta^{2,i} \in A^2$, pick $v^{2,i}\in \mathcal{V}$ such that $V_{v^{2,i}}(\eta^{2,i}) < \min_{\eta^{1,j} \in A^1}V_{v^{2,i}}(\eta^{1,j} ).$ We can take positive affine transformations for each $v^{2,i}$ such that for each $i$, $\min_{\eta^{1,j} \in A^1}V_{v^{2,i}}(\eta^{1,j} ) = a^1 > b^2 = V_{v^{2,i}}(\eta^{2,i})$. Let $\hat{\mathcal{V}}=\{v^{2,i}\}_{i=1}^{|A^2|} \subseteq \mathcal{V}$, which is finite. We can find $b^1 > a^1$ and $a^2 < b^2$ such that $U(\eta) \in [a^1, b^1]$ for all $\eta \in A^1$ and $U(\eta) \in [a^2, b^2]$ for all $\eta \in A^2$, where $U$ is defined according to (\ref{eq:finite:robust:information}) on $\hat{\mathcal{V}}$. Now, consider the case where the indifference partition $\{A^k\}_{k=1}^m$ contains more than two blocks, i.e., $m \ge 3$. Suppose by the induction hypothesis that we are able to construct for the partition $\{A^k\}_{k=1}^{m-2} \cup \{A^{m-1} \cup A^m\}$ a finite set $\hat{\mathcal{V}} \subseteq \mathcal{V}$ and find numbers $b^1 > a^1 > b^2 > a^2 > ... >b^{m-1} > a^{m-1}$ such that the function $U$ defined according to (\ref{eq:finite:robust:information}) on $\hat{\mathcal{V}}$ satisfies that for all $k \le m-1$ and all $\eta \in A^k$, $U(\eta) \in [a^k, b^k]$, and for all $\eta \in A^{m-1} \cup A^m$, $U(\eta) \in [a^{m-1}, b^{m-1}]$. Now, pick $b^m < a^{m-1}$. For each $\eta^{m,i} \in A^m$, pick $v^{m,i} \in \mathcal{V}$ such that $\min_{\eta \in A\backslash A^m} V_{v^{m,i}}(\eta) = a^1$ and $V_{v^{m,i}}(\eta^{m,i})=b^m$. Expand $\hat{\mathcal{V}}$ to include $\{v^{m,i}\}_{i=1}^{|A^m|}$ and pick $a^m$ sufficiently small, and we are done.  \end{proof}

\begin{proof}[Proof of Theorem \ref{thm:sp=garp}] It suffices to prove the first statement of the theorem. Let $p$ be consistent and pre-convex on $A$. Suppose, to the contrary, that $p$ violates SP on a finite menu $B\subseteq A$. By consistency, there is a preference $\succsim^{c}$ that extends $p$, i.e., $R \,\subseteq\, \succsim^c$ and $P \,\subseteq\, \succ^c$. Since $p$ violates SP on $B$, by Lemma \ref{lm:finite1}, there is a finite sequence of mutually disjoint finite menus $(D^k)_{k=0}^{n}$ such that (i) $D^0=B$, (ii) for all $x \in B$, there exists $y \in co(\cup_{k=0}^n D^k)$ such that $x P y$, and (iii) for all $k \in \{0,...,n-1\}$ and $x \in D^{k+1}$, there is $y \in co(\cup_{l=0}^{k} D^l)$ such that $x \neq y$ and $x R y$. Let $x^*$ be the $\succsim^{c}$-worst alternative in $B$. Since $x^* P y^1$ for some $y^1 \in  co(\cup_{k=0}^n D^k)$, by the convexity of the set $\{y: \text{not~} x^* P y\}$, there is  $z^1 \in \cup_{k=0}^n D^k$ such that $x^* P z^1$. If $z^1 \in D^0=B$, then we have a contradiction since $x^*$ is the $\succsim^{c}$-worst alternative in $B$. Since $z^1 \in D^k$ for some $k\ge 1$, we have $z^1 R y^2$ for some $y^2 \in co(\cup_{l=0}^{k-1} D^l)$ with $z^1 \neq y^2$. By the convexity of the set $\{y: \text{not~}  z^1 R y\}$, we have $z^1 R z^2$ for some $z^2 \in \cup_{l=0}^{k-1} D^l$. By a simple induction, we have $x^* P z^1 R z^2 R ... R z^m$ for some $z^m \in D^0 = B$. This contradicts  the fact that $x^*$ is the $\succsim^c$-worst alternative in $B$.   
\end{proof}

\begin{proof}[Proof of Theorem \ref{thm:generalize:exact}]
Note that for each $t$ and $x \in A^t$, $x$ is weakly better than alternatives in $B^t$ and strictly better than those in $B^t \backslash A^t$. Since $\mathbb{R}_+^n \backslash B^t$ is assumed to be convex, to apply Theorem \ref{thm:sp=garp} to prove the theorem, it suffices to show that $A^t \cup (\mathbb{R}_+^n \backslash B^t)$ is convex. First, note that for each $x \in A^t$, we show that $\{x\} \cup (\mathbb{R}_+^n \backslash B^t)$ is convex. To see this, suppose to the contrary that for some $y\notin B^t$ and $\lambda\in(0,1)$, 
$\lambda x+(1-\lambda)y\in B^t$. Choose $\epsilon >_n 0$ such that $y-\epsilon \in \mathbb{R}_+^n \backslash B^t$. Since
$x+((1-\lambda)/\lambda)\epsilon\notin B^t$, convexity of $\mathbb{R}_+^n\backslash B^t$ implies $\lambda\left(x+\frac{1-\lambda}{\lambda}\epsilon\right) +(1-\lambda)(y-\epsilon) =\lambda x+(1-\lambda)y \notin B^t,$ a contradiction.\medskip

Now, we show that for any $x,y \in A^t \cup (\mathbb{R}_+^n \backslash B^t)$ and $\lambda \in [0,1]$, $z=\lambda x+(1-\lambda )y \in A^t \cup (\mathbb{R}_+^n \backslash B^t)$. The trivial case is when $x,y \in \mathbb{R}_+^n \backslash B^t$ as  $\mathbb{R}_+^n \backslash B^t$ is convex. If $x,y \in A^t$, then since $\mathcal{O}$ is convex valued, $z \in A^t$ or $z \notin B^t$, and we are done. If $x \in A^t$ and $y \in \mathbb{R}_+^n \backslash B^t$, then the convexity of $\{x\} \cup (\mathbb{R}_+^n \backslash B^t)$ implies $z \in \{x\} \cup (\mathbb{R}_+^n \backslash B^t)$.  
\end{proof}

\begingroup
\fontsize{12pt}{11pt}\selectfont
\setlength{\bibsep}{1pt}
\bibliographystyle{aer}
\bibliography{rcpbib}
\endgroup

\newpage

\appendix
\section*{Online Appendix (for online publication only)}

\setcounter{page}{1} 

This Online Appendix to ``\textit{Revealed Convex Preference}'' presents supplementary empirical analyses, extended theoretical results, and proofs and examples omitted from the main text.

\section*{OA-1. An Empirical Application to Choice under Risk}
\label{sec:empirical}

In this section, we implement our test using the recent data set of \cite{AlosFerrerBuckenmaierGaragnaniRustichini2026}. We focus on Experiment 1, which contains 103 subjects and 37 pairs of lotteries. Each lottery has four equiprobable monetary outcomes. For every pair $(F,G)$, subjects first make repeated binary choices between the two lotteries and subsequently choose directly from a set of compound lotteries that randomize between $F$ and $G$. In particular, in the latter task, a subject chooses a mixture weight $\lambda$
and hence a compound lottery $\lambda F+(1-\lambda)G$, where $\lambda\in\left\{0,\frac14,\frac12,\frac34,1\right\}.$ We use these explicit mixture choices in our analysis. After reducing the compound lotteries, we interpret the choice in the usual revealed-preference sense: The chosen mixture is weakly preferred to every other available mixture in the same choice problem. \medskip 

We compare rationalizability and convex rationalizability under two different restrictions on preferences. First, we require the rationalizing preference to respect first-order stochastic dominance (FOSD). Second, we require it to respect second-order stochastic dominance (SOSD). To allow for a small number of choice errors, we also conduct a deletion exercise. For each subject and each integer $n\in\{0,1,2,3,4\}$, we ask whether there exists a collection of at most $n$ of the 37 choice problems such that, after deleting these problems,
the remaining data are rationalizable. We conduct this exercise separately for rationalizability without the convexity restriction (corresponding to the ``Unrestricted'' column) and convex rationalizability (corresponding to the ``Convex'' column), and separately under FOSD and SOSD. The results are reported in Table~\ref{tab:empirical_application}. Thus, for example, with at most two deleted choice problems, all 103 subjects can be rationalized under FOSD, both with and without convexity. Under SOSD, 72 subjects can be rationalized without the convexity restriction, whereas only 57 can be rationalized by a convex preference. The set of deleted choice problems is chosen separately for each subject and for each test. \medskip

Under FOSD, convexity has no additional empirical bite in this sample. With no deletions, 100 of the 103 subjects are rationalizable under FOSD, and the same 100 subjects are convex-rationalizable. The remaining three subjects can be accommodated after deleting at most one choice problem, so all 103 subjects pass both tests once one deletion is allowed. By contrast, under SOSD, only
20 subjects are rationalizable with no deletions, while 19 are
 rationalizable by convex preferences. The difference becomes substantially larger once a small number of choice problems may be deleted. For example, with up to one
deletion, 46 subjects can be rationalized by a preference respecting SOSD, whereas only 34 can be rationalized by a convex preference respecting SOSD.
Thus, while convexity imposes no additional empirical restriction under FOSD in this sample, it imposes a nontrivial additional restriction under SOSD.

\begin{table}[htbp]
\centering
\caption{Testing Rationalizability under Stochastic Dominance}
\label{tab:empirical_application}
\small
\setlength{\tabcolsep}{5pt}
\renewcommand{\arraystretch}{1.15}
\begin{tabular}{ccccc}
\hline\hline
&\multicolumn{2}{c}{FOSD}&\multicolumn{2}{c}{SOSD}\\
\cline{2-3}\cline{4-5}
Max.\ \# deleted&Unrestricted&Convex&Unrestricted&Convex\\
\hline
0&100 (97.1\%)&100 (97.1\%)&20 (19.4\%)&19 (18.4\%)\\
1&103 (100.0\%)&103 (100.0\%)&46 (44.7\%)&34 (33.0\%)\\
2&103 (100.0\%)&103 (100.0\%)&72 (69.9\%)&57 (55.3\%)\\
3&103 (100.0\%)&103 (100.0\%)&82 (79.6\%)&74 (71.8\%)\\
4&103 (100.0\%)&103 (100.0\%)&92 (89.3\%)&80 (77.7\%)\\
\hline\hline
\end{tabular}

\vspace{2mm}

\parbox{\textwidth}{\footnotesize
\textit{Notes:} The sample contains 103 subjects, each facing 37 lottery-pair choice problems in Experiment~1 of \cite{AlosFerrerBuckenmaierGaragnaniRustichini2026}. ``Max.\ \# deleted'' denotes the maximum number of choice problems that may be deleted when determining whether a subject passes the corresponding rationalizability test. FOSD and SOSD denote the two stochastic-dominance requirements imposed on the rationalizing preference. ``Unrestricted'' imposes no additional restriction on the rationalizing preference, whereas ``Convex'' additionally requires the preference to be convex. Each entry reports the number of subjects, out of 103, who pass the corresponding test; percentages are relative to the full sample.
}
\end{table}

\section*{OA-2. Omitted Proofs and Examples}
\begin{proof}[Proof of Corollary \ref{coro:porder:pure_strict}]
By (\ref{eq:independence}), we have $H=G_P$ and $$P^{*} = \{(x,y) \in H \times X: x P y \text{ or } x P \diamond \rhd y \text{ or } x \rhd y\},$$ with the revised order pair being $q = (\trianglerighteq^*, P^{*})$. By independence of $\trianglerighteq^*$ and $\rhd$,  $$x \rhd y \text{~and~} y \trianglerighteq^* z ~\Longrightarrow~ \frac{1}{2}y + \frac{1}{2}x \trianglerighteq^* \frac{1}{2}z + \frac{1}{2}x \rhd \frac{1}{2}z + \frac{1}{2}y.$$ By the compatibility of $(\trianglerighteq^*, \rhd)$, we have $\frac{1}{2}y + \frac{1}{2}x  \rhd \frac{1}{2}z + \frac{1}{2}y$, which further implies $x \rhd z$. Thus, both $\trianglerighteq^* \diamond\, \rhd \subseteq \rhd$ and  $ \rhd\, \diamond  \trianglerighteq^*  \subseteq \rhd$ hold. Since $I_X \subseteq \trianglerighteq^*$, it follows that $\trianglerighteq^* \diamond\, \rhd = \rhd$ and  $ \rhd\, \diamond  \trianglerighteq^*  = \rhd$. 
\medskip

For any finite menu $A$, we have
\[
\Pi_{\trianglerighteq^*}(A)
=
\bar{A}:=
\{x \in X: x\trianglerighteq^* y \text{ for some } y \in co(A)\}.
\]  Moreover, since $\rhd\subseteq\,\trianglerighteq^*$, $\trianglerighteq^* \diamond\, \rhd = \rhd$ and  $ \rhd\, \diamond  \trianglerighteq^*  = \rhd$ imply $
P^*\diamond\trianglerighteq^*= (P\diamond\trianglerighteq^*)\cup\rhd. $ 
It follows that $q$ satisfies SP on $A$ if and only if the following set
\begin{equation*}
\begin{split}
& \, \big\{ x \in A: x^{\downarrow}_{P^{*}} \cap \Pi_{\trianglerighteq^*}(A) = \emptyset \big\} \\
= & \, \big\{ x \in A: \{y\in X: x P \diamond \trianglerighteq^* y \text{ or } x \rhd y\} \cap co(A) =\emptyset \big\}
\end{split}
\end{equation*}
is not empty. Therefore, if $p$ admits a convex completion, by Theorem \ref{thm:porder}, condition (\ref{eq:independence:simpler}) holds for every finite menu contained in $H$. \medskip

Conversely, suppose that $q$ violates SP on some finite menu $A \subseteq H$. It suffices to show that condition (\ref{eq:independence:simpler}) fails on some finite menu $B \subseteq H$. By the preceding analysis, for every $x \in A$,
\[
\{y \in X: x P \diamond \trianglerighteq^* y \text{ or } x \rhd y\} \cap co(A) \neq \emptyset.
\]
If there is no $\hat{x}\in A$ such that $\hat{x}\rhd\hat{y}$ for some $\hat{y}\in co(A)$, then $A$ already violates condition (\ref{eq:independence:simpler}), and we are done. Otherwise, consider some such $\hat{x} \in A$. By an argument analogous to the proof of  Lemma \ref{lm:linear:reduction} and the fact that $\rhd \subseteq \trianglerighteq^*$, after removing $\hat{x}$, for every $x \in A\backslash\{\hat{x}\}$ we still have
\[
\{y \in X: x P \diamond \trianglerighteq^* y \text{ or } x \rhd y\}
\cap co(A\backslash\{\hat{x}\}) \neq \emptyset.
\]  Inductively, we can shrink the menu by consecutively removing such $\hat{x}$ and obtain a subset $B \subseteq A$ such that for every $x \in B$,
\[
\{y \in X: x P \diamond \trianglerighteq^* y\} \cap co(B) \neq \emptyset.
\]
The set $B$ cannot be empty: We cannot remove all such alternatives since for any singleton menu $\{z\}$, not $z \rhd z$. Thus, $B$ violates condition (\ref{eq:independence:simpler}).  
\end{proof}

\bigskip

\noindent \textbf{Example on the Comparison with \cite{demuynck2009general}.} This example demonstrates that the characterization condition of \cite{demuynck2009general} is sufficient but not necessary for obtaining convex completions for a given order pair. Let $x$ and $y$ be two distinct alternatives in $X$. Consider an order pair $p=(R,P)$, where  $R=I_X\cup\,\{(x,y),(x,\frac{1}{2}x+\frac{1}{2}y),(y,\frac{1}{2}x+\frac{1}{2}y),(\frac{1}{2}x+\frac{1}{2}y,y)\}$, and  $P$ is the asymmetric part of $R$. By the order pair, $x$ is strictly better than both $y$ and $\frac{1}{2}x+\frac{1}{2}y$, while the latter two are equally good. By Theorem \ref{thm:main}, $p$ admits a convex completion, since $G_P=\{x\}$ and $\Pi_R(\{x\})=\{x\}$ contains neither $y$ nor $\frac{1}{2}x+\frac{1}{2}y$. However, $p$ does not admit a completion that satisfies the stronger property of \cite{demuynck2009general}, since $xPy$ would imply that $\frac{1}{2}x+\frac{1}{2}y$ is strictly better than $y$, which contradicts the condition $(y,\frac{1}{2}x+\frac{1}{2}y) \in R$. Hence, the characterization condition of \cite{demuynck2009general} is sufficient but not necessary for convex completions.\footnote{\,See Section 3.2 of \cite{demuynck2009general} for more details of the characterization condition.}

\bigskip

\begin{figure}[!htb]
		\centering
		\begin{subfigure}{\textwidth}  
			\centering
			\begin{tikzpicture}[scale=12]
				\draw[->] (0, 0) -- (0.78, 0);
				\draw[->] (0, 0) -- (0, 0.76);
				\draw[red, thick] (0,0.4) -- (0.1,0.1);
				\draw[red, thick] (0.4, 0) -- (0.1,0.1);
				\draw[yellow, thick] (0.6, 0) -- (0,0.2);
				\draw[blue, thick] (0.3, 0) -- (0,0.7);
				\draw[green, thick] (0.5, 0) -- (0,0.5);
				\filldraw[black] (0.1, 0.1) circle (0.1pt);
				\node[align=center] at (0.12, 0.12) {$x^3$};
				\filldraw[black] (0.4, 0.1) circle (0.1pt);
				\node[align=center] at (0.42, 0.12) {$x^2$};
				\filldraw[black] (0.3, 0.1) circle (0.1pt);
				\node[align=center] at (0.32, 0.12) {$x^4$};
				\filldraw[black] (0.1, 0.466667) circle (0.1pt);
				\node[align=center] at (0.12, 0.48) {$x^1$};
				\node[align=center] at (0.05, 0.53) {$B^1$};
				\node[align=center] at (0.05, 0.42) {$B^2$};
				\node[align=center] at (0.02, 0.275) {$B^3$};
				\node[align=center] at (0.02, 0.17) {$B^4$};
				\draw[dashed] (0.1, 0.1) -- (0.4,0.1);
				\draw[dashed] (0.1, 0.1) -- (0.1, 0.466667);
				\draw[dashed] (0.4, 0.1) -- (0.1, 0.466667);
			\end{tikzpicture} 
		\end{subfigure} 
		\caption{Simplification of Checking SP}\label{fig_co-convex}
	\end{figure}
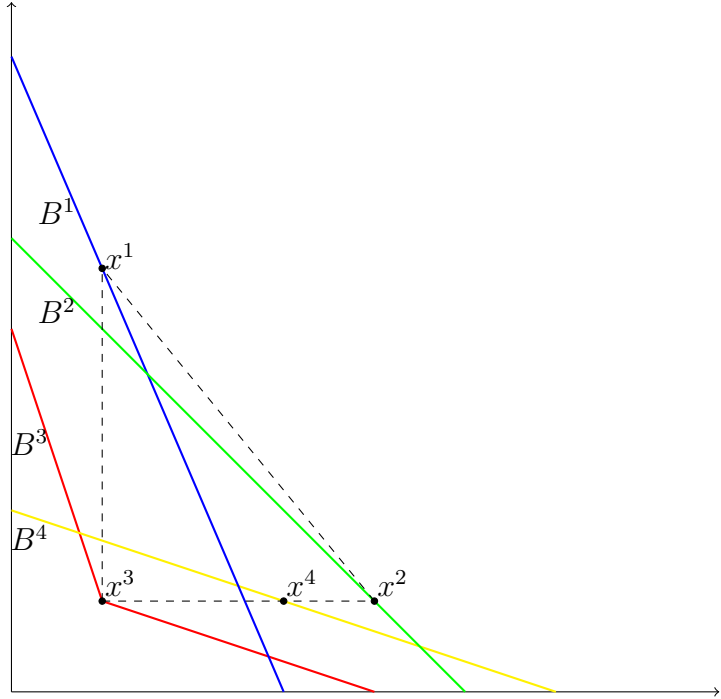

\noindent \textbf{Example on Combining GARP and SP for Testing Convexity.}  This example demonstrates when Theorem \ref{thm:sp=garp} can be applied to simplify the testing for convex preferences even when the pre-convexity condition is only locally satisfied but not globally. 
Let $X=\mathbb{R}^2_+$. Consider a consumption data set $\mathcal{O}=\{(x^t, B^t)\}_{t \in \{1,2,3,4\}}$ depicted in Figure \ref{fig_co-convex}: Each budget line is depicted as a concrete line with the corresponding chosen bundle $x^t$ on it and the budget set $B^t$ containing all bundles on or below it. First, note that $x^1$ is revealed to be strictly preferred to $x^3$ and $x^2$ is revealed to be strictly preferred to both $x^4$ and $x^3$. Hence, GARP  holds. To test whether the data set can be rationalized by a strictly increasing and convex preference, we run Algorithm I for $p=(R_{\mathcal{O}}, P_{\mathcal{O}})$ on menu $\{x^1, x^2, x^3, x^4\}$. Note that $p$ satisfies SP on $\{x^1, x^2, x^3, x^4\}$ due to $x^3$. By Algorithm I, it remains to check SP on submenus of $\{x^1, x^2, x^4\}$. However, since the remaining budget sets are linear, it follows from Theorem \ref{thm:sp=garp} and GARP that SP holds. Thus, the testing procedure is simplified.  \qed

\section*{OA-3. Testing Symmetric Preferences}  This section examines another property---symmetry---that is often imposed on preferences when the dimensions of an alternative have a natural ex ante symmetry. One important example is contingent consumption. In this context, each dimension corresponds to a possible future state of the world. For a given alternative $x$, the value of its $k$-th dimension, $x_k$, represents the consumer's contingent consumption when state $k$ is realized. When the states are regarded symmetrically ex ante, it is natural to require the consumer's preference to be invariant to permutations of the state labels. A similar interpretation arises in social choice. Suppose that $x_k$ represents the consumption or allocation assigned to individual $k$ by a social planner. If the individuals are treated symmetrically and differ only by their labels, then the planner's preference over allocations should likewise be invariant to permutations of individuals. Thus, symmetry captures anonymity with respect to the relevant dimensions, whether they represent states of the world or individuals in an allocation problem. \medskip

Formally, let $X=\mathbb{R}^{n}_{+}$ and $N=\{1,...,n\}$. For any preference $\succsim$ over $X$ and nonempty $\hat{N} \subseteq N$, we say that $\succsim$ is $\hat{N}$-\textit{symmetric} if for any bijection $\phi: \hat{N} \rightarrow \hat{N}$ and $x \in X$, we have $x \sim \phi(x)$, where the bundle $\phi(x)$ is defined such that for all $k \in \hat{N}$, $(\phi(x))_{k}=x_{\phi(k)}$, and for all $k \notin \hat{N}$, $(\phi(x))_{k}=x_{k}$. With an $\hat{N}$-symmetric preference, exchanging payoffs across dimensions in $\hat{N}$ does not affect the desirability of the bundle.  For any collection $\mathcal{S}=\{N^i\}_{i=1}^m$ of mutually disjoint nonempty subsets of $N$, we say that $\succsim$ is $\mathcal{S}$-symmetric if for every $i$, $\succsim$ is $N^i$-symmetric.   \medskip

	For a given consumption data set $\mathcal{O}=\{(x^t, B^t)\}_{t \in T}$ and a collection of mutually disjoint nonempty subsets $\mathcal{S}=\{N^i\}_{i=1}^m$ of $N$, we construct a $\mathcal{S}$-symmetric consumption data set $\mathcal{O}_{\mathcal{S}}$ such that $(x,B)\in\mathcal{O}_{\mathcal{S}}$ if and only if there exist $t\in T$ and, for each $N^i\in\mathcal{S}$, a bijection $\phi_i:N^i\rightarrow N^i$ such that $x=\phi(x^t)$ and $B=\{\phi(y):y\in B^t\}$, where $\phi$ applies $\phi_i$ to the coordinates in each $N^i$ and leaves all other coordinates unchanged. With the notations defined above, the symmetric rationalizability of a given data set can be tested as follows.

\begin{theorem}\label{thm:test:symmetry}
Consider a consumption data set $\mathcal{O}=\{(x^t, B^t)\}_{t \in T}$, with $\{x^t\}_{t \in T}$ being finite, and a collection of mutually disjoint nonempty subsets $\mathcal{S}=\{N^i\}_{i=1}^m$ of $N$. The following statements are equivalent: \smallskip

\emph{(i)} The data set $\mathcal{O}$ has a strictly increasing, convex and $\mathcal{S}$-symmetric rationalization. 

\emph{(ii)} The data set $\mathcal{O}$ has a strictly increasing, quasi-concave, upper semicontinuous and $\mathcal{S}$-symmetric utility rationalization. 
		
\emph{(iii)} The data set $\mathcal{O}_{\mathcal{S}}$ has a strictly increasing and convex rationalization.
		
\emph{(iv)} The data set $\mathcal{O}_{\mathcal{S}}$ has a strictly increasing, convex and $\mathcal{S}$-symmetric rationalization. 
\end{theorem}

By statement (iii) of the theorem, testing reduces to checking whether $(R_{\mathcal{O}_{\mathcal{S}}}, P_{\mathcal{O}_{\mathcal{S}}})$ satisfies SP on every finite menu contained in the set of chosen bundles. Algorithm I can therefore be applied to conduct the test efficiently.

\begin{proof}[Proof of Theorem \ref{thm:test:symmetry}]
It suffices to prove that (iii) implies (iv), and that (iv) implies (ii). We first show that (iii) implies (iv). Let $\Phi_{\mathcal{S}}$ denote the set of permutations that independently permute the coordinates within each $N^i\in\mathcal{S}$ and leave all other coordinates unchanged. Define a symmetric binary relation $\sim_{\mathcal{S}}$ on $X$ such that $x\sim_{\mathcal{S}}y$ if and only if $y=\phi(x)$ for some $\phi\in\Phi_{\mathcal{S}}$. To show that $\mathcal{O}_{\mathcal{S}}$ has a strictly increasing, convex and $\mathcal{S}$-symmetric rationalization, it suffices to show that the order pair $p=(R_{\mathcal{O}_{\mathcal{S}}}\cup\sim_{\mathcal{S}},>_n)$ satisfies SP on every finite menu. \medskip

	Suppose to the contrary that $p$ violates SP on some finite menu $A$. Let $B=\{\phi(x):x\in A,\ \phi\in\Phi_{\mathcal{S}}\}$. Since $\Phi_{\mathcal{S}}$ is finite, $B$ is finite. Moreover, both $B$ and the data set $\mathcal{O}_{\mathcal{S}}$ are $\mathcal{S}$-symmetric. We claim that $p$ violates SP on $B$. Indeed, for every $x\in B$, there exist $y\in A$ and $\phi\in\Phi_{\mathcal{S}}$ such that $x=\phi(y)$. Since $p$ violates SP on $A$, there exists $z\in\Pi_{R_{\mathcal{O}_{\mathcal{S}}}\cup\sim_{\mathcal{S}}}(A)$ such that $y>_n z$. By the $\mathcal{S}$-symmetry of the relation and the fact that $A\subseteq B$, we have $\phi(z)\in\Pi_{R_{\mathcal{O}_{\mathcal{S}}}\cup\sim_{\mathcal{S}}}(B)$, while $\phi(y)>_n\phi(z)$. Hence, $p$ violates SP on $B$. \medskip

	The $\mathcal{S}$-symmetry of $B$ and $\mathcal{O}_{\mathcal{S}}$ also implies that $\Pi_{R_{\mathcal{O}_{\mathcal{S}}}}(B)$ is $\mathcal{S}$-symmetric. Therefore, we have $\Pi_{R_{\mathcal{O}_{\mathcal{S}}}\cup\sim_{\mathcal{S}}}(B)
	=
	\Pi_{R_{\mathcal{O}_{\mathcal{S}}}}(B).
	$  It follows that $q=(R_{\mathcal{O}_{\mathcal{S}}},>_n)$ violates SP on $B$. This contradicts statement (iii). Hence, (iii) implies (iv). \medskip

Finally, we show that (iv) implies (ii). Let $\succsim$ be a strictly increasing, convex and $\mathcal{S}$-symmetric preference that rationalizes $\mathcal{O}_{\mathcal{S}}$. Apply the utility construction in the proof of Theorem \ref{thm:test:consumer} to $\succsim$. The construction preserves $\mathcal{S}$-symmetry, and thus the resulting utility function $u$ is $\mathcal{S}$-symmetric. Hence, $u$ is a strictly increasing, quasi-concave, upper semicontinuous and $\mathcal{S}$-symmetric utility rationalization of $\mathcal{O}$.
\end{proof}

\section*{OA-4. Identification}
We fix an order pair $p=(R, P)$ and assume that $p$ allows for a convex completion throughout this section. Given the assumption that the DM's true preference is convex, we aim to obtain more information regarding the DM's preference.  For this purpose, we have the following definition.

	\begin{definition}
		For any two alternatives $x$ and $y$, $x$ is identified to be weakly preferred to $y$ if for every convex preference $\succsim$ that extends $p$, we have $x \succsim y$, and $x$ is identified to be strictly preferred to $y$ if for every convex preference $\succsim$ that extends $p$, we have $x \succ y$.  
	\end{definition}

	Our notion of ``identified weakly preferred'' and ``identified strictly preferred'' is robust as it requires the identified preference relation to be true for every possible candidate of the DM's true preference. To identify such relations, we follow the approach developed by \cite{varian1982nonparametric}. \medskip

	Specifically, \cite{varian1982nonparametric} considers the choice environment studied by \cite{afriat1967construction}, where each observation consists of a bundle $x^t$ and a positive price vector $p^t$. He poses the following question: for a given bundle $x$ that is not chosen in any budget set of the data set, what is the set of price vectors under which $x$ could possibly be chosen without leading to a contradiction with the existing observed choices? He then suggests to take each price vector $p$, expand the data set to include $(x,p)$, and verify whether the expanded data set passes the test of rationality. \medskip

	In our problem, to determine whether $x$ \textit{must} be weakly preferred to $y$ under every possible convex preference of the DM, we need to exclude the possibility that $y$ is strictly preferred to $x$. To do so, we can include the relation $yPx$ in the strict component of the order pair. It then follows that $x$ is identified as weakly preferred to $y$  if and only if the expanded order pair violates SP on some finite menu. In a similar manner, we can identify whether $x$ must be strictly preferred to $y$. The following theorem gives the  characterization results.
	\medskip

	\begin{theorem}\label{thm:identification}
		Consider an order pair $p$ that admits a convex completion. For any $x,y \in X$, we have the following identification results.  \smallskip
		
		\emph{(i)} $x$ is identified as weakly preferred to $y$ if and only if for some finite menu $A$: $$x \in \Pi_R(A) \text{~and~} \{y\}=\{z \in A: \Pi_R(A) \cap z^{\downarrow}_P = \emptyset\}.$$

		\emph{(ii)} $x$ is identified as strictly preferred to $y$ if and only if for some finite menu $A$: $$y \notin A,\, x \in \Pi_R(A) \text{~and~} \{y\}=\{z \in A\cup \{y\}: \Pi_R(A\cup \{y\}) \cap z^{\downarrow}_P = \emptyset\}.$$
	\end{theorem}

	To understand Theorem \ref{thm:identification}, note that if $y$ is the only alternative in menu $A$ that satisfies $\Pi_R(A) \cap y^{\downarrow}_P = \emptyset$, then it is the sole candidate for the worst alternative in this menu. Hence, based on our analysis in Section \ref{subsec:main}, each alternative in $\Pi_R(A)$ is weakly better than $y$, and each alternative in $A \backslash \{y\}$ is strictly better than $y$, which implies that each alternative in $\Pi_R(A \backslash \{y\})$ is strictly better than $y$. These are precisely the two statements above.

\begin{proof}[Proof of Theorem \ref{thm:identification}]
		Consider an order pair $p=(R, P)$ and assume that it admits a convex completion. Consider any one of its convex completions $\succsim$. First note that for any finite menu $A$ and $y \in A$, if $y$ is the only alternative in $A$ such that $\Pi_R(A) \cap y^{\downarrow}_P = \emptyset$, then we must have for all $z \in \Pi_R(A)$, $z \succsim y$, and for all $z \in A\backslash \{y\}$, $z \succ y$. Thus, for all $z \in \Pi_R(A\backslash \{y\})$, we have $z \succ y$. This proves the ``if'' part of (i) and (ii). \medskip

	To see the ``only if'' part of (i), note that if $x$ is identified as weakly preferred to $y$, then $q=(R, P \cup \, \{(y,x)\})$ must violate SP on some finite menu $A$. Since $p$ satisfies SP on menu $A$, to fail SP for $q$ on $A$, $x$ must be in $\Pi_R(A)$, and $y$ must be the only alternative in $A$ such that $y^{\downarrow}_P \cap \Pi_R(A) = \emptyset$. \medskip

To see the ``only if'' part of (ii), note that if $x$ is identified as strictly preferred to $y$, then $r=(R \cup \, \{(y,x)\}, P)$ must violate SP on some finite menu $A$. Since $p$ satisfies SP on $A$, if $y \in A$ or $x \notin \Pi_R(A)$, then $\Pi_R(A)=\Pi_{R\cup\{(y,x)\}}(A)$, and we have a contradiction. Thus, we have $y \notin A$ and $x \in \Pi_R(A)$. Consider menu $A \cup \{y\}$, and we have $\Pi_R(A\cup \{y\}) = \Pi_{R\cup\{(y,x)\}}(A)$. Since $r$ violates SP on $A$, for each $z \in A$, we have $\Pi_{R\cup\{(y,x)\}}(A) \cap z^{\downarrow}_P = \Pi_R(A\cup \{y\}) \cap z^{\downarrow}_P \neq \emptyset$. Since $p$ satisfies SP on $A \cup \{y\}$, it has to be that $y$ is the only alternative in $A \cup \{y\}$ such that $\Pi_R(A\cup \{y\}) \cap y^{\downarrow}_P = \emptyset$. 
	\end{proof}

\section*{OA-5. Strictly Convex Completion} 
In this section, we provide a characterization of order pairs that allow for strictly convex completions. For this purpose, we introduce the following notations. \medskip

Given a convex set $A \subseteq X$, $x \in A$ is an extreme point of $A$ if we cannot find $z,w \in A\backslash \{x\}$ and $t \in (0,1)$ such that $tz+(1-t)w=x$. Denote by $ex(A)$ the set of all extreme points of $A$.  Given a binary relation $R$, let $Tr(R)$ be the transitive closure of $R$: $x \, Tr(R) \, y$ if there is a finite sequence of alternatives $(x^k)_{k=1}^m$ ($m \geq 2$) such that $x^1 = y$, $x^m = x$, and for all $k\in \{1, \ldots, m-1\}$, $x^{k+1} R x^k$. Given an order pair $p = (R, P)$, for any $x \in X$, let $\Gamma_R(x) = \{y : y \, Tr(R) \, x\}$, and for any  $A \subseteq X$, let $\Gamma_R(A) = \bigcup_{x \in A} \Gamma_R(x)$ and $\Lambda_R(A)=co(\Gamma_R(A))$. Note that $\Gamma_R(A) \subseteq \Lambda_R(A) \subseteq \Pi_R(A)$.

\begin{definition}
	An order pair $p$ satisfies the transitive separation property (TSP) on a finite menu $A$ if there exists $x \in A \cap  ex(\Lambda_R(A))$ such that $\Lambda_R(A) \cap x^{\downarrow}_P = \emptyset.$
\end{definition}

If an order pair $p = (R, P)$ admits a strictly convex completion $\succsim$, then $p$ satisfies TSP on every finite menu. To see this, consider a finite menu $A$. Let $x \in A$ be the $\succsim$-worst alternative. Similar to our analysis in Section \ref{subsec:main}, we can show that the set $\Lambda_R(A)$ must be contained in the upper contour set $\{y: y \succsim x\}$ of $x$ and satisfies $\Lambda_R(A) \cap x^{\downarrow}_P = \emptyset$. By the strict convexity of $\succsim$, $x$ is an extreme point of $\{y: y \succsim x\}$ and, thus, an extreme point of $\Lambda_R(A)$.\footnote{\,Suppose to the contrary that $x = tz + (1-t)w$ for some $t \in (0,1)$ and $z, w \in \{y: y \succsim x\}$ with $x \notin \{z, w\}$. By the strict convexity of $\succsim$, we have either $x \succ z$ or $x \succ w$, leading to a contradiction.}  Our next theorem asserts that the converse direction is also true.

\begin{theorem}\label{thm:sconvex}
An  order pair $p$ admits a strictly convex completion if and only if it satisfies TSP on every finite menu. 
\end{theorem}

\begin{proof}[Proof of Theorem \ref{thm:sconvex}]
	It remains to prove the sufficiency. Let an order pair $p=(R, P)$ satisfy TSP on every finite menu. We say that $p$ satisfies the strong separating property (SSP) on a finite menu $A$ if there exists $x \in A$ such that $x \in ex(\Pi_R(A))$ and $x^{\downarrow}_P \cap \Pi_R(A) = \emptyset$. \medskip

	We first show that $p$ satisfies SSP on every finite menu. Suppose to the contrary that $p$ violates SSP on a finite menu $A$. Two facts regarding extreme points are worth noting. 
	First, if $x$ is an extreme point of a convex set $C$, then for every convex set $D \subseteq C$ with $x \in D$, $x$ remains to be an extreme point of $D$. Second, if $x \in A$ is not an extreme point of $\Lambda_R(A)$, then we can find two distinct points $z,w \in \Gamma_R(A)$ and $t\in (0,1)$ such that $x=tz+(1-t)w$. By the definition of $\Gamma_R(A)$, there exist two finite sequences $(z^k)_{k=1}^n$ and $(w^k)_{k=1}^n$ such that $z^1,w^1 \in A$, $z^n=z$, $w^n=w$, and for all $k\in \{1,...,n-1\}$, $z^{k+1}R z^k$ and $w^{k+1}R w^k$. By the two facts,  following a similar proof to that of Lemma \ref{lm:finite1}, we can find a finite sequence of finite menus $\{A^k\}_{k=0}^n$ such that (i) $A^0=A$, (ii) for all $l \in \{0,...,n-1\}$ and $x \in A^{l+1}$,  $x \notin co(\cup_{k=0}^l A^k)$ and there exists $\theta(x) \in co(\cup_{k=0}^l A^k)$ with $x R \theta(x)$, and (iii) for every $x \in A \cap ex(co(\cup_{k=0}^n A^k))$, there exists $z \in co(\cup_{k=0}^n A^k)$ such that $x P z$. Let $\bar{A}$ consist of all alternatives in $A$ and every $\theta(x)$ that is not in $ex(co(\cup_{k=0}^n A^k))$. A simple induction implies that $\cup_{k=0}^n A^k \subseteq \Gamma_R(\bar{A})$. It then follows from the construction of $\bar{A}$ that for all $x \in \bar{A} \cap ex(\Lambda_R(\bar{A}))$, we have $x \in A \cap ex(co(\cup_{k=0}^n A^k))$. For such an alternative $x$, by (iii) and the fact that $co(\cup_{k=0}^n A^k) \subseteq \Lambda_R(\bar{A})$, we can find  $z \in \Lambda_R(\bar{A})$ such that $x P z$. Therefore, $p$ violates TSP on $\bar{A}$, a contradiction. \medskip

	The remainder of the proof is based on the fact that $p$ satisfies SSP on every finite menu. For a given finite menu $A$, we can construct a strictly decreasing sequence $(A^k)_{k=0}^n$ of submenus of $A$ such that $A^0=A$, $A^n=\emptyset$, and for all $0\le k,k+1 \le n$, $$A^{k+1}=A^k\backslash \{x \in A^k: x\in ex(\Pi_R(A^k)) \text{~and~} x^{\downarrow}_P \cap \Pi_R(A^k)=\emptyset\}.$$  Following a proof similar to that of Theorem \ref{thm:algorithm}, we can show that if $p$ satisfies SSP on every menu in $A$, then there is a preference $\succsim$ on $X$ such that (i) for all $x,y \in X$, $x R y$ implies $x \succsim y$, (ii) for all $x \in A$ and $y \in X$, $x P y$ implies $x \succ y$, and (iii) for all $x \in A$, $x$ is an extreme point of the set $\{y: y \succsim x\}$.  The remaining proof is similar to that of Theorem \ref{thm:main}. Let $\mathcal{R}$ be the set of all order pairs that extend $p$ and satisfy SSP on every finite menu. Endowed with the extension order $\subseteq$, $(\mathcal{R}, \subseteq)$ is a partially ordered set. We can then show that $(\mathcal{R}, \subseteq)$ has a maximal element $p^*=(R^*, P^*)$ such that $R^*$ is a convex preference, and $P^*=(R^*)^a$.\footnote{\,The only difference is to transfer the condition about the extreme points to some ``finite'' statement. Note that if $x\in A$ is not an extreme point of $\Pi_R(A)$, then we can find distinct $z,w \in \Pi_R(A)$ and $t \in (0,1)$ such that $x=tz+(1-t)w$. Since for some \textit{finite} integer $k$, we have $z,w\in \Pi_R^k(A)$, we can show that $x$ is not an extreme point of $\Pi_R^k(A)$. It then follows that the arguments we use to prove Lemmas \ref{lm:finite1} and \ref{lm:chain1} can be similarly applied here to show that $(\mathcal{R}, \subseteq)$ has a maximal element.} To see that $R^*$ is strictly convex, suppose to the contrary that for some distinct $x,y \in X$,  $x R^* y R^* tx+(1-t)y$ for some $t \in (0,1)$.  It follows that $tx+(1-t)y$ is not an extreme point of its upper contour set $\{z: z R^* tx+(1-t)y\}$, contradicting the fact that $p^*$ satisfies SSP on $\{tx+(1-t)y\}$.  
\end{proof}

In general, the condition in Theorem \ref{thm:sconvex} cannot be replaced by the requirement that the order pair satisfy TSP on every finite menu contained in $G_P$. Thus, one must either verify TSP on every finite menu or establish it by a general argument. We next consider two special cases in which the collection of menus that must be checked can be reduced, paralleling our results for convex completions. \medskip

The first case is when $p=(I_X, P \cup \rhd^*)$ for some I-partial order $\rhd^*$. By Theorem \ref{thm:sconvex}, $p$ admits a strictly convex completion if and only if for every finite menu $A$,
\begin{equation}\label{eq:strict_convex_strict_order}
\text{there is~}  x \in ex(co(A))  \text{~such that for all~} y \in co(A), \text{~not~} x  P\cup \rhd^*  y.   
\end{equation}
If condition (\ref{eq:strict_convex_strict_order}) fails for a finite menu $A$, then it also fails for its submenu $\hat{A}=ex(co(A))$. Hence, for each $x \in \hat{A}$, there is $y \in co(\hat{A})$ such that either $x P y$ or $x \rhd^* y$. By Lemma \ref{lm:linear:reduction} and a simple induction, we can find a menu $B\subseteq\hat{A}$ such that for every $x\in B$, there exists $z\in co(B)$ for which either $xPz$ or $xP\diamond\rhd^* z$.

\begin{proposition}\label{prop:sconvex:strict}
Given an order pair $p=(I_X, P\cup\rhd^*)$ where $\rhd^*$ is an I-partial order, it admits a strictly convex completion if and only if $q=(I_X, P^{**})$ satisfies TSP on every finite menu contained in $G_P$, where $P^{**}=\{(x,y) \in G_P\times X: x P y \text{~or~} x P \diamond \rhd^* y\}$.     
\end{proposition}

By an argument similar to that used for Algorithm I, determining whether the order pair $p= (I_X,P\cup\rhd^*)$ in Proposition \ref{prop:sconvex:strict} admits a strictly convex completion requires checking TSP only along a decreasing sequence of menus contained in $G_P$. Thus, when $G_P$ is finite, at most $|G_P|$ TSP checks are needed. 
\medskip

To demonstrate the applicability of Proposition \ref{prop:sconvex:strict}, we apply it to the consumer choice setting. Consider a (possibly infinite) data set $\mathcal{O} = \{(x^t, p^t)\}_{t \in T}$ such that for each $t$, $x^t \in \mathbb{R}_+^n$ is the bundle consumed by the consumer and $p^t \in \mathbb{R}_{++}^n$ is the price vector. The data set $\mathcal{O}$ is said to be \textit{strictly rationalizable} if there is a preference $\succsim$ over $\mathbb{R}_+^n$ such that for each $t$, $x^t$ is the \textit{unique} $\succsim$-optimal bundle in the budget set $B^t = \{x \in \mathbb{R}_+^n : x \cdot p^t \leq x^t \cdot p^t\}$. It is shown that $\mathcal{O}$ is strictly rationalizable if and only if it satisfies the Strong Axiom of Revealed Preference (SARP, \citep{houthakker1950revealed}).\footnote{\,In this specific setting, SARP means that for all nonempty and finite sequences of mutually distinct $\{x^{t_s}\}_{s=1}^n$ with $n \geq 2$, if for all $k \in \{1, \ldots, n-1\}$, $x^{t_{k+1}} \cdot p^{t_{k+1}} \geq x^{t_k} \cdot p^{t_{k+1}}$, then $x^{t_1} \cdot p^{t_1} < x^{t_n} \cdot p^{t_1}$.} Notably, SARP implies that for every finite menu $A \subseteq \{x^t\}_{t \in T}$, there is $x \in A$ such that for all $t \in T$ with $x = x^t$, we have $$\forall y\in A \backslash \{x\}, ~ x \cdot p^t < y \cdot p^t.$$  It follows that $x$ is an extreme point of $co(A)$, and for all $t \in T$ with $x = x^t$, the budget set $B^t$ intersects with $co(A)$ at only $x$. Hence, by Proposition \ref{prop:sconvex:strict}, SARP is also sufficient for $\mathcal{O}$ to be strictly rationalized by a strictly increasing and strictly convex preference.\footnote{\,The pair $p=(I_X, P\cup\rhd^*)$ can be defined as $P=\{(x,y):x=x^t \text{ and } y\in B^t \backslash \{x^t\} \text{~for some~} t\in T\}$ and $\rhd^*=>_n$. It follows that $G_P = \{x^t\}_{t \in T}$ and $P \diamond \rhd^* \subseteq\, P$.} \medskip

The second case we consider is when $p=(R,\rhd)$ with $\rhd$ being an I-partial order. We first introduce notations that will be used for stating our result. Recall that $\mathring{G}_R=\{x\in X:xRy\text{ for some }y\neq x\}.$ For every finite menu $A\subseteq \mathring{G}_R$, define
$$\Theta_{\rhd}(A)
=
\left\{
x\in A\cap ex(co(A)):
x_{\rhd}^{\downarrow}\cap co(A)=\emptyset
\right\}.$$ The set $\Theta_{\rhd}(A)$ contains the alternatives in $A$ that are potential candidates for the worst alternative in $co(A)$ when only $\rhd$ is taken into account. With $\Theta_{\rhd}(A)$, we can further use $R$ to identify a set of alternatives in $A$ that cannot be the worst in $co(A)$. Define $$\Omega_p(A)
=
\left\{
x\in\Theta_{\rhd}(A):\,\text{there is~} y \in x_R^{\downarrow} \backslash  \Theta_{\rhd}(A) \text{~with~} y \in co(A) \text{~or~} y_{\rhd}^{\downarrow} \cap co(A) \neq \emptyset
\right\}.$$ Alternatives in $\Omega_p(A)$ cannot be the worst in $A$: Each such alternative is either weakly better than some $y$ in $co(A)\backslash\Theta_{\rhd}(A)$ that cannot be the worst alternative in $co(A)$, or weakly better than some $y$ that is known to be strictly better than an alternative in $co(A)$. Finally, let the restriction of $R$ to $\Theta_{\rhd}(A)$ be $$R_{\rhd\mid A}
=
R\cap
\left(
\Theta_{\rhd}(A)\times\Theta_{\rhd}(A)
\right).$$ With these notations, we state the proposition as follows.

\begin{proposition}\label{prop:sconvex:weak}
Consider an order pair $p=(R,\rhd)$ such that $\rhd$ is an I-partial order. Let
\[
W_R=\{x:yRx\text{ for some }y\neq x\}.
\]
The following four statements are equivalent: \smallskip

\emph{(i)} The order pair $p$ admits a strictly convex completion. \smallskip

\emph{(ii)} The order pair $p$ satisfies TSP on every finite menu contained in $W_R$. \smallskip

\emph{(iii)} For any finite index set $N=\{1,\ldots,n\}$ and $f,g:N\rightarrow X$, if \emph{(a)} $g(N)\subseteq W_R$, \emph{(b)} $f(N)=ex(co(f(N)))$, \emph{(c)} for all $k\in N$, $f(k)Rg(k)$ and $f(k)\neq g(k)$, \emph{(d)} for all $k\in N$ and $x\in co(f(N))$, not $f(k) \rhd x$, and \emph{(e)} for all $k\in N$, there is $y\in co(f(N))$ such that $g(k)=y$ or $g(k)\rhd y$,  then there is a nonempty $Q\subseteq N$ such that $g(Q)\subseteq f(Q)$.\footnote{\,For any $Q\subseteq N$, we adopt the standard notation $f(Q)=\{f(k):k\in Q\}$ and $g(Q)=\{g(k):k\in Q\}$.} \smallskip

\emph{(iv)} For every   finite menu
$A\subseteq\mathring{G}_R$,   \begin{equation}\label{eq:sconvex:simple}
 \text{there is~} x \in \Theta_{\rhd}(A) \text{~such that~} x^{\downarrow}_{Tr(R_{\rhd\mid A})}
\cap
\Omega_p(A)
=
\emptyset.   
\end{equation} 

\end{proposition}

Statement (iii) of Proposition \ref{prop:sconvex:weak} corresponds to the No Extended Dominance axiom of \cite{bossert2009non} (henceforth BS) and reduces to BS's No Strong Inclusion axiom when $\rhd=\emptyset$. While BS restrict attention to the case where $X=\mathbb{R}^n$, $\{(x,y):xRy\text{ and }x\neq y\}$ is finite, and $\rhd$ is the strict Euclidean order, Proposition \ref{prop:sconvex:weak} demonstrates that their characterization for strictly convex completions continues to hold in the infinite case and for more general convex domains and I-partial orders. \medskip

Statement (ii) simplifies the condition in Theorem \ref{thm:sconvex} by requiring TSP to be checked only on finite menus contained in $W_R$. When $W_R$ is finite, at most $|W_R|$ rounds of TSP checks are needed to determine whether $p$ admits a strictly convex completion. \medskip

Statement (iv) provides another simplification that is particularly useful when the primitive weak preference information $R$ is derived from the DM's choices in finitely many choice problems, as it only requires checking menus contained in $\mathring{G}_R$. Importantly, statement (iv) does not assert
that it suffices to check TSP on finite menus contained in $\mathring{G}_R$.\footnote{\,This distinction is necessary. Consider four distinct alternatives $x,y,z,$ and $w$ with $x=\frac34z+\frac14w$ and $y=\frac14z+\frac34w$. Let $R=I_X\cup\{(z,x),(w,y)\}$ and $\rhd=\emptyset$. TSP fails on $\{x,y\}$, although $\{x,y\}$ is not contained in $\mathring{G}_R=\{z,w\}$.}   Rather, it provides an alternative method for verifying TSP. When
$\mathring{G}_R$ is finite, it suffices to check
(\ref{eq:sconvex:simple}) for a decreasing sequence of menus in
$\mathring{G}_R$. To see this, let $x\in B\subseteq A\subseteq
\mathring{G}_R$ and suppose that $x$ satisfies
(\ref{eq:sconvex:simple}) for $A$. Clearly,
$x\in\Theta_{\rhd}(B)$. We claim that
\[
x^{\downarrow}_{Tr(R_{\rhd\mid B})}\cap\Omega_p(B)=\emptyset.
\]
Consider any sequence $x=x^1,\ldots,x^m$ such that
$x^iR_{\rhd\mid B}x^{i+1}$ for every $i$. We can show inductively that
$x^i\in\Theta_{\rhd}(A)$ for every $i$. If
$x^i\in\Theta_{\rhd}(A)$ but
$x^{i+1}\notin\Theta_{\rhd}(A)$, then by
$x^{i+1}\in B\subseteq co(A)$ and $x^iRx^{i+1}$, we have
$x^i\in\Omega_p(A)$, contradicting
(\ref{eq:sconvex:simple}). Thus
$x^m\in\Theta_{\rhd}(A)$ and $x^m\notin\Omega_p(A)$. Now, suppose to the contrary that $x^m\in\Omega_p(B)$. Then there is
$y\in (x^m)^{\downarrow}_R\backslash\Theta_{\rhd}(B)$ such that
$y\in co(B)$ or
$y_{\rhd}^{\downarrow}\cap co(B)\neq\emptyset$. Since
$x^m\notin\Omega_p(A)$, we must have
$y\in\Theta_{\rhd}(A)$. If $y\in co(B)$, the fact that
$y\in ex(co(A))$ implies $y\in B$ and hence
$y\in\Theta_{\rhd}(B)$, a contradiction. If
$y_{\rhd}^{\downarrow}\cap co(B)\neq\emptyset$, then
$y_{\rhd}^{\downarrow}\cap co(A)\neq\emptyset$, again contradicting
$y\in\Theta_{\rhd}(A)$. This proves the claim. \medskip 

Given the above claim, we can start with $A^0=\mathring{G}_R$. Whenever $A^k$ is nonempty,
choose an alternative $x^k$ satisfying
(\ref{eq:sconvex:simple}) and let
$A^{k+1}=A^k\backslash\{x^k\}$. The preceding argument implies that
any menu that violates (\ref{eq:sconvex:simple}) cannot contain
$x^k$, and hence must be contained in $A^{k+1}$. Thus, a similar
argument to that of Algorithm I shows that statement (iv) can
be verified with at most $|\mathring{G}_R|$ checks.

\begin{proof}[Proof of Proposition \ref{prop:sconvex:weak}]
\noindent (i) $\Rightarrow$ (iii). Let $f,g$ and $N$ satisfy conditions (a)--(e) in statement (iii). Let $\succsim$ be a strictly convex preference that extends $(R,\rhd)$. We want to show that there is a nonempty $Q\subseteq N$ such that $g(Q)\subseteq f(Q)$. \medskip

Define $Q\subseteq N$ such that $k\in Q$ if and only if for all $r\in N$, $f(r)\succsim f(k)$. For each $k\in Q$, since $f(k)\succsim g(k)$, transitivity and convexity imply that every $y\in co(f(N))$ satisfies $y\succsim g(k)$. By condition (e), it follows that $g(k)\in co(f(N))$: Otherwise $g(k) \rhd y$ for some $y\in co(f(N))$, so $g(k)\succ y\succsim g(k)$, a contradiction. By the strict convexity of $\succsim$, for all $k\in Q$ and $y\in co(f(N))\backslash f(Q)$, we have $y\succ f(k)$. Hence $g(k)\in f(Q)$ for every $k\in Q$, and therefore $g(Q)\subseteq f(Q)$. \bigskip

\noindent (ii) $\Rightarrow$ (i). It suffices to show that when $p$ violates TSP on some finite menu $A$, then it violates TSP on some finite menu $B\subseteq W_R$. Consider such a menu $A$. Note that $p$ violates TSP on $A$ if and only if $A$ satisfies the following property:
\begin{equation}\label{eq:prop2}
\exists\,\text{finite }D\subseteq\Gamma_R(A):
A\subseteq D
\text{ and }
x\in A\cap ex(co(D))
\text{ implies }
x^{\downarrow}_\rhd \cap co(D)\neq\emptyset.
\end{equation}
If $A=\{x\}$, then we must have $\{x\}\subsetneq\Gamma_R(\{x\})$. Thus $x\in W_R$. By letting $B=\{x\}$, we are done.\medskip

When $|A|\geq2$ and $x\notin W_R$ for some $x\in A$, let $A^*=A\backslash\{x\}$. Pick $D$ that satisfies condition \eqref{eq:prop2}, and let $D^*=D\backslash\{x\}$. Since $x\notin W_R$, we have $D^*\subseteq\Gamma_R(A^*)$. If $x\notin ex(co(D))$, then $co(D)=co(D^*)$, and $A^*$ satisfies condition \eqref{eq:prop2}. If $x\in ex(co(D))$, consider any $y\in A^*\cap ex(co(D^*))$. If $y\in ex(co(D))$, then since $x^{\downarrow}_\rhd \cap co(D)\neq\emptyset$ and $y^{\downarrow}_\rhd \cap co(D)\neq\emptyset$, by Lemma \ref{lm:linear:reduction}, we have $y^{\downarrow}_\rhd \cap co(D^*)\neq\emptyset$. If $y\notin ex(co(D))$, we can find some $z\in co(D^*)$ and $t\in(0,1)$ such that $y=tz+(1-t)x$. It follows from Lemma \ref{lm:linear:reduction} and the independence of $\rhd$ that $y^{\downarrow}_\rhd \cap co(D^*)\neq\emptyset$. By induction, when $|A|\geq2$, we can find some finite menu $B\subseteq A$ with $B\subseteq W_R$ that satisfies condition \eqref{eq:prop2}, i.e., $p$ violates TSP on $B$. \bigskip

\noindent (iii) $\Rightarrow$ (ii). Assume to the contrary that there exists a finite menu $A\subseteq W_R$ such that, for all $x\in A\cap ex(\Lambda_R(A))$, $x^{\downarrow}_\rhd \cap\Lambda_R(A)\neq\emptyset$. Consider a finite menu $D$ with $A\subseteq D\subseteq\Gamma_R(A)$ such that, for all $x\in A\cap ex(co(D))$, $x^{\downarrow}_\rhd \cap co(D)\neq\emptyset$, and for all $y\in D\backslash A$, there is a sequence $(x^k)_{k=1}^m$ with $m\geq2$ such that $x^1\in A$, $x^m=y$, and $x^{k+1}Rx^k$ for all $k\in\{1,\ldots,m-1\}$. \medskip

Let $B=ex(co(D))$, so $co(B)=co(D)$. Let
\[
B^*=\{x\in B:x^{\downarrow}_\rhd \cap co(B)=\emptyset\}.
\]
By Lemma \ref{lm:linear:reduction}, we have $B^*\neq\emptyset$. It follows that $A\cap B^*=\emptyset$. Enumerate $B^*$ as $B^*=\{x^1,\ldots,x^n\}$. Let $N=\{1,\ldots,n\}$ and define $f:N\rightarrow X$ by $f(k)=x^k$. For each $k\in N$, find a finite sequence of alternatives $(x^{k,i})_{i=1}^{l_k}$ with $l_k\geq2$ such that $x^{k,1}\in A$, $x^{k,l_k}=x^k$, and $x^{k,i+1}Rx^{k,i}$ for all $i\in\{1,\ldots,l_k-1\}$. We further require that the sequence $(x^{k,i})_{i=1}^{l_k}$ is minimal, in the sense that no strictly shorter sequence has the same property. For each $k\in N$, let $g(k)=x^{k,l_k-1}$. The minimality of $(x^{k,i})_{i=1}^{l_k}$ implies $f(k)Rg(k)$ and $f(k)\neq g(k)$, so condition (c) holds. \medskip

We verify the other conditions in statement (iii). Since $f(k)Rg(k)$ and $f(k)\neq g(k)$ for every $k\in N$, we have $g(N)\subseteq W_R$, so condition (a) holds. Condition (b) holds since $B=ex(co(B))$ and $B^*\subseteq B$. By the definition of $B^*$, condition (d) holds. Finally, note that $g(k)\in co(B)$ for all $k\in N$. Either $g(k)\in co(B^*)$ or there are $t\in(0,1]$, $z\in co(B\backslash B^*)$ and $w\in co(B^*)$ such that $g(k)=tz+(1-t)w$. In the latter case, by Lemma \ref{lm:linear:reduction}, there is $w^*\in co(B^*)$ with $g(k) \rhd w^*$. Thus condition (e) holds.\medskip

To derive a contradiction, suppose that for some nonempty $Q\subseteq N$, $g(Q)\subseteq f(Q)$. Pick $k\in Q$ such that $l_k\leq l_{k'}$ for all $k'\in Q$. Since $x^{k,l_k-1}\in f(Q)$, there is $k'\in Q$ such that $x^{k'}=x^{k,l_k-1}$. The minimality of $(x^{k',i})_{i=1}^{l_{k'}}$ implies $l_{k'}\leq l_k-1$, contradicting $l_k\leq l_{k'}$. \medskip



Now consider any nonempty $Q\subseteq N$, and choose $j=  \min Q$. It follows that $g(j)=y^{i_j}$ is either in $X\backslash A$ or in $\{x^{i_1},...,x^{i_{j-1}}\}$. Hence, $ g(j) \notin f(Q)$, and thus $g(Q)\not\subseteq f(Q)$, contradicting statement (iii).  \bigskip

\noindent (i) $\Rightarrow$ (iv). Let $\succsim$ be a strictly convex completion of $p$. We have argued that for any finite menu $A \subseteq \mathring{G}_R$, the $\succsim$-worst alternative in $A$ must be in $\Theta_{\rhd}(A)$ and cannot be in $\Omega_p(A)$. Thus, if $x$ is the $\succsim$-worst alternative in $A$, then there is no $y \in \Omega_p(A)$ such that $x Tr(R_{\rhd \mid A}) y$, that is, (\ref{eq:sconvex:simple}) holds. \bigskip


\noindent (iv) $\Rightarrow$ (iii).
Let $f,g$ and $N$ satisfy conditions (a)--(e) in statement (iii), and let
$A=f(N)$. By condition (c), $A\subseteq \mathring{G}_R$. Conditions
(b) and (d) imply that $\Theta_{\rhd}(A)=A.$ 
By statement (iv), choose $x$ that satisfies (\ref{eq:sconvex:simple}) and let $B=x^{\downarrow}_{Tr(R_{\rhd\mid A})}.$ Since $R$ is reflexive, $B$ is nonempty and $B\subseteq A$. We first show that for all $y \in B$, $$\Phi_p(y, A):=  \left\{
z\in y_R^{\downarrow}\backslash\{y\}:
z\in co(A)
\text{~or~}
z_{\rhd}^{\downarrow}\cap co(A)\neq\emptyset
\right\}$$ is a subset of $B$. Take
$y\in B$ and $z\in\Phi_p(y,A)$. If
$z\notin A=\Theta_{\rhd}(A)$, then
$y\in\Omega_p(A)$, contradicting $y\in x^{\downarrow}_{Tr(R_{\rhd\mid A})}$ and  $x^{\downarrow}_{Tr(R_{\rhd\mid A})}
\cap\Omega_p(A)=\emptyset.$  Hence $z\in A=\Theta_{\rhd}(A)$. Since $yRz$, we have
$y\,R_{\rhd\mid A}\,z$, and therefore $x\,Tr(R_{\rhd\mid A})\,z.$ Thus $z\in B$, and hence $\Phi_p(y,A)\subseteq B$ for every $y\in B.$  Let
$Q=\{k\in N:f(k)\in B\}.$
Since $B\neq\emptyset$ and $B\subseteq A=f(N)$, we have
$Q\neq\emptyset$ and $f(Q)=B$. For every $k\in Q$, conditions
(c) and (e) imply $g(k)\in\Phi_p(f(k),A).$ Since $f(k)\in B$ and $\Phi_p(f(k),A)\subseteq B$, it follows that $g(k)\in B=f(Q).$ 
Therefore $g(Q)\subseteq f(Q),$ which proves statement (iii). \end{proof}

\end{document}